\documentclass[preprint]{techrepelsarticle}

\usepackage{url}
\usepackage{amsfonts}
\usepackage{algorithm}
\usepackage{algorithmic}
\usepackage{xspace}
\usepackage{epstopdf}
\usepackage{amstext}
\usepackage{amssymb}
\usepackage{amsmath}
\usepackage{fancyvrb}
\usepackage{ifmtarg} 
\usepackage{wrapfig}
\usepackage{lscape}
\usepackage{gensymb}
\usepackage{rotating}
\usepackage{times}

\usepackage[english]{babel}

\newtheorem{example}{Example}
\newtheorem{definition}{Definition}

\newtheorem{theorem}{Theorem}
\newtheorem{lemma}{Lemma}
\newtheorem{proposition}{Proposition}

\newproof{pf}{Proof}

\usepackage{ifthen}
\newcommand{\brifnotempty}[1]{\ifthenelse{\equal{#1}{}}{}{ \br{#1}}}
\newenvironment{lemma*}[2][]
	{\pagebreak[2] \par \noindent \textbf{Lemma~\ref{#2}\brifnotempty{#1}.}\it}{\par}
\newenvironment{theorem*}[2][]
	{\pagebreak[2] \par \noindent \textbf{Theorem~\ref{#2}\brifnotempty{#1}.}\it}{\par}
\newenvironment{proposition*}[2][]
	{\pagebreak[2] \par \noindent \textbf{Proposition~\ref{#2}\brifnotempty{#1}.}\it}{\par}
\newenvironment{corollary*}[2][]
	{\pagebreak[2] \par \noindent \textbf{Corollary~\ref{#2}\brifnotempty{#1}.}\it}{\par}

\newcommand{\logic}[0]{\ensuremath{\mathit{L}}\xspace}

\newcommand{\SKB}[0]{\ensuremath{\mathit{KB}}\xspace}
\newcommand{\TKB}[0]{\ensuremath{\mathsf{KB}}\xspace}
\newcommand{\KB}[0]{\ensuremath{\mathit{kb}}\xspace}
\newcommand{\KBStr}[0]{\ensuremath{\mathcal{KB}}\xspace}

\newcommand{\ConForM}[1]{\ensuremath{{\sf Con}_{#1}}\xspace}

\newcommand{\SBelS}[0]{\ensuremath{\mathit{BS}}\xspace}
\newcommand{\BelS}[0]{\ensuremath{B}\xspace}
\newcommand{\TBelS}[0]{\ensuremath{{\sf B}}\xspace}
\newcommand{\Bel}[0]{\ensuremath{\mathit{b}}\xspace}

\newcommand{\BelForM}[1]{\ensuremath{{\sf Bel}_{#1}}\xspace}

\newcommand{\acc}[0]{\ensuremath{\mathbf{acc}}\xspace}

\newcommand{\SOp}[1][OP]{\ensuremath{\mathit{#1}}\xspace}
\newcommand{\Op}[1][op]{\ensuremath{\mathit{#1}}\xspace}
\newcommand{\EOp}[0]{\ensuremath{\mathbf{op}}\xspace}
\makeatletter
\newcommand{\Nxt}[1]{
  \@ifmtarg{#1}{
  \ensuremath{\mathbf{next}}\xspace}{\ensuremath{\mathbf{next}(#1)}\xspace}
}
\makeatother

\newcommand{\mng}[0]{\ensuremath{\mathbf{mng}}\xspace}

\newcommand{\Ctxt}[0]{\ensuremath{\mathit{C}}\xspace}
\newcommand{\TCtxt}[0]{\ensuremath{\mathsf{C}}\xspace}
\newcommand{\CtxtIn}[0]{\ensuremath{n}\xspace}

\newcommand{\TIL}[0]{\ensuremath{\mathsf{IL}}\xspace}
\newcommand{\IL}[0]{\ensuremath{\mathit{IL}}\xspace}
\newcommand{\ILIn}[0]{\ensuremath{k}\xspace}

\newcommand{\naf}[0]{\ensuremath{\mathbf{not}}\xspace}
\newcommand{\BRHd}[1]{\ensuremath{\mathrm{hd}({#1})}\xspace}
\newcommand{\BRBd}[1]{\ensuremath{\mathrm{bd}({#1})}\xspace}
\newcommand{\BRBA}[2]{\ensuremath{#1{:}#2}\xspace}
\newcommand{\BRSA}[2]{\ensuremath{#1{::}#2}\xspace}

\newcommand{\rMCS}[0]{rMCS\xspace}
\newcommand{\rMCSs}[0]{rMCSs\xspace}

\newcommand{\aMCS}[0]{aMCS\xspace}
\newcommand{\aMCSs}[0]{aMCSs\xspace}

\newcommand{\TBR}[0]{\ensuremath{\mathsf{BR}}\xspace}
\newcommand{\SBR}[0]{\ensuremath{\mathit{BR}}\xspace}

\newcommand{\inpt}[0]{\ensuremath{{I}}\xspace}

\newcommand{\Tinpt}[0]{\ensuremath{{\sf I}}\xspace}

\newcommand{\inptForM}[1]{\ensuremath{{\sf Inp}_{#1}}\xspace}

\newcommand{\inptStr}[0]{\ensuremath{\mathcal{I}}\xspace}

\newcommand{\app}[0]{\ensuremath{\mathbf{app}}\xspace}

\newcommand{\upd}[0]{\ensuremath{\mathbf{upd}}\xspace}

\newcommand{\EqStr}[0]{\ensuremath{\mathcal{B}}\xspace}

\newcommand{\Patient}[0]{Dave\xspace}

\newcommand{\term}[1]{\ensuremath{\mathit{#1}}\xspace}

\newcommand{\pred}[1]{\ensuremath{\mathrm{#1}}\xspace}

\newcommand{\lit}[2]{\ensuremath{\pred{#1}(\term{#2})}\xspace}

\newcommand{\op}[1]{\ensuremath{\mathsf{#1}}\xspace}

  \newcommand{\problemA}[0]{\ensuremath{Q^\forall}\xspace}
  \newcommand{\problemE}[0]{\ensuremath{Q^\exists}\xspace}

  \newcommand{\pref}[0]{\ensuremath{\prec}\xspace}

  \newcommand{\cons}[1]{\ensuremath{cons(#1)}\xspace}
  
  \newcommand{\addCons}[0]{\ensuremath{\op{addC}}\xspace}

   \newcommand{\redfun}{\mathbf{red}}
   
   \newcommand{\geql}{\mathbf{GE}}
   
   \newcommand{\wfs}{\mathsf{WF}}
   \newcommand{\wfsStr}{\mathcal{WF}}

   \newcommand{\RBS}[0]{\ensuremath{RB}\xspace}

\newcommand{\Repair}[0]{\ensuremath{\mathcal{R}}\xspace}

\newcommand{\define}[1]{\emph{#1}}

\newcommand{\ra}{\rightarrow}
\newcommand{\la}{\leftarrow}

\newcommand{\tuple}[1]{\ensuremath{\tupleLeft {#1} \tupleRight}\xspace}
\newcommand{\tupleLeft}[0]{\ensuremath{\langle}\xspace}
\newcommand{\tupleRight}[0]{\ensuremath{\rangle}\xspace}
\newcommand{\set}[1]{\{#1\}}

\newcommand{\iec}[0]{i.e.,\xspace}
\newcommand{\cf}[0]{cf.\xspace}
\newcommand{\egc}[0]{e.g.,\xspace}
\newcommand{\wrt}[0]{with respect to\xspace}

\newcommand{\bi}[0]{\begin{itemize}}
\newcommand{\ei}[0]{\end{itemize}}
\newcommand{\be}[0]{\begin{enumerate}}
\newcommand{\ee}[0]{\end{enumerate}}

\newcommand\vartextvisiblespace[1][.5em]{
  \makebox[#1]{
    \kern.07em
    \vrule height.3ex
    \hrulefill
    \vrule height.3ex
    \kern.07em
  }
}

\newcommand{\TMsym}[0]{\ensuremath{\mathbb{S}}\xspace}
\newcommand{\TMstates}[0]{\ensuremath{\mathbb{Q}}\xspace}
\newcommand{\TMblank}[0]{\mathtt{\vartextvisiblespace}\xspace}
\newcommand{\TMalph}[0]{\ensuremath{\Gamma}\xspace}
\newcommand{\TMinput}[0]{\ensuremath{\Sigma}\xspace}
\newcommand{\TMfunc}[0]{\ensuremath{\delta}\xspace}
\newcommand{\MTM}[0]{\ensuremath{M_{\mathrm{TM}}}\xspace}

\newcommand{\larsAlph}[0]{\ensuremath{\mathcal{A}}\xspace}

\newcommand{\abox}[0]{\ensuremath{\mathcal{A}}\xspace}

\newcommand{\m}[1]{
  \begingroup
  \def\FV@Space{ }
  \mathcode`\_="8000 
  \dous 
  $#1$
  \endgroup
}
\newcommand{\dous}{
  \begingroup\lccode`~=`\_ \lowercase{\endgroup\let~}\sb
}

\begin{document}

\begin{frontmatter}
\title{Reactive Multi-Context Systems: \\ Heterogeneous Reasoning in Dynamic Environments}

\author[leipzig]{Gerhard~Brewka}
\ead{brewka@informatik.uni-leipzig.de}

\author[leipzig]{Stefan~Ellmauthaler}
\ead{ellmauthaler@informatik.uni-leipzig.de}

\author[lisbon]{Ricardo~Gon\c{c}alves}
\ead{rjrg@fct.unl.pt}

\author[lisbon]{Matthias~Knorr\corref{cor1}}
\ead{mkn@fct.unl.pt}

\author[lisbon]{Jo\~{a}o~Leite}
\ead{jleite@fct.unl.pt}

\author[leipzig]{J\"org~P\"uhrer\corref{cor1}}
\ead{jpuehrer@dbai.tuwien.ac.at}

\address[leipzig]{Institute of Computer Science, Leipzig University, Germany}
\address[lisbon]{NOVA LINCS \& Departamento de Inform\'{a}tica,
Universidade NOVA de Lisboa, Portugal}

  \begin{abstract}
Managed multi-context systems (mMCSs) allow for the integration of heterogeneous knowledge sources in a modular and very general way.
They were, however, mainly designed for static scenarios and are therefore not well-suited for dynamic environments in which continuous reasoning over such heterogeneous knowledge with constantly arriving streams of data is necessary.
In this paper, we introduce reactive multi-context systems (rMCSs), a framework for reactive reasoning in the presence of heterogeneous knowledge sources and data streams. 
We show that rMCSs are indeed well-suited for this purpose by illustrating how several typical problems arising in the context of stream reasoning can be handled using them, by showing how inconsistencies possibly occurring in the integration of multiple knowledge sources can be handled, and by arguing that the potential non-determinism of rMCSs can be avoided if needed using an alternative, more skeptical well-founded semantics instead with beneficial computational properties. 
We also investigate the computational complexity of various reasoning problems related to rMCSs.
Finally, we discuss related work, and show that rMCSs do not only generalize mMCSs to dynamic settings, but also capture/extend relevant approaches w.r.t.\ dynamics in knowledge representation and stream reasoning. 

  \end{abstract}
  \begin{keyword}
Heterogeneous knowledge, Stream reasoning, Knowledge integration, Reactive systems, Dynamic systems
  \end{keyword}

\end{frontmatter}

\section{Introduction}

Fueled by initiatives such as the Semantic Web, Linked Open Data, and the Internet of Things, among others,
the wide and increasing availability of machine-processable data and knowledge
has prepared the ground and called for a new class of dynamic, rich, knowledge-intensive applications.
Such new applications require automated reasoning based on the integration of several heterogeneous knowledge bases -- possibly overlapping, independently developed, and written in distinct languages with different semantic assumptions -- together with data/event streams produced by sensors and detectors, to be able to support automation and problem-solving, to enforce traceable and correct decisions, and to facilitate the internalization of relevant dynamic data and knowledge into such heterogeneous knowledge bases.

Consider a scenario where \Patient, an elderly person suffering from dementia, lives alone in an apartment equipped with various sensors, e.g., smoke detectors, cameras, and body sensors measuring relevant body functions (e.g., pulse, blood pressure, etc.). An assisted living application in such a scenario could leverage the information continuously received from the sensors, together with \Patient's medical records stored in a relational database, a biomedical health ontology with information about diseases, their symptoms and treatments, represented in some description logic, some action policy rules represented as a non-monotonic logic program, to name only a few, and use it to detect relevant events, suggest appropriate action, and even raise alarms, while keeping a history of relevant events and \Patient's medical records up to date, thus allowing him to live on his own despite his condition.
After detecting that \Patient left the room while preparing a meal, the system could alert him in case he does not return soon, or even turn the stove off in case it detects that \Patient fell asleep, not wanting to wake him up because his current treatment/health status values rest over immediate nutrition. Naturally, if \Patient is not gone long enough, and no sensor shows any potential problems (smoke, gas, fire, etc.), then the system should seamlessly take no action.

Given the requirements posed by novel applications such as the one just described, the
availability of a vast number of knowledge bases -- written using many different formalisms -- and the relevance of streams of data/events produced by sensors/detectors, modern research in knowledge representation and reasoning faces two fundamental problems: dealing with the \emph{\bf integration} of heterogeneous data and knowledge, and dealing with the \emph{\bf dynamics} of such novel knowledge-based systems.

\paragraph{\bf Integration}

The first problem stems from the availability of knowledge bases written in many different languages and formats developed over the last decades,
from the rather basic ones, such as relational
databases or the more recent triplestores, to the more expressive ones,
such as ontology languages (e.g., description logics), temporal and modal
logics, non-monotonic logics, or logic programs under answer set
semantics, to name just a few. Each of these formalisms was developed
for different purposes and with different design goals in mind.
Whereas some of these formalisms could be combined to form a new, more expressive
formalism, with features from its constituents -- such as dl-programs \cite{DBLP:journals/ai/EiterILST08} and Hybrid MKNF \cite{Motik2010} which, to different extent, combine description logics and logic programs under answer set semantics -- in general this is simply not
feasible, either due to the mismatch between certain
assumptions underlying their semantics, or because of the high price to pay,
often in terms of complexity, sometimes even in terms of decidability.
It is nowadays widely accepted that there simply is no such thing as a single universal,
general purpose knowledge representation language.

What seems to be needed is a principled way of integrating knowledge expressed in different formalisms.
Multi-context systems (MCSs) provide a general framework for this kind of integration. The basic idea underlying MCSs is to leave the diverse
formalisms and knowledge bases untouched, and to use so-called bridge rules to model the flow of information among different parts of the system.
An MCS consists of reasoning units -- called contexts for historical reasons \cite{GiunchigliaS94} -- where each unit is equipped with a collection of bridge rules. In a nutshell, the bridge rules allow contexts to ``listen" to other contexts, that is to take into account beliefs held in other contexts.

Bridge rules are similar to
logic programming rules (including default negation), with an important difference: they provide means
to access other contexts in their bodies. Bridge rules not only allow for a fully declarative specification of the information
flow, but they also allow information to be modified instead of being just passed along as is. Using bridge rules we may
translate a piece of information into the language/format of another context,
pass on an abstraction of the original information, leaving out unnecessary details,
select or hide information, add conclusions to a context based on the absence of information in another one,
and even use simple encodings of preferences among parent contexts.

MCSs went through several development steps until they reached their present form. Advancing
work in \cite{giunchiglia93,mccarthy87} aiming to integrate different inference systems,
monotonic heterogeneous multi-context systems were
defined in \cite{GiunchigliaS94}, with reasoning
within as well as across monotonic contexts. 
The first, still limited attempts to include non-monotonic
reasoning were done in \cite{roeser05,brrose07a},
where default negation in the rules is used to allow
for reasoning based on the {\em absence} of information from a
context.

The non-monotonic MCSs of \cite{BrewkaE07} substantially generalize
previous approaches, by accommodating {\em heterogeneous}\/ and both
{\em monotonic} and {\em non-monotonic}\/ contexts. Hence, they are
capable of integrating, among many others, ``typical'' monotonic logics like
description logics or temporal logics, and non-monotonic formalisms like
Reiter's default logic, logic programs under answer set semantics, circumscription,
defeasible logic, or theories in autoepistemic logic.
The semantics of nonmonotonic MCSs is defined in terms of equilibria: a belief set for each context that is acceptable for its knowledge base augmented by the heads of
its applicable bridge rules.

More recently, the so-called managed MCSs (mMCSs) \cite{BrewkaEFW11} addressed a limitation of MCSs in the way they integrate knowledge between contexts.
Instead of simply \emph{adding} the head of an applicable bridge rule to the context's knowledge
base, which could cause some inconsistency, mMCSs allow for operations other than addition, such as, for instance, \emph{revision} and \emph{deletion}, hence
dealing  with the problem of consistency management within contexts.

\paragraph{\bf Dynamics}

The second problem stems from the shift from static knowledge-based systems that
assume a one-shot computation, usually triggered by a user query,
to open and dynamic scenarios where there is a need to react and evolve in the presence of incoming information.

Indeed, traditional knowledge-based systems -- including the different variants of MCSs mentioned above -- focus entirely on static situations, which is the right thing for applications such as for instance expert systems, configuration or planning problems, where the available background knowledge changes rather slowly, if at all, and where all that is needed is the solution of a new instance of a known problem. However, the new kinds of applications we consider are becoming more and more important, and these require continuous online reasoning, including observing and reacting to events.

There are some examples of systems developed with the purpose of reacting to streams of incoming information, such as Reactive ASP \cite{GebserGKS11,GebserGKOSS12}, C-SPARQL \cite{BarbieriBCVG10}, Ontology Streams \cite{LecueP13} and ETALIS \cite{AnicicRFS12}, to name only a few.
However, they are very limited in the kind of knowledge that can be represented, and the kind of reasoning allowed, hence unsuitable to address the requirements of the applications we envision,
such as those that need to integrate heterogeneous knowledge bases.
Additionally, reacting to the streams of incoming information is only part of the dynamic requirements of our applications. In many cases, the incoming information is processed only once, perhaps requiring complex reasoning using various knowledge bases to infer the right way to react, and does not have to be dealt with again -- e.g., concluding that nothing needs to be done after determining that the tachycardia is caused by the decongestant recently taken by \Patient. In other cases, it is important that these observations not only influence the current reaction of the system -- do nothing in the previous example -- but, at the same time, be able to change the knowledge bases in a more permanent way, i.e., allowing for the internalization of knowledge. For example, relevant observations regarding \Patient's health status should be added to his medical records, such as for example that he had an episode of tachycardia caused by a decongestant, and, in the future, maybe even revise such episode if it is found that \Patient had forgotten to take the decongestant after all.
Other more sophisticated changes in the knowledge bases include, for example, an update to the biomedical health ontology whenever new treatments are found or the revision of the policy rules whenever some exceptions are found. EVOLP~\cite{AlferesBLP02} extends logic programming under answer set semantics with the possibility to specify its evolution, through successive updates, in reaction to external observations. It is nevertheless limited to a single knowledge representation formalism and to a single operation (update).

\paragraph{\bf Our Approach}

In this paper, we aim to address both challenges in a single, uniform approach. We develop a system that allows us to integrate heterogeneous knowledge bases with streams of incoming information and to use them for continuous online reasoning, reacting, and evolving the knowledge bases by internalizing relevant knowledge. In a nutshell, our work follows the multi-context systems tradition, but adds what is needed to also deal with dynamics.

To this end, we introduce \emph{reactive Multi-Context Systems (rMCSs)}. These systems build upon mMCSs and thus provide their functionality for integrating heterogeneous knowledge sources, admitting also operations for manipulating their knowledge bases.
In addition, rMCSs can handle continuous streams of input data. Equilibria remain the fundamental underlying semantic notion, but the focus now lies on the dynamic evolution of the systems. Given an initial configuration of knowledge bases, that is, an initial knowledge base for each context, a specific input stream will lead to a corresponding stream of equilibria, generated by respective updates of the knowledge bases.

Contrary to existing MCSs which possess only one type of bridge rules modeling the information flow to be taken into account when computing equilibria (or the operations that need to be applied in case of mMCSs), rMCSs have an additional, different type of bridge rules, distinguished by the occurrence of the operator \textbf{next} in the head. Whereas the head of a managing bridge rule in mMCS specifies how a specific knowledge base should temporarily be changed for the purpose of determining the current equilibrium, the head of a new kind of bridge rule in rMCS with the operator \textbf{next} in the head specifies how the knowledge base should permanently be changed as the system \emph{moves} to the next state. These new bridge rules allow us to decouple the computation of equilibria from specifying the evolution of knowledge base configurations.

It is often the case that the information \emph{currently} available from some stream needs to be taken into account by some context when computing the semantics at that point. The way such information is taken into account is specified by a bridge rule with a management operation in its head, just as in mMCSs. For example, we may have some default value for some variable stored in a context, but, under some conditions ($\pred{cond1}$), prefer to use the value that is read from some sensor (observed in some stream) whenever one is available. This could be specified by the bridge rule (where $\op{set}$ is a management operation with the obvious meaning):
\begin{align*}
\op{set(\term{\pred{value(\term{X})}})} & \la \BRSA{st}{\pred{meter(\term{X})}},\pred{cond1}.
\end{align*}
Whenever $\pred{meter(\term{X})}$ is observed in the stream, and $\pred{cond1}$ holds, the $\pred{value(\term{X})}$ would be matched, but only while $\pred{meter(\term{X})}$ is observed and $\pred{cond1}$ holds, i.e., the management operation's effects are not permanent. In subsequent states, if there was no observation for $\pred{meter(\term{X})}$, or $\pred{cond1}$ no longer holds, then $\pred{value(\term{X})}$ would return to its original value. However, sometimes we may wish to specify some permanent change to the knowledge base of some context that persists beyond the state at which the pre-conditions were true, which is why a new form of bridge rule is needed. In the previous example, if under certain conditions ($\pred{cond2}$) we want to internalize the current observation in a permanent way, i.e., permanently update the default value, then the following new kind of bridge rule would be specified:
\begin{align*}
\Nxt{\op{set(\term{\pred{value(\term{X})}})}} & \la \BRSA{st}{\pred{meter(\term{X})}},\pred{cond2}.
\end{align*}
The intuitive reading of this bridge rule is that whenever at some state both $\pred{meter(\term{X})}$ is true in the stream $st$ and condition $\pred{cond2}$ holds, then the bridge rule's context should be changed according to the operation $\op{set(\term{\pred{value(\term{X})}})}$ when moving to the next state. These two different behaviors could not be achieved with a single kind of bridge rule since one of them aims at temporarily affecting the system at current state while the other one aims at permanently affecting the system from the subsequent state. Interestingly, the two kinds could be combined to specify that some current effect should be made permanent.

This new kind of bridge rule further allows the specification of the evolution of some context based on current conditions, which was not possible in mMCS. Suppose, for example, a single context that maintains a counter $ctr(.)$ keeping track of how often the context was updated. With the new kind of bridge rule, we could simply write:
\begin{align*}
\Nxt{\op{set(\term{\pred{ctr(\term{N+1})}})}} & \la \pred{ctr(\term{N})}.
\end{align*}
Assuming that the management operation $\op{set}$ has the intuitively expected meaning, using the bridge rule
\begin{align*}
\op{set(\term{\pred{ctr(\term{N+1})}})} & \la \pred{ctr(\term{N})}.
\end{align*}
would not work, as it would lead to the non-existence of equilibria. In mMCSs, one might think of a workaround that would overload the management function, but that would also require that the language of the context be extended, which might not be possible and is clearly at odds with the very idea of declarative knowledge representation. Additional examples showing the benefits of the \textbf{next} operator will be discussed in Sect.~\ref{sec_next}.

The main contributions of this paper can be summarized as follows:
\begin{itemize}
\item We extend multi-context systems with the concepts relevant to deal with dynamics, ultimately leading to the notion of \emph{equilibria streams}.
\item We introduce a new, second type of bridge rules using the operator \textbf{next}. This allows us to separate the computation of equilibria from the specification of the context evolution which provides a lot more modeling flexibility and leads to much more natural problem representations.
\item We study various forms of inconsistency, generalize existing inconsistency handling techniques to rMCSs and introduce several new ones.
\item  For a subclass of rMCSs we define well-founded semantics as a skeptical, deterministic alternative to the equilibria-based semantics. The new semantics gives rise to so-called grounded equilibria streams.
\item We investigate the complexity of various problems related to rMCSs and provide a number of complexity results.
\end{itemize}

The paper is organized as follows. In Section~\ref{sec-rmcs}, we introduce reactive MCSs, our framework for reactive reasoning in the presence of heterogeneous knowledge sources. In particular, we show how to integrate data streams into mMCSs and how to model the dynamics of our systems, based on two types of bridge rules. Section~\ref{sec-modeling} illustrates how several typical problems arising in the context of stream reasoning can be handled using our framework. Reasoning based on multiple knowledge sources that need to be integrated faces the problem of potential inconsistencies. Section~\ref{sec-inconsMan} discusses various methods for handling inconsistencies, with a special focus on non-existence of equilibria. In particular, we show how methods developed for mMCSs can be generalized to rMCSs. Nondeterminism in rMCSs is discussed in Section~\ref{sec-wfs}. One way of avoiding nondeterminism is by applying an alternative, skeptical semantics. We show how such a semantics, called well-founded semantics, can be defined for rMCSs, and what the effect of using this semantics instead of the original one is. The complexity of various problems related to rMCSs is investigated in Section~\ref{sec:complexity}. Section~\ref{sec-related} discusses related work, with a special focus on two of the most relevant approaches w.r.t.\ stream reasoning, namely LARS (Logic-based framework for Analyzing Reasoning over Streams) \cite{BeckDEF15} and STARQL~\cite{OptiqueD5.1}. Section~\ref{sec-conclusions} concludes and points out directions for future work.

This paper combines and unifies the results of \cite{BrewkaEP14} and \cite{GoncalvesKL14}, two papers by different subsets of the authors describing independent adaptations of multi-context systems for dynamic environments. The approach developed here generalizes these earlier approaches and substantially improves on the presentation of the underlying concepts.
 
\section{Reactive Multi-Context Systems}\label{sec-rmcs}

Reactive multi-context systems (rMCSs) make use of basic ideas from managed multi-context systems (mMCSs)~\cite{BrewkaEFW11} which extend multi-context systems (MCSs) as defined by Brewka and Eiter~\cite{BrewkaE07} by management capabilities.
In particular, similar to mMCSs, we will make use of a management function
and bridge rules that allow for conflict resolution between contexts as well as
a fine-grained declarative specification of the information flow between contexts.
To not unnecessarily burden the reader with repetitive material on these common components, we abstain from recalling the details of mMCSs first.
It will be clear from the presentation when new concepts/ideas specific to \rMCSs\ will be presented.

\subsection{Specifying the Components of an \rMCS}\label{subsec:SynrMCS}

Similar as for previous notions of MCSs, we build on an abstract notion of a \emph{logic}, which is a
triple $\logic=\tuple{\SKB,\SBelS, \acc}$, where
$\SKB$ is the \emph{set of
admissible knowledge bases} of $\logic$, 
$\SBelS$ is the \emph{set of possible belief sets},
whose elements are called \emph{beliefs}; and
$\acc: \SKB \to 2^{\SBelS}$ is a function describing the
semantics of $\logic$ by assigning to each knowledge base a set of \emph{acceptable
belief sets}.\footnote{To ease readability, throughout the paper, we will often use the following convention when writing symbols: single entities are lower-case, while sets of entities and structures with different components are upper-case; in addition, sequences of those are indicated in {\sf sans serif}, while notions with a temporal dimension are written in calligraphic letters (only upper-case, such as $\mathcal{S}$ or $\mathcal{I}$); finally, operators and functions are {\bf bold}.}

\begin{example}\label{ex-logics}
We illustrate how different formalisms we use in examples throughout the paper can be represented by the notion of a logic.

First, consider the case of 
description logics (DLs) as (commonly decidable) fragments of first-order logic \cite{DLhandbook}. Given a DL language $\mathcal{L}$, we consider the logic $\logic_d=\tuple{\SKB_d,\SBelS_d, \acc_d}$ where $\SKB_d$ is the set of all well-formed DL knowledge bases over $\mathcal{L}$, also called ontologies, $\SBelS_d$ is the set of deductively closed subsets of $\mathcal{L}$, and $\acc_d$ maps every $\KB\in\SKB_d$ to $\{E\}$, where $E$ is the set of formulas in $\mathcal{L}$ entailed by $\KB$.

As an example for a non-deterministic formalism, consider logic programs under the answer set semantics~\cite{GelfondL91}.
Given a set of ground, i.e., variable-free, atoms $A$, we use the logic $\logic_a=\tuple{\SKB_a,\SBelS_a, \acc_a}$ such that $\SKB_a$ is the set of all logic programs over $A$.
The set of possible belief sets is given by the set $\SBelS_a=2^A$ of possible answer sets
and the function $\acc_a$ maps every logic program to the set of its answer sets.

Given a set $E$ of entries, a simple logic for storing elements from $E$ can be realized by the logic $\logic_s=\tuple{\SKB_s,\SBelS_s, \acc_s}$, such that
$\SKB_s=\SBelS_s=2^E$, and $\acc_s$ maps every set $E'\subseteq E$ to $\{E'\}$.
Such $\logic_s$ can, e.g., be used to represent a simple database logic. We will call a logic of this type a storage logic.
\end{example}

In addition to a logic that captures language and semantics of a formalism to be integrated in an \rMCS,
a context also describes how a knowledge base belonging to the logic can be manipulated.

\begin{definition}[Context]
A context is a triple $\Ctxt=\tuple{\logic,\SOp, \mng}$ where
\begin{itemize}
  \item $\logic=\tuple{\SKB,\SBelS, \acc}$ is a logic,
  \item $\SOp$ is a \emph{set of operations},
  \item $\mng:2^{\SOp} \times \SKB \rightarrow \SKB$ is a \emph{management function}.
  \end{itemize}
\end{definition}
\noindent
For an indexed context $\Ctxt_i$ we will write $\logic_i=\tuple{\SKB_i,\SBelS_i, \acc_i}$, $\SOp_i$, and $\mng_i$ to denote its components.
Note that we leave the exact nature of the operations in \SOp unspecified -- they can be seen as mere labels whose semantics is determined by the management function -- and that
we use a deterministic management function instead of a non-deterministic one, unlike mMCS~\cite{BrewkaEFW11}.\footnote{Note that it is straightforward to adapt our definitions to use non-deterministic management functions. However, as they are not essential to our approach, we here refrain from doing so to keep notation simpler.}

\begin{example}\label{ex-context}
Consider the assisted living scenario from the Introduction in which we want to 
recognize potential threats
caused, e.g., by overheating of the stove.
We use the context $\Ctxt_{st}$ to monitor the stove.
Its logic $\logic_{st}$ is a storage logic taking $E=\set{\pred{pw},
\lit{tm}{cold},\lit{tm}{hot}}$ as the set of entries, representing the stove's power status (on if $\pred{pw}$ is present, and off otherwise) and a qualitative value for its temperature (cold/hot).
The current temperature and power state of the stove should be kept up to date in a knowledge base over $\logic_{st}$ using the following operations:
\[
\SOp_{st}=\{\op{setPower(\term{off}),
 \op{setPower(\term{on})}},\op{setTemp(\term{cold})}, \op{setTemp(\term{hot})}
\}
\]

\noindent
The semantics of the operations is given, for $\SOp'\subseteq \SOp_{st}$, by $\mng_{st}(\SOp',\KB)=$
\[
\begin{array}{l@{}l}
\{\pred{pw}\mid& \op{setPower(\term{on})}\in \SOp' \lor \\&(\pred{pw}\in \KB \land \op{setPower(\term{off})}\not\in \SOp') \}\cup\\
                 \{\lit{tm}{t}\mid& \op{setTemp(\term{t})}\in \SOp'\}.
\end{array}
\]
We assume a single constantly operating temperature sensor which triggers exactly one of the $ \op{setTemp}$ operations.
Thus, the current value is simply inserted and we do not need to care about persistence of the temperature value or conflicting information.
The power information, on the other hand, is based on someone toggling a switch, which requires persistence of the fluent $\pred{pw}$.
Note that $\mng_{st}$ ensures that the stove is considered on whenever it is switched on, and also when it is not being switched off and already considered on in the given knowledge base $\KB$. The second alternative implements persistence. Also note that whenever both conflicting $\op{setPower}$ operations are in $\SOp'$, $\op{setPower(\term{on})}$ ``wins", that is, $\pred{pw}$ will be in the knowledge base. This is justified by the application: a potential overheating of the stove is considerably more harmful than unnecessarily turning off the electricity.

Assume we have a knowledge base $\KB=\{\lit{tm}{cold}\}$.
Then, an update with the set $\SOp=\{\op{setPower(\term{on})},\op{setTemp(\term{hot})}\}$ of operations would result in the knowledge base
$\mng_{st}(\SOp,\KB)=\{\pred{pw},\lit{tm}{hot}\}$.
\end{example}

Contexts exchange information using 
\define{bridge rules} that are rules similar in spirit to those in logic programming and that determine which operations from $\SOp_i$ to apply to $\KB_i$ in a context $\Ctxt_i$ depending on beliefs held in other contexts, and, in our approach also on input from the outside.
To keep the approach as abstract as possible, we only require that inputs be elements of some formal \emph{input language} $\IL$. Moreover, we allow for situations where input comes from different sources with potentially different input languages and thus consider tuples $\tuple{\IL_1,\ldots,\IL_\ILIn}$ of input languages.
 \begin{definition}[Bridge Rule]
    Let $\TCtxt = \tuple{\Ctxt_1,\ldots,\Ctxt_\CtxtIn}$ be a tuple of contexts and $\TIL=\tuple{\IL_1,\ldots,\IL_\ILIn}$ a tuple of input languages.
    A \emph{bridge rule for $\Ctxt_i$ over $\TCtxt$ and \TIL}, $i\in\{1,\ldots,\CtxtIn\}$, is of the form \begin{align}
      \label{bridgerule}
      \EOp \la& a_1,\ldots, a_j,\naf\ a_{j+1},\ldots,\naf\ a_m
    \end{align}
    such that $\EOp = \Op$ or $\EOp=\Nxt{\Op}$ for $\Op\in \SOp_i$,
    $j\in \{0,\ldots,m\}$, 
    and every \emph{atom} $a_\ell$, $\ell\in \{1, \ldots,m\}$, is one of the following: 
    \begin{itemize}
    \item a \emph{context atom} \BRBA{c}{\Bel} with $c\,{\in}\, \{1,\ldots,\CtxtIn\}$ and $\Bel\in \BelS$ for some $\BelS\in\SBelS_{c}$, or
    \item an \emph{input atom} \BRSA{s}{b} with $s\in\set{1,\ldots,\ILIn}$
      and $b \in \IL_s$.
    \end{itemize}
For a bridge rule $r$ of the form (\ref{bridgerule}) $\BRHd{r}$ denotes $\EOp$, the \emph{head of $r$}, while $\BRBd{r}=\{a_1,\ldots, a_j,\naf\ a_{j+1},\ldots,\naf\ a_m\}$ is the \emph{body of $r$}. A \emph{literal} is either an atom or an atom preceded by $\naf$, and we differentiate between \define{context literals} and \define{input literals}.
\end{definition}
\noindent
Roughly, a set of bridge rules for $\Ctxt_i$ describes which operations to apply to its knowledge base $\KB_i$, depending on whether currently available beliefs and external inputs match the literals in the body.
Intuitively, rules with head $\Op$ affect the computation of the semantics at the current time instant, while rules with head $\Nxt{\Op}$ affect the computation of the knowledge base(s) at the next time instant.
We define their precise semantics later in Section~\ref{subsec:SemrMCS} and proceed by defining reactive multi-context systems.

  \begin{definition}[Reactive Multi-Context System]
    A \emph{reactive Multi-Context System \\(\rMCS)} is a tuple $M= \tuple{\TCtxt, \TIL, \TBR}$, where
    \begin{itemize}
    \item $\TCtxt=\tuple{\Ctxt_1,\ldots,\Ctxt_\CtxtIn}$ is a tuple of contexts;
    \item $\TIL=\tuple{\IL_1,\ldots,\IL_\ILIn}$ is a tuple of input languages;
    \item $\TBR=\tuple{\SBR_1,\ldots,\SBR_\CtxtIn}$ is a tuple such that each $\SBR_i$, $i\in\{1,\ldots,\CtxtIn\}$, is a set of bridge rules for $\Ctxt_i$ over $\TCtxt$ and \TIL.
    \end{itemize}
  \end{definition}

\begin{example}\label{ex-rmcs}
Consider \rMCS $M_{ex{\ref{ex-rmcs}}}= \tuple{\tuple{\Ctxt_{st}}, \tuple{\IL_{ex{\ref{ex-rmcs}}}}, \tuple{\SBR_{ex{\ref{ex-rmcs}}}}}$ with $\Ctxt_{st}$ as in Example~\ref{ex-context}.\footnote{Throughout the paper we use labels (such as $st$ in $\Ctxt_{st}$) instead of numerical indices in our examples.}
The input language $\IL_{ex{\ref{ex-rmcs}}}=\{\pred{switch}\}$ is used to report whether the power switch of the stove has been turned.
The bridge rules in $\SBR_{ex{\ref{ex-rmcs}}}$ are given by
\begin{align*}
\Nxt{\op{setPower(\term{on})}} & \la \BRSA{ex{\ref{ex-rmcs}}}{\pred{switch}}, \naf\ \BRBA{st}{\pred{pw}}.\\
\Nxt{\op{setPower(\term{off})}} & \la \BRSA{ex{\ref{ex-rmcs}}}{\pred{switch}}, \BRBA{st}{\pred{pw}}.
\end{align*}
and react to switching the stove on or off:
depending on the current power state of the stove that is stored in the knowledge base of $\Ctxt_{st}$, whenever the switch is activated, the bridge rules derive an update of the knowledge base where the power state is reversed.
\end{example}

\subsection{Reacting to External Inputs - Semantics of \rMCSs}\label{subsec:SemrMCS}

To define the semantics of \rMCSs, we first focus on the static case of a single time instant, and only subsequently introduce the corresponding dynamic notions for reacting to inputs changing over time.

We start with the evaluation of bridge rules, for which we need to know current beliefs and current external information.
The former is captured by the notion of a \define{belief state} denoted by a tuple of belief sets -- one for each context -- similar as in previous work on multi-context systems.
\begin{definition}[Belief State]
Let $M= \tuple{\tuple{\Ctxt_1,\ldots,\Ctxt_\CtxtIn}, \TIL, \TBR}$ be an \rMCS.
Then, a \emph{belief state for $M$} is a tuple $\TBelS=\tuple{\BelS_1,\ldots,\BelS_\CtxtIn}$ such that $\BelS_i\in \SBelS_{i}$, for each  $i\in\{1,\ldots,\CtxtIn\}$.
We use $\BelForM{M}$ to denote the set of all belief states for~$M$.
\end{definition}
To also capture the current external information, we introduce the notion of an \define{input}.
\begin{definition}[Input]
Let $M= \tuple{\TCtxt, \tuple{\IL_1,\ldots,\IL_\ILIn}, \TBR}$ be an \rMCS.
Then an \emph{input for $M$} is a tuple $\Tinpt=\tuple{\inpt_1,\ldots,\inpt_\ILIn}$ such that $\inpt_i\subseteq \IL_i$, $i\in\{1,\ldots,\ILIn\}$. 
The \emph{set of all inputs for $M$} is denoted by $\inptForM{M}$.
\end{definition}

We are now ready to define when literals (in bridge rule bodies) are satisfied.
\begin{definition}[Satisfaction of Literals]
 Let  $M=\tuple{\TCtxt, \TIL, \TBR}$ be an \rMCS
  such that $\TCtxt=\tuple{\Ctxt_1,\ldots,\Ctxt_\CtxtIn}$ and $\TIL=\tuple{\IL_1,\ldots,\IL_\ILIn}$.
Given an input $\Tinpt=\tuple{\inpt_1,\ldots,\inpt_\ILIn}$ for $M$ and a belief state $\TBelS=\tuple{\BelS_1,\ldots,\BelS_\CtxtIn}$ for $M$, we define the satisfaction of literals as:

\begin{itemize}

\item $\tuple{\Tinpt,\TBelS} \models a_\ell$ if $a_\ell$ is of the form \BRBA{c}{\Bel} and $\Bel\in \BelS_c$;

\item $\tuple{\Tinpt,\TBelS} \models a_\ell$ if $a_\ell$ is of the form \BRSA{s}{b} and $b\in \inpt_s$;

\item $\tuple{\Tinpt,\TBelS} \models \naf\ a_\ell$ if $\tuple{\Tinpt,\TBelS} \not\models a_\ell$.

\end{itemize}

Let $r$ be a bridge rule for $\Ctxt_i$ over $\TCtxt$ and \TIL. Then

\begin{itemize}

\item $\tuple{\Tinpt,\TBelS} \models \BRBd{r}$ if $\tuple{\Tinpt,\TBelS} \models l$ for every $l\in \BRBd{r}$.

\end{itemize}

\end{definition}
If $\tuple{\Tinpt,\TBelS}\models \BRBd{r}$, we say that $r$ is \define{applicable under \Tinpt and $\TBelS$}.
The operations encoded in the heads of applicable bridge rules in an \rMCS determine which knowledge base updates should take place.
We collect them in two disjoint sets.
\begin{definition}[Applicable Bridge Rules]
 Let  $M=\tuple{\TCtxt, \TIL, \TBR}$ be an \rMCS such that $\TCtxt=\tuple{\Ctxt_1,\ldots,\Ctxt_\CtxtIn}$ and $\TBR=\tuple{BR_1,\ldots,BR_\CtxtIn}$. Given an input $\Tinpt$ for $M$ and a belief state $\TBelS$ for $M$, we define, for each $i\in\{1,\ldots,\CtxtIn\}$, the sets
\begin{itemize}
\item  $\app^{now}_i(\Tinpt,\TBelS) = \{\BRHd{r} \mid r\in \SBR_i, \tuple{\Tinpt,\TBelS} \models \BRBd{r}, \BRHd{r}\in \SOp_i\}$;

\item  $\app^{next}_i(\Tinpt,\TBelS) = \{\Op \mid  r\in \SBR_i, \tuple{\Tinpt,\TBelS} \models \BRBd{r},\BRHd{r}=\Nxt{\Op}\}$.
\end{itemize}
\end{definition}
Intuitively, the operations in $\app^{now}_i(\Tinpt,\TBelS)$ are used for non-persistent updates of the knowledge base that influence the semantics of an \rMCS for a single point in time. The operations in $\app^{next}_i(\Tinpt,\TBelS)$
on the other hand are used for changing knowledge bases over time. They are not used for computing the current semantics
but are applied in the next point in time depending on the current semantics.
This continuous change of knowledge bases over time is the reason why,
unlike in previous work on MCSs, we do not consider knowledge bases as part of the contexts to which they are associated
but store them in a separate configuration structure defined next.
\begin{definition}[Configuration of Knowledge Bases]
Let $M=\tuple{\TCtxt, \TIL, \TBR}$ be an \rMCS such that $\TCtxt=\tuple{\Ctxt_1,\ldots,\Ctxt_\CtxtIn}$.
A \emph{configuration of knowledge bases for $M$} is a tuple $\TKB=\tuple{\KB_1,\ldots,\KB_\CtxtIn}$ such that $\KB_i\in \SKB_{i}$, for each  $i\in\{1,\ldots,\CtxtIn\}$.
We use $\ConForM{M}$ to denote the set of all configurations of knowledge bases for~$M$.
\end{definition}

The semantics of an \rMCS for a single time instant is given by its \define{equilibria}.

\begin{definition}[Equilibrium]\label{def:Equilibrium}
Let $M=\tuple{\tuple{\Ctxt_1,\ldots,\Ctxt_\CtxtIn}, \TIL, \TBR}$ be an \rMCS, $\TKB=\tuple{\KB_1,\ldots,\KB_\CtxtIn}$ a configuration of knowledge bases for $M$, and $\Tinpt$ an input for $M$. Then, a belief state $\TBelS=\tuple{\BelS_1,\ldots,\BelS_\CtxtIn}$ for $M$ is an \emph{equilibrium} of $M$ given \TKB and \Tinpt if, for each $i\in \{1,\ldots,\CtxtIn\}$, we have that
\[\BelS_i\in \acc_{i}(\KB'),\text{ where }\KB'=\mng_i(\app^{now}_i(\Tinpt,\TBelS),\KB_i).\]
\end{definition}

\begin{example}\label{ex-eq}
Consider \rMCS $M_{ex{\ref{ex-rmcs}}}$ from Example~\ref{ex-rmcs}
with the configuration of knowledge bases for $M_{ex{\ref{ex-rmcs}}}$, $\TKB=\tuple{\KB_{st}}=\tuple{\emptyset}$, 
representing that the stove is turned off, input $\Tinpt=\tuple{\{\pred{switch}\}}$ for $M_{ex{\ref{ex-rmcs}}}$ and the belief state $\TBelS=\tuple{\emptyset}$ for $M_{ex{\ref{ex-rmcs}}}$.
As both bridge rules in $\SBR_{ex{\ref{ex-rmcs}}}$ use the \Nxt{} operator, we have
$\app^{now}_{st}(\Tinpt,\TBelS)=\emptyset$ and consequently, following the definition of $\mng_{st}$ in Example~\ref{ex-context}, $\KB_{st}$ remains unchanged, \iec
$\mng_{st}(\app^{now}_{st}(\Tinpt,\TBelS),\KB_{st})=\KB_{st}$.
Thus, $\acc_{st}(\mng_{st}(\app^{now}_{st}(\Tinpt,\TBelS),\KB_{st}))=\{\emptyset\}$.
Thus, \TBelS is an equilibrium of $M_{ex{\ref{ex-rmcs}}}$
given \TKB and \Tinpt.
\end{example}

Based on an equilibrium at the current time instant, we can compute an updated configuration of knowledge bases 
using the update function as introduced next.

\begin{definition}[Update Function]
Let $M=\tuple{\TCtxt, \TIL, \TBR}$ be an \rMCS such that $\TCtxt=\tuple{\Ctxt_1,\ldots,\Ctxt_\CtxtIn}$, $\TKB=\tuple{\KB_1,\ldots,\KB_\CtxtIn}$ a configuration of knowledge bases for $M$, $\Tinpt$ an input for $M$, and $\TBelS$ a belief state for $M$. Then, the \emph{update function for $M$} is defined as $\upd_M(\TKB,\Tinpt,\TBelS)=\tuple{\KB_1',\ldots,\KB'_\CtxtIn}$, such that, for each $i\in \{1\ldots,\CtxtIn\}$, $\KB'_i=\mng_i(\app^{next}_i(\Tinpt,\TBelS),\KB_i)$.
\end{definition}

We can finally show how an \rMCS behaves in the presence of external information that changes over time.
For this purpose, we assume that an \rMCS receives data in a stream of inputs, i.e., an input for each time instant, and we represent individual time instants by natural numbers.
These can be interpreted as logical time instants that do not necessarily represent specific physical time points, nor do we require
that every pair of consecutive natural numbers represents equidistant physical time spans.

\begin{definition}[Input Stream]
  Let $M = \tuple{\TCtxt, \TIL, \TBR}$ be an \rMCS such that $\TIL=\tuple{\IL_1,\ldots,\IL_\ILIn}$.
  An \define{input stream for} $M$ (until $\tau$) is a function $\inptStr:[1..\tau]\to \inptForM{M}$ where $\tau\in\mathbb{N}\cup\{\infty\}$.
\end{definition}
\noindent
We will omit the term ``until $\tau$'' whenever the limit of the stream is irrelevant.
Clearly, an input stream for $M$ until $\tau$ also fully determines an input stream for $M$ until $\tau'$ for every $1\leq\tau'<\tau$.
For any input stream $\inptStr$ and $t\in [1..\tau]$, we will use $\inptStr^t$ to denote $\inptStr(t)$, i.e., the input $\tuple{\inpt_1,\ldots,\inpt_\ILIn}$ for $M$ at time $t$.
We also term \emph{stream} the restriction of an input stream $\inptStr$ to a single input language $\IL_i$, i.e., a function $\inptStr_i:[1..\tau]\to 2^{\IL_i}$ that is fully determined by $\inptStr$.

Note that $\inptStr^t$ encapsulates (input) data for every input language of $M$.
Hence, we assume that, at every time instant, we receive input from every external source of the \rMCS.
This synchronicity is required since the evaluation of a bridge rule may depend on the availability of information from multiple streams.
One possibility for modeling external sources that do not continuously provide information
is setting $\inptStr_s^t$ to the empty set for representing a lack of input from the source with language $\IL_s$ at time $t$.

The semantics of an \rMCS over time is given by its \emph{equilibria streams} for a given initial configuration
of knowledge bases and an input stream for the system.
\begin{definition}[Equilibria Stream]\label{def:StreamOfEquilibria}
Let $M=\tuple{\TCtxt, \TIL, \TBR}$ be an \rMCS, $\TKB$ a configuration of knowledge bases for $M$, and $\inptStr$ an input stream for $M$ until $\tau$ where
$\tau\in\mathbb{N}\cup\{\infty\}$. 
Then, an \emph{equilibria stream of $M$ given $\TKB$ and $\inptStr$} is a function ${\EqStr:[1..\tau]\to \BelForM{M}}$ such that

\begin{itemize}
\item $\EqStr^t$ is an equilibrium of $M$ given $\KBStr^t$ and $\inptStr^t$, 
where $\KBStr^t$ is inductively defined as
\begin{itemize}
\item $\KBStr^1=\TKB$
\item $\KBStr^{t+1}=\upd_M(\KBStr^{t},\inptStr^{t},\EqStr^{t})$. 
\end{itemize}
\end{itemize}
We will also refer to the function $\KBStr:[1..\tau]\to \ConForM{M}$ as the \emph{configurations stream of $M$ given $\TKB$, $\inptStr$, and $\EqStr$}.
\end{definition}
Note that the limit $\tau$ of an equilibria stream is aligned with that of the given input stream.
Following the definition, it is easy to see that if we have an equilibria stream $\EqStr$ of $M$ given $\TKB$ and $\inptStr$, then the substream of $\EqStr$ of size $\tau'$, with $\tau'\leq \tau$, is an equilibria stream of $M$ given $\TKB$ and $\inptStr'$, where $\inptStr'$ is the substream of $\inptStr$ of size $\tau'$.
This implies that, conversely, each extension of the input stream can only lead to equilibria streams that extend those obtained given the original input stream.

\begin{example}\label{ex-seq}
Recall $M_{ex{\ref{ex-rmcs}}}$, $\TKB$, and $\TBelS$  from Example~\ref{ex-eq}, as well as
an input stream $\inptStr$ until $3$ with $\inptStr^{1}=\tuple{\{\pred{switch}\}}$,
$\inptStr^{2}=\tuple{\emptyset}$, and
$\inptStr^{3}=\tuple{\{\pred{switch}\}}$.
There is an equilibria stream \EqStr of $M_{ex{\ref{ex-rmcs}}}$ given $\TKB$ and $\inptStr$.
Note that the input $\inptStr^{1}$ coincides with input \Tinpt from Example~\ref{ex-eq}.
As $\TBelS$ is the only equilibrium of $M_{ex{\ref{ex-rmcs}}}$
given \TKB and \Tinpt, we have that $\EqStr^1=\TBelS$.

As we have $\app^{next}_{st}(\Tinpt,\TBelS)=\{\op{setPower(\term{on})}\}$,
the update function provides the following configuration of knowledge bases for time instant $2$ (with $\KBStr^1=\TKB$):
\[
\KBStr^2=
\upd_{M_{ex{\ref{ex-rmcs}}}}(\KBStr^1,\inptStr^{1},\EqStr^{1})=
\tuple{\mng_{st}(\app^{next}_{st}(\Tinpt,\TBelS),\KB)}=
\tuple{\{\pred{pw}\}}.
\]
Thus, switching the power state at time $1$ leads to an updated knowledge base indicating that the stove is on at time $2$.
The table in Figure~\ref{fig:stovePowerStreams} summarizes the equilibria stream and the configurations stream given $\TKB$ and $\inptStr$. 
\begin{figure}
\renewcommand*{\arraystretch}{1.2}
\begin{footnotesize}
\[
\begin{array}{|c|c|c|c|c|}
\hline
t & \KBStr^t & \inptStr^{t} & \EqStr^t & \app^{next}_{st}(\inptStr^t,\EqStr^t)\\
\hline
1 & \tuple{\emptyset} & \tuple{\{\pred{switch}\}} & \tuple{\emptyset} & \{\op{setPower(\term{on})}\}\\
\hline
2 & \tuple{\{\pred{pw}\}} & \tuple{\emptyset}           & \tuple{\{\pred{pw}\}} & \emptyset\\
\hline
3 & \tuple{\{\pred{pw}\}} & \tuple{\{\pred{switch}\}} & \tuple{\{\pred{pw}\}} & \{\op{setPower(\term{off})}\}\\
\hline
\end{array}
\]
\end{footnotesize}
\caption{Streams and applicable operations for $M_{ex{\ref{ex-rmcs}}}$}
\label{fig:stovePowerStreams}
\end{figure}

\end{example}

\subsection{Rule Schemata}
For convenience, we will represent sets of bridge rules using rule schemata.
Intuitively, a rule schema is a parametrized bridge rule and each instance is a bridge rule.
It will be convenient to use not only parameters,
but also conditions to additionally constrain the set of instances.
We will use them, \egc for allowing arithmetic operations and comparison relations in bridge rules.

Given a tuple $\TCtxt=\tuple{\Ctxt_1,\ldots,\Ctxt_\CtxtIn}$ of contexts and
a tuple $\TIL$ of input languages,
let $\mathcal{A}$ be the alphabet of symbols occurring in all possible bridge rules
for all contexts $\Ctxt_i\in\{\Ctxt_1,\ldots,\Ctxt_\CtxtIn\}$ over $\TCtxt$ and \TIL.
We call a string over $\mathcal{A}$ an \emph{instantiation term} for $\TCtxt$ and $\TIL$.
Then, an \emph{instantiation condition} for $\TCtxt$ and $\TIL$ is a predicate $id(T_1,\dots,T_o)$, where $T_1,\dots,T_o$
are strings over $P\cup\mathcal{A}$ and where $P = \{P_1, \ldots, P_k\}$ is a set of symbols (called \emph{parameters})
such that $P\cap\mathcal{A}=\emptyset$.
A \emph{rule schema} $R$ for $\TCtxt$ and $\TIL$ with parameters $P$ is of the form
\begin{align}\label{schema}
H \la& A_1,\ldots, A_j, \naf\ A_{j+1},\ldots,\naf\ A_m, D_1, \ldots, D_s
\end{align}
such that $H$, $A_1$, ..., $A_m$ are strings over $P\cup\mathcal{A}$ and $D_i$ is an instantiation condition for each $i\in\{1,\ldots s\}$.

A bridge rule $r = \EOp \la a_1,\ldots, a_j,\naf\ a_{j+1},\ldots,\naf\ a_m$ for $\Ctxt_i$ over $\TCtxt$ and \TIL
is an \emph{instance} of a rule schema $R$ for $\TCtxt$ and \TIL with parameters $P$ of the form (\ref{schema}) if $r$ results from $H \la A_1,\ldots, A_j, \naf\ A_{j+1},\ldots,\naf\ A_m$ by a uniform substitution $\sigma$ of parameters with instantiation terms, and
for each $D_i=id(T_1,\dots,T_o)$, $i\in\{1,\ldots s\}$, the predicate $id(T_1\sigma,\dots,T_o\sigma)$ holds.

We adopt the convention from logic programming that parameters start with uppercase letters.
Moreover, we will denote instantiation conditions representing comparison and arithmetic operations
using standard infix notation.
For example, the rule schema
\begin{align}\label{form-schema}
\op{add(\mathit{tmp(S,X)})} \la &\BRSA{S}{temp = X}, 45 \leq X \leq 65.
\end{align}
expresses that temperature values received on stream $S$ should be added to the knowledge base (tagged with the stream index) if they lie between $45\degree$C and $65\degree$C.

Note that we assume that instantiation conditions only evaluate to true if
their arguments are substituted by appropriate instantiation terms,
for example when $X$ is substituted by numeric values in the temperature example.

\subsection{Example Scenario}\label{subsec:scenario}

So far, in our examples we have only shown \rMCSs with a single context, to ease the presentation.
We conclude this section with an extensive example of the assisted-living scenario thereby illustrating the usage of an \rMCS with several contexts (and input languages) and the flow of information within these.

\begin{figure}[t!]\centering
\includegraphics[width=0.9\textwidth]{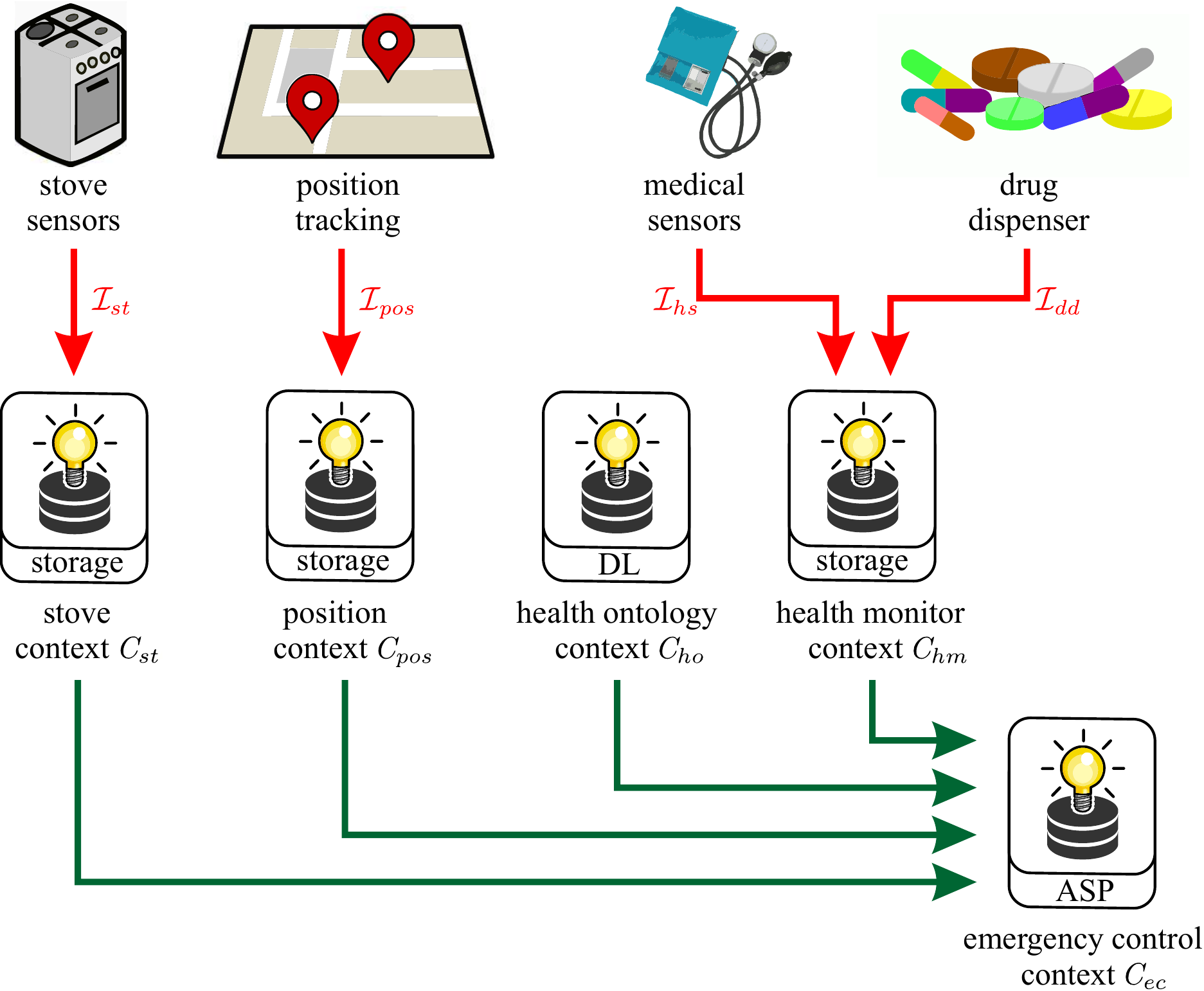}
\caption{Structure of $M_{al}$}
\label{fig:running_ex}
\end{figure}

The \rMCS for our assisted living scenario, $M_{al}=\tuple{\TCtxt,\TIL,\TBR}$, as illustrated in Figure~\ref{fig:running_ex}, with $\TCtxt=\tuple{\Ctxt_{st},\Ctxt_{pos},\Ctxt_{hm},\Ctxt_{ho},\Ctxt_{ec}}$ has five contexts: $\Ctxt_{st}$ is the stove context, $\Ctxt_{pos}$ keeps track of the current position of \Patient, $\Ctxt_{hm}$ monitors \Patient's health status, $\Ctxt_{ho}$ is a medical ontology that the system may consult, and $\Ctxt_{ec}$ is the emergency control context that is responsible for detecting potential threats and taking actions, such as raising an alarm.
The five contexts are connected by their corresponding sets of bridge rules $\TBR=\tuple{\SBR_{st},\SBR_{pos},\SBR_{hm},\SBR_{ho},\SBR_{ec}}$, and 
four input languages $\TIL=\tuple{\IL_{st},\IL_{pos},\IL_{hs},\IL_{dd}}$ represent the sensors used in the assisted living scenario.

The stove context $\Ctxt_{st}$ has already been described in Examples \ref{ex-context} and \ref{ex-rmcs}.
We adjust $\IL_{st}=\IL_{ex{\ref{ex-rmcs}}}\cup\{{\lit{tmp}{T}}\mid T\in \mathbb{Z}\}$ to allow for temperature sensor input data in $\IL_{st}$ and by adding further bridge rules for context $\Ctxt_{st}$, \iec
$\SBR_{st}$ extends $\SBR_{ex{\ref{ex-rmcs}}}$ (with $st$ replacing $ex{\ref{ex-rmcs}}$ in input literals) by the instances of the following bridge rule schemata
\begin{align*}
\op{setTemp(\term{cold})} & \la \BRSA{st}{\lit{tmp}{T}},\term{T}\leq 45.\\
\op{setTemp(\term{hot})} & \la \BRSA{st}{\lit{tmp}{T}}, 45 < \term{T}.
\end{align*}
that classify the current temperature value as cold or hot.
These bridge rules do not use the \Nxt{} operator and hence influence the computation of equilibria rather than the update of knowledge bases for the next step, i.e., temperature information does not persist in the knowledge base, but continuously computed from the input stream.

The input language $\IL_{pos}$ is given by
\[\set{\lit{enters}{kitchen},\lit{enters}{bathroom},\lit{enters}{bedroom}}\]
and non-empty sensor (stream) input for $\inpt_{pos}$ signals when \Patient changes rooms.
This information is used in context $\Ctxt_{pos}$ to keep track of \Patient 's position.
This context also uses a storage logic with $E=\{\pred{pos}(\term{P})\mid \term{P}\in\{\term{kitchen},\term{bathroom},\term{bedroom}\}\}$, and its corresponding set of bridge rules $\SBR_{pos}$ is given by the schema
\begin{align*}
\Nxt{\op{setPos(\term{P})}} &\la \BRSA{pos}{\lit{enters}{P}}.\\
\op{setPos(\term{P})} &\la \BRSA{pos}{\lit{enters}{P}}.
\end{align*}
The management function writes \Patient 's new position into the knowledge base whenever he changes rooms and keeps the previous
position, otherwise, which is why we have a bridge rule with the operator $\Nxt{}$, to ensure that, as for $\op{setPower(\term{P})}$ for $\Ctxt_{st}$, the current position is stored.
In addition, here we also have an identical copy of the bridge rule without $\Nxt{}$ with the rationale that the information is available immediately.
Intuitively, this will allow us to avoid situations in which the conditions for an alarm would be met including the absence of \Patient, although he is just entering the room in question.

The input language $\IL_{hs}$ is used to provide stream data on the health sensor readings from \Patient while $\IL_{dd}$ serves to provide data on the (usage of the) drug dispenser.
The information from both sensors is used in context $\Ctxt_{hm}$ which serves as a health monitor and also builds on a storage logic with 
\[E=\{\lit{status}{asleep},\lit{m}{drugA},\lit{bp}{high},\lit{bp}{normal}\}.\]
Here, $\IL_{hs}$ contains $\{\pred{asleep}\}\cup \{\lit{bpReading}{R}\mid R$ a blood pressure value$\}$ while $\IL_{dd}$ contains $\{\lit{dispensed}{drugA}\}$ and the bridge rules in $\SBR_{hm}$ are given by:
\begin{align*}
\op{setStatus(\pred{asleep})} & \la \BRSA{hs}{\pred{asleep}}.\\
\Nxt{\op{setBP(\term{high})}} & \la \BRSA{hs}{\lit{bpReading}{R}},\term{R}\geq 140/90.\\
\Nxt{\op{setBP(\term{normal})}} & \la \BRSA{hs}{\lit{bpReading}{R}},\term{R}< 140/90.\\
\Nxt{\op{setMed(\lit{m}{drugA})}} & \la \BRSA{dd}{\lit{dispensed}{drugA}}.
\end{align*}
While the sleep status can be monitored permanently and new data arrives at every time instant, blood pressure measurements and taking medications are only taken occasionally, so this data needs to be stored until new information arrives.
Here, we abstract from the details on the duration of effects of medications.
The management function $\mng_{hm}(\SOp,\KB)$ can then be defined quite similarly to $\mng_{st}(\SOp,\KB)$.

The context $\Ctxt_{ho}=\tuple{\logic_{ho},\SOp_{ho},\mng_{ho}}$ represents a biomedical health ontology, which uses the DL logic described in Example~\ref{ex-logics}.
It contains information on diseases, treatments, and so on, and, it can, e.g., be used to consult for possible effects of medications by context $\Ctxt_{ec}$ as explained below.
In our case, $\KB_{ho}$ is simplified to
\begin{align*}
drugA &: \exists \pred{contains}.\pred{ephedrine}\\
\exists \pred{contains}.\pred{ephedrine} & \sqsubseteq \pred{causesHighBP}
\end{align*}
which allows one to derive that $drugA$ causes high blood pressure.
In our simple setting, the management function $\mng_{ho}$ leaves the knowledge base $\KB_{ho}$ unchanged.

$\Ctxt_{ec}=\tuple{\logic_{ec},\SOp_{ec},\mng_{ec}}$ is the context for detecting emergencies.
It is implemented as an answer set program, hence the acceptable belief sets of $\logic_{ec}$ are the answer sets
of its knowledge bases as outlined in Example~\ref{ex-logics}.
The bridge rules in $\SBR_{ec}$ do not refer to stream input but query other contexts:
\begin{align*}
\op{add(\lit{oven}{on,hot})} & \la \BRBA{st}{\pred{pw}},\BRBA{st}{\lit{tm}{hot}}.\\
\op{add(\lit{humanPos}{P})} & \la \BRBA{pos}{\lit{pos}{P}}.\\
\op{add(\lit{status}{asleep})} & \la \BRBA{hm}{\pred{asleep}}.\\
\op{add(\pred{highBP})} & \la \BRBA{hm}{\lit{bp}{high}}.\\
\op{add(\pred{highBPMed})} & \la \BRBA{ho}{\lit{causesHighBP}{D}}, \BRBA{hm}{\lit{m}{D}}.
\end{align*}
Note that in the first bridge rule, we only import information for the case that will possibly be problematic, namely if the oven is turned on and hot, whereas, e.g., in the second, we use a schema to keep track of the position in all cases.
The answer set program $\KB_{ec}$ is given by the rules:
\begin{align*}
\lit{turnOff}{stove} & \la \lit{oven}{on,hot}, \lit{status}{asleep}.\\
\lit{alert}{stove} & \la \lit{oven}{on,hot}, \naf\ \lit{humanPos}{kitchen}, \naf\ \lit{status}{asleep}.\\
\lit{call}{medAssist} & \la \pred{highBP},  \naf\ \pred{highBPMed}.
\end{align*}

\begin{sidewaysfigure}
\renewcommand*{\arraystretch}{1.2}
\begin{scriptsize}
\[
\begin{array}{|c|c|c|c|c|c|}
\hline
t & \KBStr^t & \inptStr^{t} & \app^{now}_{i}(\inptStr^t,\EqStr^t) & \EqStr^t & \app^{next}_{i}(\inptStr^t,\EqStr^t)\\
\hline
1 & \langle\emptyset,\{\lit{pos}{kitchen}\}, & \langle\{\lit{tmp}{19},\pred{switch}\},\emptyset,  & \langle\{\op{setTemp(\term{cold})}\},\emptyset,\emptyset,\emptyset, & \langle\{\lit{tm}{cold}\},\{\lit{pos}{kitchen}\}, & \langle\{\op{setPower(\term{on})}\},\emptyset,\\
   & \{\lit{bp}{normal}\},\emptyset,\emptyset\rangle  &  \{\lit{bpReading}{135/86}\},\emptyset\rangle & \{\op{add(\lit{humanPos}{kitchen})}\}\rangle                                              &   \{\lit{bp}{normal}\},\emptyset,  & \{\op{setBP(\term{normal})}\},\emptyset,\emptyset\rangle\\
   & & & & \{\lit{humanPos}{kitchen}\}\rangle & \\
\hline
2 & \langle\{\pred{pw}\},\{\lit{pos}{kitchen}\}, & \langle\{\lit{tmp}{21}\},\emptyset,  & \langle\{\op{setTemp(\term{cold})}\},\emptyset,\emptyset,\emptyset, & \langle\{\lit{tm}{cold},\pred{pw}\}, & \langle\emptyset,\emptyset,\\
   & \{\lit{bp}{normal}\},\emptyset,\emptyset\rangle  & \emptyset,\{\lit{dispensed}{drugA}\}\rangle & \{\op{add(\lit{humanPos}{kitchen})}\}\rangle                                              &   \{\lit{pos}{kitchen}\},\{\lit{bp}{normal}\},  & \{\op{setMed(\lit{m}{drugA})}\},\\
   & & & & \emptyset,\{\lit{humanPos}{kitchen}\}\rangle & \emptyset,\emptyset\rangle \\
\hline
3 & \langle\{\pred{pw}\},\{\lit{pos}{kitchen}\}, & \langle\{\lit{tmp}{27}\},\{\lit{enters}{bedroom}\},  & \langle\{\op{setTemp(\term{cold})}\}, & \langle\{\lit{tm}{cold},\pred{pw}\}, & \langle\emptyset,\{\op{setPos(\term{bedroom})}\},\\
   & \{\lit{bp}{normal},\lit{m}{drugA}\},  & \{\lit{bpReading}{138/89}\},\emptyset\rangle & \{\op{setPos(\term{bedroom})}\},\emptyset,\emptyset,                                             &   \{\lit{pos}{bedroom}\},\{\lit{bp}{normal}\},  & \{\op{setBP(\term{normal})}\},\emptyset,\emptyset\rangle\\
   & \emptyset,\emptyset\rangle & & \{\op{add(\lit{humanPos}{bedroom})},  & \emptyset,\{\lit{humanPos}{bedroom}, & \\
   & & & \op{add(\pred{highBPMed})}\}\rangle & \pred{highBPMed}\}\rangle &  \\
\hline
4 & \langle\{\pred{pw}\},\{\lit{pos}{bedroom}\}, & \langle\{\lit{tmp}{36}\},\emptyset,  & \langle\{\op{setTemp(\term{cold})}\},\emptyset,\emptyset,\emptyset, & \langle\{\lit{tm}{cold},\pred{pw}\}, & \langle\emptyset,\emptyset,\\
   & \{\lit{bp}{normal},\lit{m}{drugA}\},  & \{\lit{bpReading}{148/97}\},\emptyset\rangle & \{\op{add(\lit{humanPos}{bedroom})}, &   \{\lit{pos}{bedroom}\},\{\lit{bp}{normal}\},  & \{\op{setBP(\term{high})}\},\emptyset,\emptyset\rangle\\
   & \emptyset,\emptyset\rangle & & \op{add(\pred{highBPMed})}\}\rangle & \emptyset,\{\lit{humanPos}{bedroom}, & \\
   & & & & \pred{highBPMed}\}\rangle & \\
\hline
5  & \langle\{\pred{pw}\},\{\lit{pos}{bedroom}\}, & \langle\{\lit{tmp}{43}\},\emptyset,  & \langle\{\op{setTemp(\term{cold})}\},\emptyset,\emptyset,\emptyset, & \langle\{\lit{tm}{cold},\pred{pw}\}, & \langle\emptyset,\emptyset,\\
   & \{\lit{bp}{high},\lit{m}{drugA}\},  & \{\lit{bpReading}{146/95}\},\emptyset\rangle & \{\op{add(\lit{humanPos}{bedroom})}, &   \{\lit{pos}{bedroom}\},\{\lit{bp}{high}\},  & \{\op{setBP(\term{high})}\},\emptyset,\emptyset\rangle\\
   & \emptyset,\emptyset\rangle & & \op{add(\pred{highBPMed})}, & \emptyset,\{\lit{humanPos}{bedroom}, & \\
   & & & \op{add(\pred{highBP})}\}\rangle & \pred{highBPMed}, \pred{highBP}\}\rangle & \\
\hline
6  & \langle\{\pred{pw}\},\{\lit{pos}{bedroom}\}, & \langle\{\lit{tmp}{51}\},\emptyset,  & \langle\{\op{setTemp(\term{hot})}\},\emptyset, & \langle\{\lit{tm}{hot},\pred{pw}\}, & \langle\emptyset,\emptyset,\emptyset,\emptyset,\emptyset\rangle\\
   & \{\lit{bp}{high},\lit{m}{drugA}\},  & \{\pred{asleep}\},\emptyset\rangle & \{\op{setStatus(\pred{asleep})}\},\emptyset, &  \{\lit{pos}{bedroom}\},\{\lit{bp}{high}\}, & \\
   & \emptyset,\emptyset\rangle & & \{\op{add(\lit{humanPos}{bedroom})}, & \emptyset,\{\lit{humanPos}{bedroom},  & \\
   & & & \op{add(\pred{highBPMed})}, & \pred{highBPMed}, \pred{highBP}, & \\
   & & & \op{add(\pred{highBP})}, & \lit{oven}{on,hot}, \lit{status}{asleep}, & \\
   & & & \op{add(\lit{oven}{on,hot})}, &  \lit{turnOff}{stove}\}\rangle & \\
   & & & \op{add(\lit{status}{asleep})}\}\rangle & & \\
\hline
\end{array}
\]
\end{scriptsize}
\caption{Streams and applicable operations for the example scenario given by $M_{al}$}
\label{fig:fullRun}
\end{sidewaysfigure}

The management function of $\Ctxt_{ec}$, $\mng_{ec}(\SOp,\KB)$, that adds information from the bridge rules temporarily
as input facts to the context's knowledge base can be defined similar to the previous contexts.

An example run of this entire \rMCS is given in Figure~\ref{fig:fullRun}.
Note that we have omitted from $\KBStr^t$ and $\EqStr^t$ all information which is fixed to ease the reading, which is why, e.g., $\TKB_{ho}$ and $\TKB_{ec}$ in the figure are always empty.
The presented situation begins with \Patient in the kitchen, where he turns on the stove at $t=1$.
The effect of this is visible at $t=2$, where he takes his medication (also in the kitchen).
Note that blood pressure readings are not constantly done, and therefore at $t=2$ no reading occurs.
Subsequently, \Patient leaves for the bedroom to rest (thus forgetting about the stove), and this change of position is available right away and stored.
The next blood pressure reading at $t=4$ is high, which is stored in the corresponding knowledge base from $t=5$ onwards, at which time point a medical assistent would be called if the medication most likely causing this was not registered.
Finally, at $t=6$, the oven has become hot, but since \Patient has fallen asleep, the context for detecting emergencies initiates turning off the stove to prevent possible problems.
If \Patient was not asleep yet, an alarm would have been sounded reminding him of the stove instead. 

Finally, note that in our example scenario, currently the effects of taken medications do not wear off. 
Such extension can be handled by e.g., incorporating explicit time, one of the modeling features of \rMCSs described in the next section.
 
\section{Modeling with \rMCSs}\label{sec-modeling}

The running example demonstrates that \rMCSs can jointly integrate different formalisms and deal with a dynamic environment.
In this section, we want to focus on aspects relevant to the knowledge engineer, namely how to model a certain scenario using an \rMCS.
To this end, we elaborate on different generic modeling techniques for rMCSs that we consider helpful in typical target applications.
For concrete implementations, these techniques can still be refined and tailored towards the specific needs of the problem domain at hand.
First, in Section~\ref{sec_incorp_stream_data}, we discuss bridge rules as a device for translating stream data to a knowledge base language.
When to use the \Nxt{} operator in the head of bridge rules is addressed in Section~\ref{sec_next}.
In Section~\ref{sec:time}, we discuss how we deal with time on the object level, including the use of timestamps as well as external and logical clocks.
Then, in Section~\ref{sec:data}, another technique is presented that allows for managing conflicting data in input streams on the modeling level, \egc contradictory sensor measurements.
An important issue for systems that are continuously online is when to keep data and when to remove it.
In Section~\ref{sec:forget}, we discuss how to do this and provide techniques for dynamically adjusting what information is kept in the system.
These techniques allow us, \egc to modify the size of sliding windows for stream data depending on the current situation.
While the computation of acceptable belief sets may be easy for some contexts, it might be expensive for others that have to perform complex reasoning.
In practice, it will therefore be wise to only evaluate these contexts if necessary.

\subsection{Incorporating Stream Data}\label{sec_incorp_stream_data}
Bridge rules are responsible for providing a context with information from streams and other contexts.
As we deal with heterogeneous context languages and since input languages may differ from context languages,
one important aspect of \rMCSs (and earlier types of MCSs) is that bridge rules can be seen as a translation device between the different languages:
bridge rule bodies use the languages of the source contexts or streams whereas a bridge rule head is an
operation that produces a knowledge base in the target language (via the management function).
But bridge rules do not necessarily pass information the way it is used in the body.
Rather, we can model bridge rules such that we they focus on information relevant to the target context
and, \egc translate sensor data into a form that is convenient for reasoning.
Consider the bridge rule schemas shown in Section~\ref{subsec:scenario} for $\Ctxt_{st}$:
\begin{align*}
\op{setTemp(\term{cold})} & \la \BRSA{ex{\ref{ex-rmcs}}}{\lit{tmp}{T}},\term{T}\leq 45.\\
\op{setTemp(\term{hot})} & \la \BRSA{ex{\ref{ex-rmcs}}}{\lit{tmp}{T}}, 45 < \term{T}.
\end{align*}
In this case, the bridge rules only communicate whether the stove is hot or cold, abstracting away the exact temperature value coming from the sensor. This is in contrast to rule schema $(\ref{form-schema})$ discussed in the previous section where parameter $X$ appears in the rule head whereby a concrete temperature value is added to the knowledge base.

\subsection{Operational and Declarative Bridge Rules}\label{sec_next}
For gaining a better understanding
of when to use 
\Nxt{} in the head of a bridge rule,
reconsider the example of turning a switch on and off as in Examples~\ref{ex-context} and~\ref{ex-rmcs}.
Remember that the bridge rules
\begin{align*}
\Nxt{\op{setPower(\term{on})}} & \la \BRSA{ex{\ref{ex-rmcs}}}{\pred{switch}}, \naf\ \BRBA{st}{\pred{pw}}.\\
\Nxt{\op{setPower(\term{off})}} & \la \BRSA{ex{\ref{ex-rmcs}}}{\pred{switch}}, \BRBA{st}{\pred{pw}}.
\end{align*}
were used so that an occurrence of an input atom \pred{switch} causes the knowledge base to contain an inverted power state at the next time point.
In order to highlight the difference, assume we would instead use the bridge rules
\begin{align*}
\op{setPower(\term{on})} & \la \BRSA{ex{\ref{ex-rmcs}}}{\pred{switch}}, \naf\ \BRBA{st}{\pred{pw}}.\\
\op{setPower(\term{off})} & \la \BRSA{ex{\ref{ex-rmcs}}}{\pred{switch}}, \BRBA{st}{\pred{pw}}.
\end{align*}
without \Nxt{}. We refer to the version of $M_{ex{\ref{ex-rmcs}}}$ from Example~\ref{ex-rmcs} with these modified bridge rules by $M'_{ex{\ref{ex-rmcs}}}$.
Consider the configuration $\TKB=\tuple{\KB_{st}}$ of knowledge bases
with $\KB_{st}=\emptyset$ and the input $\Tinpt=\tuple{\{\pred{switch}\}}$ as in Example~\ref{ex-eq}.
Indeed, $M'_{ex{\ref{ex-rmcs}}}$ has no equilibrium given \TKB and \Tinpt:
If we take belief state $\TBelS=\tuple{\emptyset}$
as in Example~\ref{ex-eq}, we have
$\app^{now}_{st}(\Tinpt,\TBelS)=\{\op{setPower(\term{on})}\}$ (instead of $\emptyset$ as in Example~\ref{ex-eq}).
Consequently, following the definition of $\mng_{st}$ in Example~\ref{ex-context}, we get $\mng_{st}(\app^{now}_{st}(\inpt,\TBelS),\KB_{st})=\{\lit{pw}{on}\}$.
But then \TBelS is not contained in $\acc_{st}(\mng_{st}(\app^{now}_{st}(\Tinpt,\TBelS),\KB_{st}))=\{\{\lit{pw}{on}\}\}$, thus, \TBelS is not an equilibrium of $M'_{ex{\ref{ex-rmcs}}}$ given \TKB and \Tinpt. This is in line with the intuition that without \Nxt{}, turning the switch should affect the current equilibrium.
However, $\TBelS'=\tuple{\{\lit{pw}{on}\}}$ is also not an equilibrium of $M'_{ex{\ref{ex-rmcs}}}$.
Since $\app^{now}_{st}(\Tinpt,\TBelS')=\{\op{setPower(\term{off})}\}$, we get
$\mng_{st}(\app^{now}_{st}(\Tinpt,\TBelS'),\KB_{st})=\emptyset$.
But then $\acc_{st}(\mng_{st}(\app^{now}_{st}(\Tinpt,\TBelS'),\KB_{st}))=\{\emptyset\}$
does not contain $\TBelS'$, consequently also $\TBelS'$ is not an equilibrium of $M'_{ex{\ref{ex-rmcs}}}$ given \TKB and \Tinpt.
The two bridge rules without \Nxt{} prevent stability of a belief state required for an equilibrium: believing that the power is on (respectively off) causes an update of the knowledge base that the power is off (respectively on) which is in turn inconsistent to the belief.

The reason why the change considered here does not work is that the two types of bridge rule are meant to be used for different purposes.
The bridge rules using \Nxt{} are responsible for changing knowledge bases over time.
An \rMCS without such bridge rules cannot alter its initial configuration of knowledge bases.
Thus, these rules come with an \emph{operational} flavor.
Bridge rules without \Nxt{} on the other hand have a pure \emph{declarative} nature and their purpose is to semantically integrate the contexts of an \rMCS.

Still, as we have seen for $M_{al}$ in Section~\ref{subsec:scenario}, it sometimes makes sense to use essentially the same bridge rules with and without \Nxt{}:
the bridge rule
\begin{align*}
\Nxt{\op{setPos(\term{P})}} &\la \BRSA{pos}{\lit{enters}{P}}.\end{align*}
ensures that the information about the position of \Patient persists in the knowledge base, whereas
\begin{align*}
\op{setPos(\term{P})} &\la \BRSA{pos}{\lit{enters}{P}}.
\end{align*}
provides this information for computing equilibria in the current time instant.

\subsection{Integration of Time}\label{sec:time}
The structure of input streams and equilibria streams implicitly induces a discrete logical time for \rMCSs.
In order to operate in dynamic environments, in many cases it is necessary to deal with
explicit physical time or logical time, and both can be achieved on the level of modeling.
To this end, it is necessary to have access to explicit time points on this level, \iec in the bridge rules and knowledge bases.
A natural way to make such time points explicitly available is the use of an external clock that provides the current time via an input stream $\inptStr_c$.
Thus, every input for the \rMCS contains information about the current time which can then be queried in bridge rules.

Consider a context $C_1$ that requires information about the development of temperature values from an incoming sensor over time.
The bridge rule schema
\begin{align*}
\Nxt{\op{add(\lit{tmpAtTime}{Temp,T})}} & \la \BRSA{tmp}{\lit{tmp}{Temp}},\BRSA{c}{\lit{now}{T}}.\\
\end{align*}
can be used to add atoms of form $\lit{tmpAtTime}{Temp,T}$ to the context, where $Temp$ is the current temperature and $T$ stands for the time of the sensor reading.
This setting also allows for querying, e.g., whether a temperature of $45\degree$C was exceeded within the last $10$ minutes, as expressed in the following bridge rule schema:
\begin{align*}
\op{add(\pred{recentlyHot})}  \la & \BRBA{1}{\lit{tmpAtTime}{Temp,T'}},\\&\term{Temp}>45,\BRSA{c}{\lit{now}{T}},T'\geq T-10.\\
\end{align*}

Another possibility is to use logical time synchronized with the length of the equilibria stream, \egc whenever the use of an external clock is not an option.
We can obtain timestamps from the computation of knowledge base updates by
using one context $C_{\mathit{clock}}$ that keeps information about the current logical time that uses the following bridge rule schemas and whose initial knowledge base is assumed to be empty:
\begin{align*}
\op{setTime(\lit{now}{0})} & \la \naf\ \BRBA{clock}{\pred{timeAvailable}}.\\
\Nxt{\op{add(\pred{timeAvailable})}} & \la \BRBA{clock}{\lit{now}{0}}.\\
\Nxt{\op{setTime(\lit{now}{T+1})}} & \la \BRBA{clock}{\lit{now}{T}}.
\end{align*}
The first rule is used for initialization ensuring that if no time information is yet available, the logical time is set to the value $0$.
The third rule increments the current time by one and stores the updated value in the knowledge base of the next time instant.
Finally, the second rule, ensures that once this value is greater than $0$, the first rule can no longer be applied.

\subsection{Handling Inconsistent Stream Data}\label{sec:data}

In many situations, inconsistencies may occur when dealing with multiple external sources of data.
Consider, \egc an array of sensors that measure interdependent properties among which there may be a sensor that provides more fine-grained results, but which is also less reliable, so that it sometimes provides values that conflict with the data from the remaining sensors. In this case, we would like to use such a measure of reliability to consistently accommodate relevant sensor data.
Next, we present a technique for integrating possibly inconsistent stream data into a context of an \rMCS.
Let $M = \tuple{\TCtxt, \TIL, \TBR}$ be an \rMCS with $\TIL=\tuple{\IL_1,\ldots,\IL_\ILIn}$, and $\Ctxt_i\in \TCtxt$ a context whose aim is to receive and consistently accommodate (potentially conflicting) information from the streams.
To deal with possible inconsistencies, $\Ctxt_i$ has bridge rules of the form
\[\addCons(D,\mathit{j}) \la \BRSA{j}{D}.\]
for $j\in\{1,\ldots,\ILIn\}$, where the operation $\addCons$ is meant to consistently add the information of sensor $j$ to the context.
To address possible inconsistencies among sensors,
we foresee a management function $\mng_i$ that operates based on a total preference relation~\pref over the available sensors. The second argument of the $\addCons$ operation provides information about the source of a piece of information and thus a way of imposing preferences on the information to be added. Without loss of generality assume $\IL_1 > \ldots > \IL_\ILIn$, that is, input language $\IL_1$ has highest priority.
Moreover, a notion of consistency needs to be specific to the context and so we assume a property \cons{\KB} that holds if the knowledge base \KB is consistent (according to such domain specific notion of consistency).

Given a set of operations $\SOp$, we define the sets of input data from each sensor
${inp}_j'=\{d\mid \addCons(d,\mathit{j}) \in \SOp\}$
for $j\in\{1,\dots,\ILIn\}$.

We then assume that ${inp}^c_0(\SOp) = \emptyset$ and let
\begin{equation*}
{inp}^c_j(\SOp) =\begin{cases}
 {inp}^c_{j-1}(\SOp) \cup {inp}_j' & \text{ if }\cons{{inp}^c_{j-1}(\SOp) \cup {inp}_j'}\\
{inp}^c_{j-1}(\SOp) & \text{ otherwise.}
\end{cases}
\end{equation*}
Finally, we define $\mng_i(\SOp,\KB_i) = \KB_i \cup {inp}^c_{\ILIn}(\SOp)$.

The intuitive idea is that, starting with the input stream with highest priority, data from each sensor is only incorporated into the knowledge base if such data is consistent with the data already collected from the input streams with higher priority.

Note that by considering ${inp}^c_0(\SOp) = \emptyset$, the solution only considers inconsistency of data on the streams. For considering inconsistency between the streams and also the knowledge base $\KB_i$ of the context, we can set ${inp}^c_0(\SOp) = \KB_i$.

We can also easily consider the incorporation of meta-information about sensors whose readings are considered inconsistent. This only requires a small change in the definition of ${inp}^c_j(\SOp)$ to ${inp}^c_{j-1}(\SOp) \cup \{incons(j)\}$ in case a conflict occurs.
Such meta information can then be further leveraged by, e.g., initiating a control of the sensor if such measurements fall outside of the expected parameters of such sensor.

In all, this shows how the management function can solve conflicts due to inconsistent stream data based on preferences among the streams.
Of course, many more strategies for integrating inconsistent stream data can be thought of.
For example, in absence of a global ranking between streams, one way to ensure consistency is to select maximally consistent subsets of stream data.
A corresponding management function
could then be defined such that $\mng_i(\SOp,\KB_i) = \KB_i \cup\ {inp}_{mx}$, where ${inp}_{mx}$ is a maximal set where
${inp}_{mx}\subseteq\{d\mid \addCons(d,\mathit{j})\in\SOp, j\in\{1,\ldots,m\}\}$
and $\cons{{inp}_{mx}}$ holds.

The strategies above show how we can deal with contradictory information in the processed data by means of modeling.
In Section~\ref{sec-inconsMan}, we address inconsistency on the level of the formalism caused
by nonexistence of equilibria or inconsistent belief states.

\subsection{Selective Forgetting and Data Retention}\label{sec:forget}

As argued in Section~\ref{sec:time} for the example where we were constantly adding sensor information to some knowledge base, sometimes it is necessary to forget (part of this) knowledge again.
In this section, we show how \rMCSs\ can model scenarios where there is a need to dynamically adjust the size of the stored stream history.
We do that by considering an extension of our running example. Recall from $M_{al}$ in Section~\ref{subsec:scenario} that context $\Ctxt_{ec}$ is a context for detecting emergencies. In a situation where the stove is hot and Dave is asleep, $\lit{turnOff}{stove}$ is derived, signaling that the stove must be turned off. In case the stove is hot, but Dave is neither asleep nor in the kitchen, then an alarm is raised, signaling a potential emergency. In the latter case, we do not want to immediately turn the stove off, since it may well be the case that the absence of Dave from the kitchen is short. A situation is only considered a real emergency if Dave is absent from the kitchen for a (predefined) long period of time.
To model such situation, we use a context $\Ctxt_{stE}$, which collects timestamped alerts raised by context $\Ctxt_{ec}$ and uses these to check if a real stove-related emergency has occurred. Since we need to reason about time as in Section~\ref{sec:time}, we consider an input stream $\inptStr_c$ that provides the current time.
The possible knowledge bases of $\Ctxt_{stE}$ contain elements of the form $\lit{alert}{stove,t}$ where $t\in \mathbb{N}$, one element of the form $\lit{winE}{t}$, which defines the limit size of the time window between two alerts above which an emergency should be raised, and possibly one element of the form $\lit{emergency}{stove}$ that signals the existence of a stove-related emergency.
The set of bridge rules of $\Ctxt_{stE}$ is determined by the following bridge rule schemas:
\begin{align*}
\Nxt{\op{add(\lit{alert}{stove,T})}}  \la\ &\BRSA{c}{\lit{now}{T}}, \BRBA{ec}{\lit{alert}{stove}}.\\
\Nxt{\op{del(\lit{alert}{stove,T})}}  \la\ &\BRBA{stE}{\lit{alert}{stove,T}},  \naf\ \BRBA{ec}{\lit{alert}{stove}}.\\
\op{add(\lit{emergency}{stove})}  \la\ & \BRSA{c}{\lit{now}{T}},\BRBA{ec}{\lit{alert}{stove}}, \BRBA{stE}{\lit{alert}{stove,T'}},\\ & \BRBA{stE}{\lit{winE}{Y}},
|T-T'|\geq Y.
\end{align*}

The first rule adds a timestamped stove alert whenever such alert is active on context $\Ctxt_{ec}$.
The second rule removes all currently stored stove alerts whenever no such alert is coming from context $\Ctxt_{ec}$. This guarantees that the knowledge base of $\Ctxt_{stE}$ does not accumulate unnecessary information.
The last bridge rule triggers a real emergency alert whenever, in the history of alerts kept in the knowledge base of $\Ctxt_{stE}$, there is an alert whose timestamp differs from the current time more than the acceptable emergency window.

Using context $\Ctxt_{stE}$, we have shown how to model scenarios where tracking the stream history is triggered by alerts of possible emergencies.
We now also consider the case where such alerts trigger the change of the window size of stream history to be kept.
Consider a scenario with several potential emergencies, which can be just suspected or confirmed. Based on the status of the emergencies at each time point, we may need to adapt the size of the stream history that is kept.
We generically model such scenario with an \rMCS\ with a context $C_d$, which is used for emergency detection in such a dynamic environment, and an input language $\IL_s$, which represents the possible observations. Assume there are $m$ potential emergencies $e_1, \ldots, e_m$ we want the context to handle. The role of $C_d$ is to check, based on the observations made, whether one or more of the emergencies $e_i$ are suspected or confirmed. Based on information about potential emergencies, $C_d$ adjusts the time window of the observations that are kept. This is the basis for intelligent forgetting based on dynamic windows.

The only assumptions we make about how $C_d$ works internally are:

\begin{itemize}
  \item $C_d$ may signal that emergency $e_i$ is suspected ($\lit{susp}{e_i}$) or confirmed ($\lit{conf}{e_i}$).
  \item $C_d$ has information for each different observation $\term{p}$ about default, respectively actual window sizes, $\lit{defWin}{p,w}$, $\lit{win}{p,w}$, and
  \item $C_d$ has information about the window size for each observation relevant for a particular emergency, $\lit{rel}{p,e,w}$.
\end{itemize}
The set of bridge rules for $C_d$ includes the following rules.
\begin{align*}
\Nxt{\op{set}(\lit{win}{P,X})}&\la\BRBA{d}{\lit{defWin}{P,X}}, \naf\ \BRBA{d}{\lit{susp}{E}}. \\
\Nxt{\op{set}(\lit{win}{P,Y})}&\la \BRBA{d}{\lit{rel}{P,E,Y}}, \BRBA{d}{\lit{susp}{E}}. \\
\op{alarm}(E) &\la \BRBA{d}{\lit{conf}{E}}.
\end{align*}

 The operation $\op{set}$ sets the window size to a new value, deleting the old one, while $\op{alarm}$ is an operation that adds information to the context knowledge base signaling an alarm. Since an observation can be relevant for more than one emergency, it may be the case that the management function has to deal with operations $\op{set}(\lit{win}{p,w})$ with the same $p$ but with different values of $w$. In that case, in order to avoid loosing observations relevant for some suspected emergency, the management function takes the largest value of $w$ as the window size for $p$.

Finally, the following bridge rule schemas define addition and deletion of observations from some stream $s$. The deletion of observations are performed in accordance with the determined window sizes:
\begin{align*}
\Nxt{\op{add(\lit{P}{T})}} & \la \BRSA{t}{\lit{now}{T}}, \BRSA{s}{\pred{P}}.\\
\Nxt{\op{del(\lit{P}{T'})}} & \la \BRBA{d}{\lit{P}{T'}}, \BRSA{t}{\lit{now}{T}}, \BRBA{d}{\lit{win}{P,Z}}, T' < T-Z.
\end{align*}

The management function just performs additions and deletions on the context knowledge base.
Since additions always include the (current) time of addition, deletions always refer to an earlier point in time, thus these two operators can never occur simultaneously.
 
\section{Inconsistency Management}\label{sec-inconsMan}

The occurrence of inconsistencies within frameworks that aim at integrating knowledge from different sources cannot be neglected, even more so in dynamic settings where knowledge changes over time. There are many reasons why \rMCSs may fail to have an equilibria stream.
These include the absence of an acceptable belief set for one of its contexts given its current knowledge base at some point in time, some occurring conflict between the operations in the heads of bridge rules, or simply because the input stream is such that the configuration of the flow of information within the \rMCS, namely its bridge rules, prevent the existence of such an equilibria stream. In a real world situation, an \rMCS without an equilibria stream is essentially useless. Not only can it not be used at the first time point equilibria ceased to exist, but it also cannot recover, even if what caused the problem was the particular input at that time point, which is bound to subsequently change into some other input that would no longer cause any trouble. This is so because an equilibria stream requires the existence of an equilibrium at every time point.

In this section, we address the problem of inexistent equilibria streams, also known as \emph{global inconsistency}. We begin by defining a notion of coherence associated with individual contexts which allows us to first establish sufficient conditions for the existence of equilibria streams, and then abstract away from problems due to specific incoherent contexts and focus on those problems essentially caused by the way the flow of information in \rMCSs is organized through its bridge rules. We introduce the notion of a \emph{repair}, which modifies an \rMCS by changing its bridge rules at some particular point in time in order to obtain some equilibria stream, which we dub \emph{repaired equilibria stream}. We establish sufficient conditions for the existence of repaired equilibria streams and briefly discuss different possible strategies to define such repairs. However, repaired equilibria streams may not always exist either, e.g., because some particular context is incoherent. To deal with such situations, we relax the concept of equilibria stream and introduce the notion of \emph{partial equilibria stream}, which essentially allows the non-existences of equilibria at some time points. It turns out that \emph{partial equilibria streams} always exist thus solving the problem of global inconsistency for \rMCSs.

In related work, the problem of global inconsistency has been addressed in the context of mMCSs \cite{EiterFSW14} by establishing sufficient conditions for the existence of equilibria.
We also follow that idea, but among the two notions established in \cite{EiterFSW14}, \emph{diagnosis} and \emph{explanation}, the former corresponding to rules that need to be altered to restore consistency, and the latter corresponding to combinations of rules that cause inconsistency, we focus on the former.
This is justified by the fact that the two notions turn out to be dual of each other, and somehow correspond to our notion of repair. The main difference here is that we opt to not allow the (non-standard) strengthening of bridge-rule to restore consistency, and, of course, the fact that our repairs need to take into account the dynamic nature of \mbox{\rMCSs}.

We start by introducing two notions of global consistency differing only on whether we consider a particular input stream or all possible input streams.

\begin{definition}
Let $M$ be an \rMCS, $\TKB$ a configuration of knowledge bases for $M$, and $\inptStr$ an input stream for $M$.
Then, $M$ is \emph{consistent} with respect to $\TKB$ and $\inptStr$ if there exists an equilibria stream of $M$ given $\TKB$ and $\inptStr$.
$M$ is \emph{strongly consistent} with respect to $\TKB$ if, for every input stream $\inptStr$ for $M$, $M$ is consistent with respect to $\TKB$ and $\inptStr$.
\end{definition}

Obviously, for a fixed configuration of knowledge bases, strong consistency implies consistency w.r.t.\ any input stream, but not vice-versa.

Unfortunately, verifying strong consistency is in general highly complex since it requires checking all possible equilibria streams.
Nevertheless, we can establish conditions that ensure that an \rMCS\ $M$ is strongly consistent with respect to a given configuration of knowledge bases $\TKB$, hence guaranteeing the existence of an equilibria stream independently of the input. It is based on two notions -- \emph{totally coherent contexts} and \emph{acyclic \rMCSs} -- that together are sufficient to ensure (strong) consistency.

Total coherence imposes that each knowledge base of a context always has at least one acceptable belief set.

\begin{definition}
A context $\Ctxt_i$ is \emph{totally coherent} if $\acc_i(\KB)\not=\emptyset$, for every $\KB\in \SKB_i$. 
\end{definition}

The second notion describes cycles between contexts which may be a cause of inconsistency.
Acyclic \rMCSs are those whose bridge rules have no cycles.

\begin{definition}
Given an \rMCS $M=\tuple{\tuple{\Ctxt_1,\ldots,\Ctxt_\CtxtIn}, \TIL, \TBR}$, $\triangleleft_M$ is the binary relation over contexts of $M$ such that $(\Ctxt_i,\Ctxt_j)\in\triangleleft_M$ if  there is a bridge rule $r\in\SBR_i$ and $\BRBA{j}{\Bel}\in\BRBd{r}$ for some ${\Bel}$. If $(\Ctxt_i,\Ctxt_j)\in\triangleleft_M$, also denoted by $\Ctxt_i\triangleleft_M\Ctxt_j$, we say that $\Ctxt_i$ depends on $\Ctxt_j$ in $M$, dropping the reference to $M$ whenever unambiguous.
\end{definition}

\begin{definition}
An \rMCS $M$ is \emph{acyclic} if the transitive closure of $\triangleleft_M$ is irreflexive.
\end{definition}

We can show that acyclicity and total coherence together are indeed sufficient to ensure strong consistency.

\begin{proposition}\label{prop:ConditionsForStrongConsistence}
Let $M=\tuple{\tuple{\Ctxt_1,\ldots,\Ctxt_\CtxtIn}, \TIL, \TBR}$ be an acyclic \rMCS such that every $\Ctxt_i$, $1\leq i\leq n$, is totally coherent, and $\TKB$ a configuration of knowledge bases for $M$. Then, $M$ is strongly consistent with respect to $\TKB$.
\end{proposition}

Nevertheless, these conditions are rather restrictive since there are many useful cyclic \rMCSs which only under some particular configurations of knowledge bases and input streams may have no equilibria streams.  

To deal with these, and recover an equilibria stream, one possibility is to repair the \rMCSs by locally, and selectively, eliminating some of its bridge rules. 
Towards introducing the notion of \emph{repair}, given an \rMCS\  $M=\tuple{\tuple{\Ctxt_1,\ldots,\Ctxt_\CtxtIn}, \TIL, \TBR}$, we denote by $br_M$ the set of all bridge rules of $M$, i.e., $br_M=\bigcup_{1\leq i\leq \CtxtIn} \SBR_i$. Moreover, given a set $R\subseteq br_M$, denote by $M[R]$ the \rMCS obtained from $M$ by restricting the bridge rules to those not in $R$.

\begin{definition}[Repair]\label{def:Repair}

Let $M=\tuple{\TCtxt, \TIL, \TBR}$ be an \rMCS, $\TKB$ a configuration of knowledge bases for $M$, and $\inptStr$ an input stream for $M$ until $\tau$ where
$\tau\in\mathbb{N}\cup\{\infty\}$. 
Then, a repair for  $M$ given $\TKB$ and $\inptStr$ is a function ${\Repair:[1..\tau]\to 2^{br_M}}$ such that
there exists a function ${\EqStr:[1..\tau]\to \BelForM{M}}$ such that
\begin{itemize}
\item $\EqStr^t$ is an equilibrium of $M[\Repair^t]$ given $\KBStr^t$ and $\inptStr^t$, 
where $\KBStr^t$ is inductively defined as
\begin{itemize}
\item $\KBStr^1=\TKB$
\item $\KBStr^{t+1}=\upd_{M[\Repair^{t}]}(\KBStr^{t},\inptStr^{t},\EqStr^{t})$. 
\end{itemize}
\end{itemize}
We refer to $\EqStr$ as a repaired equilibria stream of  $M$ given $\TKB$, $\inptStr$ and $\Repair$.
\end{definition}

Note the generality of this notion, which considers to be a \emph{repair} essentially any sequence of bridge rules (defined by the repair function $\Repair$) that, if removed from the \rMCS at their corresponding time point, will allow for an equilibrium at that time point. This may include repairs that unnecessarily eliminate some bridge rules, and even the \emph{empty repair} i.e. the repair $\Repair_{\emptyset}$ such that $\Repair^{t}_{\emptyset}=\emptyset$ for every $t$, whenever $M$ already has an equilibria stream given $\TKB$ and $\inptStr$. This ensures that the set of repaired equilibria streams properly extends the set of equilibria streams, since equilibria streams coincide with repaired equilibria streams given the empty repair. 

\begin{proposition}\label{prop:emptyRepair}
Every equilibria stream of $M$ given $\TKB$ and $\inptStr$ is a repaired equilibria stream of $M$ given $\TKB$, $\inptStr$ and the empty repair $\Repair_{\emptyset}$.
\end{proposition}

It turns out that for \rMCSs composed of totally coherent contexts, repaired equilibria streams always exist.

\begin{proposition}\label{prop:existenceRepairedEquilibria}
Let $M=\tuple{\tuple{\Ctxt_1,\ldots,\Ctxt_\CtxtIn}, \TIL, \TBR}$ be an \rMCS such that  each $\Ctxt_i$, $i\in\{1,\ldots,n\}$, is totally coherent, $\TKB$ a configuration of knowledge bases for $M$, and $\inptStr$ an input stream for $M$ until $\tau$. Then, there exists ${\Repair:[1..\tau]\to 2^{br_M}}$ and ${\EqStr:[1..\tau]\to \BelForM{M}}$ such that $\EqStr$ is a repaired equilibria stream given $\TKB$, $\inptStr$ and $\Repair$.
\end{proposition}

Whenever repair operations are considered in the literature, e.g., in the context of databases \cite{DBLP:conf/pods/ArenasBC99}, there is a special emphasis on seeking repairs that are somehow minimal, the rational being that we want to change things as little as possible to regain consistency. In the case of repairs of \rMCS, it is easy to establish an order relation between them, based on a comparison of the bridge rules to be deleted at each time point.

\begin{definition}
Let $\Repair_a$ and $\Repair_b$ be two repairs for some \rMCS $M$ given a configuration of knowledge bases for $M$, $\TKB$ and $\inptStr$, an input stream for $M$ until $\tau$. We say that $\Repair_a\leq\Repair_b$ if $\Repair_a^{i}\subseteq\Repair_b^{i}$ for every $i\leq\tau$, and that $\Repair_a <\Repair_b$ if $\Repair_a\leq\Repair_b$ and $\Repair_a^{i}\subset\Repair_b^{i}$ for some $i\leq\tau$.
\end{definition}

This relation can be straightforwardly used to check whether a repair is minimal, and we can restrict ourselves to adopting minimal repairs. However, there may be good reasons to adopt non-minimal repairs, e.g., so that they can be determined \emph{as we go}, or so that \emph{deleted} bridge rules are not reinstated, etc. Even though investigating specific types of repairs falls outside the scope of this paper, we nevertheless present and briefly discuss some possibilities. 

\begin{definition}[Types of Repairs]
Let $\Repair$ be a repair for some \rMCS $M$ given $\TKB$ and $\inptStr$. We say that $\Repair$ is a:
\begin{description}
\item[Minimal Repair] if there is no repair $\Repair_a$ for $M$ given $\TKB$ and $\inptStr$ such that $\Repair_a <\Repair$.
\item[Global Repair] if $\Repair^{i}=\Repair^{j}$ for every $i,j\leq\tau$. 
\item[Minimal Global Repair] if $\Repair$ is global and there is no global repair $\Repair_a$ for $M$ given $\TKB$ and $\inptStr$ such that $\Repair_a <\Repair$.
\item[Incremental Repair] if $\Repair^{i}\subseteq\Repair^{j}$ for every $i\leq j\leq\tau$.
\item[Minimally Incremental Repair] if $\Repair$ is incremental and there is no incremental repair $\Repair_a$ and  $j\leq\tau$ such that $\Repair_a^{i}\subset\Repair^{i}$ for every $i\leq j$.
\end{description}
\end{definition}

\emph{Minimal repairs} perhaps correspond to the ideal situation in the sense that they never unnecessarily remove bridge rules. In some circumstances, it may be the case that if a bridge rule is somehow involved in some inconsistency, it should not be used at any time point, leading to the notion of \emph{global repair.} Given the set of all repairs, checking which are global is also obviously less complex than checking which are minimal. A further refinement -- \emph{minimal global repairs} -- would be to only consider repairs that are minimal among the global ones, which would be much simpler to check than checking whether it is simply minimal. Note that a minimal global repair is not necessarily a minimal repair. One of the problems with these types of repairs is that we can only check whether they are of that type globally, i.e., we can only check once we know the entire input stream $\inptStr$. This was not the case with \emph{plain} repairs, as defined in Definition~\ref{def:Repair}, which could be checked \emph{as we go}, i.e., we can determine what bridge rules to include in the repair at a particular time point by having access to the input stream $\inptStr$ up to that time point only. This is important so that \rMCSs can be used to effectively react to their environment. The last two types of repairs defined above allow for just that. \emph{Incremental repairs} essentially impose that removed bridge rules cannot be reused in the future, i.e., that the set of removed bridge rules monotonically grows with time, while \emph{minimally incremental repairs} further impose that only minimal sets of bridge rules can be added at each time point.
Other types of repairs could be defined, e.g., by defining some priority relation between bridge rules, some distance measure between subsets of bridge rules and minimize it when considering the repair at consecutive time points, among many other options, whose investigation we leave for future work.  Repairs could also be extended to allow for the strengthening of bridge rules, besides their elimination, generalizing ideas from \cite{EiterFSW14} and \cite{BrewkaEFW11} where the extreme case of eliminating the entire body of bridge rules as part of a repair is considered.
 
Despite the existence of repaired equilibria streams for large classes of systems, two problems remain: first, computing a repair may be excessively complex, and second, there remain situations where no repaired equilibria stream exists, namely when the \rMCS contains contexts that are not totally coherent. The second issue could be dealt with by ensuring that for each non-totally coherent context there would be some bridge rule with a management operation in its head that would always restore consistency of the context, and that such rule could always be \emph{activated} through a repair (for example, by adding a negated reserved atom to its body, and another bridge rule with that atom in its head and an empty body, so that removing this latter rule through a repair would activate the management function and restore consistency of the context). But this would require special care in the way the system is specified, and its analysis would require a very complex analysis of the entire system including the specific behavior of management functions. In practice, it would be quite hard -- close to impossible in general -- to ensure the existence of repaired equilibria streams, and we would still be faced with the first problem, that of the complexity of determining the repairs.

A solution to this problem is to relax the notion of equilibria stream so that it does not require an equilibrium at every time point. This way, if no equilibrium exists at some time point, the equilibria stream would be undefined at that point, but possibly defined again in subsequent time points. This leads to the following notion of \emph{partial equilibria stream}.\footnote{The notion of partial equilibria has also been used in the static case in \cite{DaoTranEFK15}, but in a quite different sense, namely to focus only on those contexts that are relevant for determining the semantics of a given context $C$.}

\begin{definition}[Partial Equilibria Stream]\label{def:PartialStreamOfEquilibria}
Let $M=\tuple{\TCtxt, \TIL, \TBR}$ be an \rMCS, $\TKB=\tuple{\KB_1,\ldots,\KB_\CtxtIn}$ a configuration of knowledge bases for $M$, and $\inptStr$ an input stream for $M$ until $\tau$ where
$\tau\in\mathbb{N}\cup\{\infty\}$. 
Then, a \emph{partial equilibria stream of $M$ given $\TKB$ and $\inptStr$} is a partial function ${\EqStr:[1..\tau]\nrightarrow \BelForM{M}}$ such that

\begin{itemize}
\item $\EqStr^t$ is an equilibrium of $M$ given $\KBStr^t$ and $\inptStr^t$, 
where $\KBStr^t$ is inductively defined as
\begin{itemize}
\item $\KBStr^1=\TKB$
\item $  \KBStr^{t+1}=\begin{cases}
    \upd_M(\KBStr^{t},\inptStr^{t},\EqStr^{t}),& \text{if $\EqStr^{t}$ is not undefined}.\\
    \KBStr^{t}, & \text{otherwise}.
  \end{cases}$
\end{itemize}
\item or $\EqStr^t$ is undefined.
\end{itemize}

\end{definition}

As expected, this is a proper generalization of the notion of equilibria stream:

\begin{proposition}\label{prop:equilibriaStreamPartialEquilibria}
Every equilibria stream of $M$ given $\TKB$ and $\inptStr$ is a partial equilibria stream of $M$ given $\TKB$ and $\inptStr$.
\end{proposition}

And it turns out that partial equilibria streams always exist.

\begin{proposition}\label{prop:partialEquilibriaExistence}
Let $M$ be an \rMCS, $\TKB$ a configuration of knowledge bases for $M$, and $\inptStr$ an input stream for $M$ until $\tau$. Then, there exists ${\EqStr:[1..\tau]\nrightarrow \BelForM{M}}$ such that $\EqStr$ is a partial equilibria stream given $\TKB$ and $\inptStr$.
\end{proposition}

One final word to note is that partial equilibria streams not only allow us to deal with situations where equilibria do not exist at some time instants, but they also open the ground to consider other kinds of situations where we do not wish to consider equilibria at some time point, for example because we were not able to compute them on time, or simply because we do not wish to process the input at every time point, e.g., whenever we just wish to sample the input with a lower frequency than it is generated. If we wish to restrict that partial equilibria streams only relax equilibria streams when necessary, i.e., when equilibria do not exist at some time point, we can further impose the following condition on Definition~\ref{def:PartialStreamOfEquilibria}:
\[
\EqStr^t\text{ is undefined} \implies \text{ there is no equilibrium of $M$ given $\KBStr^t$ and $\inptStr^t$}.
\]

\section{Non-Determinism and Well-Founded Semantics}\label{sec-wfs}

Reactive MCSs as considered so far are non-deterministic for two reasons.
On the one hand, we allow for contexts whose semantics may return multiple belief sets for the same knowledge base, and on the other hand, the flow of information between contexts established by bridge rules may be the source of non-determinism.
As this leads to multiple equilibria and thus to exponentially many equilibria streams, in practice, this may be undesired, which is why it is important to study under which conditions non-determinism can be avoided.

The first source of non-determinism solely depends on the choice of the contexts occurring in the \rMCS, i.e., it can be avoided when determining/designing the \rMCS in question.
The second source, however, requires to consider the entire \rMCS when eliminating the non-determinism.  
This can be achieved by introducing preferences on equilibria using, e.g., preference functions as proposed by Ellmauthaler~\cite{Ellmauthaler2013a}.
One might also adopt language constructs for expressing preferences in ASP such as optimization statements~\cite{GKKOST10} or weak constraints~\cite{BuccafurriLR97}, which essentially assign a quality measure to equilibria, or, more recently, add a pre-order on the contexts \cite{MuWW16}. 
Alternatively, and inspired by notions developed for MCSs \cite{BrewkaE07}, we may consider restrictions on \rMCSs such that these non-determinisms do not occur in the first place.
In the remainder of this section, we will focus on such restrictions leading to an alternative well-founded semantics for \rMCSs.

As a first step towards this objective, only equilibria that are subset-minimal will be considered, also with the aim of avoiding unnecessary self-justifications of information resulting from the interaction of various contexts.
Then, grounded equilibria as a special case for so-called reducible MCSs will be presented for which the existence of minimal equilibria can be effectively checked.
Subsequently, a well-founded semantics for such reducible MCSs will be defined under which an approximation of all grounded equilibria can be computed deterministically.
In the following, we transfer these notions from static MCSs in \cite{BrewkaE07} to dynamic \rMCSs and discuss under which (non-trivial) conditions they actually can be applied.

We start with the notion of minimal equilibria.
Formally, given an \rMCS $M=\tuple{\tuple{\Ctxt_1,\ldots,\Ctxt_\CtxtIn}, \TIL, \TBR}$, a configuration of knowledge bases \TKB for $M$ and an input \Tinpt for $M$, an equilibrium $\TBelS=\tuple{\BelS_1, \ldots, \BelS_\CtxtIn}$ of $M$ given $\TKB$ and $\Tinpt$ is \emph{minimal} if there is no equilibrium $\TBelS' = \tuple{\BelS_1', \ldots, \BelS_\CtxtIn'}$ of $M$ given $\TKB$ and $\Tinpt$ such that $\BelS_i' \subseteq \BelS_i$ for all $i$ with $1 \leq i \leq \CtxtIn$ and $\BelS_j' \subsetneq \BelS_j$ for some $j$ with $j\in \{1 \ldots \CtxtIn\}$.

This notion of minimality avoids to some extent unnecessary self-justifications in equilibria (streams). 
A simple example of such self-justification is shown in Example~2 in \cite{BrewkaE07}, which can be transcribed to our setting using a simple \rMCS without input languages, a single propositional context $\Ctxt_1$ with empty $\KB$ and $\SBR_1$ only containing $\op{add}(a) \la \BRBA{1}{a}$.
Then both $\tuple{\emptyset}$ and $\tuple{\{a\}}$ are equilibria.
The latter is considered dubious as $a$ is justified by itself and minimality would avoid such self-justification.
Note that this does not always work.
\begin{example}\label{ex-selfJust}
A single storage context with empty $\KB$ and bridge rules $\op{add}(b)\la \naf\ \BRBA{1}{a}.$ and $\op{add}(a) \la \BRBA{1}{a}.$ has two minimal equilibria $\tuple{\set{a}}$ and $\tuple{\set{b}}$, the former being self-justified.
\end{example}
Still, avoiding non-minimality whenever possible certainly is a decent guiding principle, and, as we will see later, it can be avoided altogether under certain conditions. 

The problem is that checking for minimality commonly raises the computational complexity of determining equilibria.
To avoid that, we now formalize conditions under which minimal equilibria can be effectively checked.

For that purpose, we start by introducing the notion of monotonic logics.
Namely, a logic $\logic=\tuple{\SKB,\SBelS, \acc}$ is \emph{monotonic} if 
\begin{enumerate}
	\item $\acc(\KB)$ is a singleton set for each $\KB \in \SKB$, and

	\item $\BelS \subseteq \BelS'$ whenever $\KB \subseteq \KB'$,
		$\acc(\KB) = \set{\BelS}$, and $\acc(\KB') = \set{\BelS'}$.
\end{enumerate}
In other words, $\logic$ is monotonic if $\acc$ is deterministic and monotonic.

Maybe not surprisingly, the logics of non-monotonic formalisms such as $\logic_a$ for the answer set semantics in Example~\ref{ex-logics} do not satisfy this definition.
We therefore proceed with introducing the notion of a reducible logic, which covers monotonic logics, but also includes logics that can be reduced to monotonic ones given a belief set of the logic.

Logic $\logic=\tuple{\SKB,\SBelS, \acc}$ is \emph{reducible} 
iff, for some $\SKB^* \subseteq \SKB$ and some reduction function $\redfun : \SKB \times \SBelS\rightarrow \SKB^*$, 
\begin{enumerate}
	\item $\tuple{\SKB^*,\SBelS, \acc}$ is monotonic, 
	\item for each $\KB \in \SKB$, and all $\BelS, \BelS' \in \SBelS$:
	\begin{itemize}
	    \item $\redfun(\KB, \BelS) = \KB$ whenever $\KB \in \SKB^*$,
	    \item $\redfun(\KB, \BelS) \subseteq \redfun(\KB, \BelS')$ whenever $\BelS'\subseteq \BelS$, and
	    \item $\acc(\redfun(\KB, \BelS)) = \set{\BelS}$ iff $\BelS \in \acc(\KB)$.
	\end{itemize}    
\end{enumerate}
Intuitively, $\logic$ is reducible, if a) the restriction of $\logic$ to $\SKB^*$ is monotonic, and b) there is a reduction function which should not reduce $\KB$ further if it already is in $\SKB^*$, which is antitonic, and by means of which acceptability of a belief set can be checked by the reduction (see also \cite{BrewkaE07}).

In a reducible \rMCS, the logics of all contexts have to be reducible.
Additionally, we require $\redfun$ and $\mng$ to be applicable in arbitrary order for all contexts, and that sequences of increasing sets of operations are monotonic in the following sense.
Given \rMCS $M=\tuple{\tuple{\Ctxt_1,\ldots,\Ctxt_\CtxtIn}, \TIL, \TBR}$ and, for some $i\in \{1,\ldots n\}$, $\KB\in\SKB_i$, we say that $\SOp\subseteq\SOp_i$ in context $\Ctxt_i$ is \emph{monotonic w.r.t.\ $\KB$}, if $\KB\subseteq \mng(\SOp,\KB)$.
Intuitively, a set of monotonic operations will not remove content from the knowledge base to which it is applied.
\begin{definition}\label{def:reducible}
Let $M=\tuple{\tuple{\Ctxt_1,\ldots,\Ctxt_\CtxtIn}, \TIL, \TBR}$ be an \rMCS, $\TKB$ a configuration of knowledge bases for $M$ and $S_i=\{\BRHd{r}\in\SOp_i \mid  r\in \SBR_i\}$ for all $i\in\{1,\ldots,\CtxtIn\}$.
Then, $M$ is \emph{reducible given $\TKB$} iff, for all $i\in\{1,\ldots,\CtxtIn\}$, 
\begin{enumerate}
\item $\logic_i$ is reducible; 
\item for all $\SOp\subseteq S_i$, all $\KB\in \SKB_i$, and all $\BelS_i\in \SBelS_i$:
\[\redfun_i(\mng_i(\SOp,\KB), \BelS_i) = \mng_i(\SOp,\redfun_i(\KB, \BelS_i));\]
\item for all $\SOp^1\subseteq \ldots \subseteq \SOp^m\subseteq S_i$ and all $j\in \{1,\ldots,m\}$: $\SOp^j$ is monotonic w.r.t.\ $\KB_i^j$ where
\begin{enumerate}
\item $\KB_i^1=\KB_i$;
\item $\KB_i^h=\mng_i(\SOp^{h-1},\KB_i^{h-1})$ for $h\in\{2,\ldots,m\}$.
\end{enumerate}
 \end{enumerate}
\end{definition}
The second condition essentially requires that no $\mng_i$ introduces formulas that are affected by $\redfun_i$. 
The third condition in addition checks whether any sequence of increasing sets of (applicable) operations is monotonic, i.e., whether some $\KB_i^m$ can be obtained in a step-wise, monotonic iteration.

\begin{example}\label{ex-reducible}
Recall $M_{al}$ from Section~\ref{subsec:scenario}, for which we want to check whether it is reducible.
First, all contexts but $\Ctxt_{ec}$ build on a monotonic logic, so their logics are automatically reducible.
This also means that the second condition is trivially satisfied for these contexts using the identity function for $\redfun_i$.
At the same time, $\logic_{ec}$ is reducible using the well-known Gelfond-Lifschitz reduct \cite{GelfondL91}, and clearly none of the bridge rules in $\SBR_{ec}$ can introduce any knowledge base formula (for any $\KB\in \SKB_{ec}$) which is affected by such $\redfun_{ec}$.
More care needs to be taken regarding the third condition: in fact, e.g., applying $\op{setTemp(\term{hot})}$ to $\{\lit{tm}{cold}\}$ yields $\{\lit{tm}{hot}\}$, which is clearly not monotonic.
This can however be avoided if it is ensured that $\KB_{st}$ does not contain information on the temperature.
Then, any increasing sequence of operations is monotonic w.r.t.\ the corresponding $\KB_i^j$ (cf.\ Example~\ref{ex-context}).
The same is not true for $\SBR_{pos}$ which is why we have to omit the bridge rules $\op{setPos(\term{P})} \la \BRSA{pos}{\lit{enters}{P}}.$ in $\SBR_{pos}$, thus only preventing the information of Dave changing the room from being available right away.
With this slight adjustment, the \rMCS $M_{al}$ is in fact reducible.
\end{example}

Following notions from logic programming, we introduce definite \rMCSs.
\begin{definition}\label{def:definiteMCS}
Let $M=\tuple{\tuple{\Ctxt_1,\ldots,\Ctxt_\CtxtIn}, \TIL, \TBR}$ be a reducible \rMCS given $\TKB$. 
Then, $M$ is \emph{definite given $\TKB$} iff, for all $i\in\{1,\ldots,\CtxtIn\}$, 
\begin{enumerate}
    \item no $r\in \TBR_i$ contains $\naf$;
	\item for all $\BelS\in\SBelS_i$: $\KB_i = \redfun(\KB_i,\BelS)$.
\end{enumerate}
\end{definition}
Thus, in a definite \rMCS (given $\TKB$), the bridge rules are monotonic and all knowledge bases are already reduced. 
This suggests the following iteration.

\begin{definition}\label{def:definiteIteration}
Let $M$ be a definite \rMCS given $\TKB$, and $\Tinpt$ an input for $M$.
For all $i\in\{1\ldots \CtxtIn\}$, let $\KB_i^0=\KB_i$ and define, for each successor ordinal $\alpha+1$,
\[
\KB_i^{\alpha+1} = \mng(\app_i^{now}(\Tinpt, \TBelS^\alpha),\KB_i^{\alpha}), 
\]
where $\TBelS^\alpha=\tuple{\BelS^\alpha_1,\ldots,\BelS^\alpha_n}$ and $\acc_i(\KB_i^\alpha)=\{\BelS^\alpha_i\}$ for any ordinal $\alpha$.
Furthermore, for each limit ordinal $\alpha$, define $\KB^\alpha_i=\bigcup_{\beta\leq\alpha}\KB^\beta_i$, and let $\KB^\infty_i=\bigcup_{\alpha>0}\KB_i^\alpha$.
\end{definition}
\noindent
We next show two properties of the iteration defined in Definition~\ref{def:definiteIteration} that will prove useful subsequently.

\begin{lemma}\label{lemma:iterationProps}
Let $M$ be a definite \rMCS given $\TKB$, and $\Tinpt$ an input for $M$.
The following holds for the iteration in Definition~\ref{def:definiteIteration} for all ordinals $\alpha$:
\begin{enumerate}
\item $M$ is definite given $\TKB^\alpha$;\label{itProps1}
\item for any ordinal $\beta$ with $\beta\leq\alpha$ we have $\KB^\beta_i\subseteq \KB^\alpha_i$ for $i\in\{1,\ldots,\CtxtIn\}$.\label{itProps2}
\end{enumerate}
\end{lemma}

Lemma~\ref{lemma:iterationProps} guarantees that the iteration is well-defined and monotonic and we can show that it yields an equilibrium.
In fact, it is the unique minimal equilibrium of $M$ given $\TKB$ and $\Tinpt$.

\begin{proposition}\label{prop:GEuniqueForDefinite}
Let $M$ be a definite \rMCS given $\TKB$, and $\Tinpt$ an input for $M$. 
Then $\tuple{\BelS^\infty_1,\ldots,\BelS^\infty_n}$ is the unique minimal equilibrium of $M$ given $\TKB$ and $\Tinpt$.
We call it \emph{grounded equilibrium of $M$ given $\TKB$ and $\Tinpt$}, denoted by $\geql(M,\TKB,\Tinpt)$.
\end{proposition}

\noindent
As pointed out in \cite{BrewkaE07}, for many logics, $\KB^\infty_i=\KB^\omega_i$ holds and the iteration even stops after finitely many steps.
This is also the case for the slightly revised scenario in Example~\ref{ex-reducible}.
Moreover, as a consequence of Proposition~\ref{prop:GEuniqueForDefinite}, no self-justification can occur in the grounded equilibrium of definite \rMCSs.

This fixpoint iteration cannot be applied to arbitrary reducible \rMCSs right away as, e.g., $\naf$ in the bridge rule bodies and non-reduced knowledge bases may allow the removal of already derived information in the iteration.
To counter that, we introduce the notion of a reduct for \rMCSs where, for a bridge rule of the form (\ref{bridgerule}), $\BRBd{r}^+=\{a_1,\ldots, a_j\}$) and $\BRBd{r}^-=\{a_{j+1},\ldots,a_m\}$.

\begin{definition}\label{def:reduct}
Let $M= \tuple{\TCtxt, \TIL, \tuple{\SBR_1,\ldots,\SBR_\CtxtIn}}$ be a reducible \rMCS given $\TKB$, $\Tinpt$ an input for $M$, and $\TBelS =\tuple{\BelS_1, \dotsc, \BelS_\CtxtIn}$ a belief state of $M$. 
The \emph{$(\Tinpt,\TBelS)$-reduct of $M$ and $\TKB$} is obtained as $M^{(\Tinpt,\TBelS)} = \tuple{\TIL, \TCtxt, \tuple{\SBR_1^{(\Tinpt,\TBelS)},\ldots,\SBR_\CtxtIn^{(\Tinpt,\TBelS)}}}$ where $\SBR_i^{(\Tinpt,\TBelS)} =\{\BRHd{r}\la \BRBd{r}^+ \mid  r\in \TBR_i, 
\BRHd{r}=\Op$ with $\Op\in\SOp_i, \tuple{\Tinpt, \TBelS} \not\models a_\ell$ for all $a_\ell\in \BRBd{r}^-\}$ 
and as $\TKB^{(\Tinpt,\TBelS)}=\tuple{\KB_1^{\BelS_1},\ldots,\KB_n^{\BelS_n}}$ where $\KB_i^{\BelS_i}=\redfun_i(\KB_i,\BelS_i)$.
\end{definition}
\noindent
Note that bridge rules with an operation under next in its head are ignored here, as these do not affect the computation of the equilibria anyway.

For all reducible \rMCSs $M$ given $\TKB$, all inputs $\Tinpt$ for $M$, and all belief states $\TBelS$ for $M$, the $(\Tinpt,\TBelS)$-reduct of $M$ and $\TKB$ is definite.
In this case, we can check whether $\TBelS$ is a grounded equilibrium for $M$ given $\TKB$ and $\Tinpt$ in the usual manner.

\begin{definition}\label{def:groundedRed}
	Let $M=\tuple{\TCtxt, \TIL, \TBR}$ be a reducible \rMCS given $\TKB$, and $\Tinpt$ an input for $M$. 
	A belief state $\TBelS$ of $M$ is a 
	\emph{grounded equilibrium of $M$} given $\TKB$ and $\Tinpt$ iff $\TBelS = \geql(M^{(\Tinpt,\TBelS)},\TKB^{(\Tinpt,\TBelS)},\Tinpt)$.
\end{definition}
Grounded equilibria of reducible \rMCSs given some $\TKB$ are also minimal.

\begin{proposition}\label{prop:grEqIsMinEq}
Every grounded equilibrium of a reducible \rMCS $M$ given $\TKB$ and an input $\Tinpt$ is a minimal equilibrium of $M$ given $\TKB$ and $\Tinpt$.
\end{proposition}

\noindent
Grounded equilibria of reducible \rMCSs are not unique equilibria in general.
Consider, e.g., again Example~\ref{ex-selfJust}.
Then $\tuple{\{b\}}$ is a grounded equilibrium, while $\tuple{\{a\}}$ is not.
Still, as checking for grounded equilibria relies on the unique grounded equilibrium of the reduct, we know that no self-justification can occur in grounded equilibria, which is also why $\tuple{\{a\}}$ is filtered.

We can now introduce grounded equilibria streams provided that \rMCS $M$ given $\TKB$ is reducible for each $\TKB$ occurring in $\KBStr$.
Unfortunately, checking for reducibility cannot be done a priori as the $\TKB$ will only be known at runtime (due to 3.\ of Definition~\ref{def:reducible}).
To complete the picture, we also spell out under which conditions an initially reducible \rMCS given $\TKB$ remains reducible.

\begin{definition}
Let $M=\tuple{\TCtxt, \TIL, \TBR}$ be a reducible \rMCS given $\TKB$.
Then $M$ is \emph{persistently reducible given $\TKB$} iff for all sequences $\SOp^1,\ldots \SOp^m$ with $m>0$ and $\SOp^j\subseteq\{\Op \mid  \Nxt{\Op}\in\BRHd{r}, r\in \SBR_i\}$, $M$ is reducible given $\TKB^j$ where $\TKB^1=\TKB$ and $\TKB^j=\tuple{\KB_1^j,\ldots,\KB_\CtxtIn^j}$ for $j>1$ with $\KB_i^j=\mng(\SOp^{j-1},\KB_i^{j-1})$.
\end{definition}
Thus, any reducible \rMCS (given some concrete $\TKB$) is persistently reducible if applying any arbitrary sequence of the operations under \Nxt{} occurring in the bridge rules yields again a reducible \rMCS.

It should be noted that checking for reducible \rMCSs is not trivial in general and even less so for persistently reducible \rMCSs. 
Still, certain kinds of contexts clearly satisfy the restrictions including those occurring in the example scenario in Example~\ref{ex-reducible}.
Namely, they either store certain pieces of information, but only do change them using \Nxt{} and not otherwise, or their $\KB$ is initially empty and this is never changed using \Nxt{}, so information is only stored temporarily for the computation of the current equilibrium.
The scenario is thus persistently reducible for any knowledge base configuration which does not store pieces of information in the initial $\KB_i$ for contexts of the latter kind.

This allows us to introduce grounded equilibria streams.

\begin{definition}
Let $M=\tuple{\TCtxt, \TIL, \TBR}$ be a persistently reducible \rMCS given $\TKB$, $\inptStr$ an input stream for $M$ until $s$, $\EqStr=\tuple{\EqStr^1,\ldots,\EqStr^s}$ an equilibria stream of $M$ given $\TKB$ and $\inptStr$, and $\KBStr$ the configurations stream of $M$ given $\TKB$, $\inptStr$, and $\EqStr$. 
Then, $\EqStr$ is a \emph{grounded equilibria stream} of $M$ given $\TKB$ and $\inptStr$ iff, for each $t\in\{1\ldots s\}$, $\EqStr^t=\geql(M^{(\Tinpt,\TBelS)},(\KBStr^t)^{(\Tinpt,\TBelS)},\inptStr^t)$.
\end{definition}
A grounded equilibria stream of some $M$ given $\TKB$ and $\inptStr$ can thus be understood as a stream of grounded equilibria.
It is thus straightforward to see that each equilibrium in a grounded equilibria stream is minimal w.r.t.\ the knowledge base configuration and input of its time point.

We remark that the notion of persistently reducible \rMCSs substantially differs from reducible MCSs on which the former are founded.
On the one hand, permitting operations in $\mng$ beyond simple addition, as used already for mMCSs, requires the non-trivial tests for reducible \rMCSs, and in addition, persistent reducibility needs to be verified for the dynamic aspect of \rMCSs.
Thus, these new notions also extend the work in \cite{BrewkaE07} from MCSs to mMCSs.

For reducible \rMCSs in general, determining grounded equilibria is not deterministic yet, since we have to guess and check the grounded equilibrium in each step.
This is why we now introduce the well-founded semantics for reducible \rMCSs $M$ following the ideas in \cite{BrewkaE07}.
Its definition is based on the operator $\gamma_{M,\TKB,\Tinpt}(\TBelS)=\geql(M^{(\Tinpt,\TBelS)},\TKB^{(\Tinpt,\TBelS)},\Tinpt)$, provided $\SBelS_i$ for each logic $\logic_i$ in all the contexts of $M$ has a least element $\BelS^*_i$ (w.r.t.\ subset inclusion).
Such \rMCSs are called \emph{normal}.

It can be shown that $\gamma_{M,\TKB,\Tinpt}$ is antitonic which means that applying $\gamma_{M,\TKB,\Tinpt}$ twice yields a monotonic operator.
Hence, by the Knaster-Tarski theorem, $(\gamma_{M,\TKB,\Tinpt})^2$ has a least fixpoint which determines the well-founded semantics.

\begin{definition}
Let $M$ be a normal, reducible \rMCS given $\TKB$, and $\Tinpt$ an input for $M$.
The \emph{well-founded model of $M$ given $\TKB$ and $\Tinpt$}, denoted $\wfs(M,\TKB,\Tinpt)$, is the least fixpoint of $(\gamma_{M,\TKB,\Tinpt})^2$.
\end{definition} 

Starting with the belief state $\TBelS^*=\tuple{\BelS^*_1,\ldots,\BelS^*_n}$, this fixpoint can be iterated, establishing the relation between $\wfs(M,\TKB,\Tinpt)$ and grounded equilibria of~$M$.

\begin{proposition}\label{prop:relWFSvsGrEq}
Let $M=\tuple{\TCtxt, \TIL, \TBR}$ be a normal, reducible \rMCS given $\TKB$, $\Tinpt$ an input for $M$, $\wfs(M,\TKB,\Tinpt)=\tuple{W_1,\ldots W_n}$, and $\TBelS=\tuple{\BelS_1,\ldots,\BelS_n}$ a grounded equilibrium of $M$ given $\TKB$ and $\Tinpt$.
Then $W_i\subseteq \BelS_i$ for $i\in\{1\ldots\CtxtIn\}$.
\end{proposition}

The well-founded model of $M$ can thus be viewed as the belief state representing what is accepted in all grounded equilibria, even though $\wfs(M,\TKB,\Tinpt)$ may itself not necessarily be a grounded equilibrium.
Yet, if the \rMCS is acyclic (i.e., no cyclic dependencies over bridge rules exist between beliefs in the \rMCS, see Sect.~\ref{sec-inconsMan}), then the grounded equilibrium of $M$ given $\TKB$ and $\Tinpt$ is unique and identical to the well-founded model.
This is indeed the case for the scenario in Example~\ref{ex-reducible}.

The well-founded semantics can be generalized to streams as follows.

\begin{definition}
Let $M=\tuple{\TCtxt, \TIL, \TBR}$ be a normal, persistently reducible \rMCS given $\TKB$, and $\inptStr$ an input stream for $M$ until $\tau$. 
The \emph{well-founded stream of $M$ given $\TKB$ and $\inptStr$} is a function ${\wfsStr:[1..\tau]\to \BelForM{M}}$ such that
\begin{itemize}
\item $\wfsStr^t$ is the well-founded model of $M$ given $\KBStr^t$ and $\inptStr^t$, 
where $\KBStr^t$ is defined as
\begin{itemize}
\item $\KBStr^1=\TKB$
\item $\KBStr^t=\upd_M(\KBStr^{t-1},\inptStr^{t-1},\wfsStr^{t-1})$, for $t>1$. 
\end{itemize}
\end{itemize} 
\end{definition}

Clearly, Proposition~\ref{prop:relWFSvsGrEq} also generalizes to the well-founded stream (of $M$ given $\TKB$ and $\Tinpt$) including all the made observations.
That is, the well-founded stream may not coincide with any (grounded) equilibria stream, unless the \rMCS in question is acyclic, in which case the grounded equilibria stream is unique and does coincide with the well-founded stream.
Again, this can be verified for the scenario in Example~\ref{ex-reducible}.
 
\section{Complexity}\label{sec:complexity}
In this section, we want to analyze the complexity of answering queries over equilibria streams of \rMCSs.
As our framework has many degrees of freedom, we need to impose some restrictions in order to get meaningful complexity results. 
In particular, we are only interested in decision problems with input of finite size, since otherwise the decision problem would unfortunately be undecidable right away.
Thus, in the following, we only consider \emph{finite} \rMCSs, i.e., we do not consider rule schemas (which stand potentially for an infinite amount of bridge rules), and assume that all knowledge bases in the given \rMCS are finite.
Also, we restrict our considerations to finite input streams.

We start by introducing the two reasoning problems we consider.
\begin{definition}\label{def:complDecProb}
The problem $\problemE$,
respectively $\problemA$,
is deciding whether for
a given finite \rMCS $M$, a belief $\Bel$ for the $k$-th context of $M$, a configuration of knowledge bases $\TKB$ for $M$, and an input stream $\inptStr$ until $\tau$,
it holds that
$\Bel\in \BelS_k$  for some $\EqStr^t=\tuple{\BelS_1,\dots,\BelS_\CtxtIn}$,  ($1\leq t\leq \tau$), for some, respectively all, equilibria stream(s) $\EqStr$ given $\TKB$ and $\inptStr$.
\end{definition}
As the complexity of an \rMCS depends on that of its individual contexts,
we introduce the notion of \define{context complexity} (\cf \cite{EiterFSW14}).
To do so, we need to focus on relevant parts of belief sets by means of projection.
Intuitively, among all beliefs, we only need to consider belief $\Bel$ that we want to query and the beliefs that
contribute to the application of bridge rules for deciding  $\problemE$ or $\problemA$.
Given $M$, $\Bel$, $k$, and $\inptStr$ as in Definition~\ref{def:complDecProb},
the \define{set of relevant beliefs} for a context $\Ctxt_i$ of $M$ is given by
\[
\begin{array}{r@{}l}
\RBS_i(M,\BRBA{k}{\Bel})=&\set{\Bel' \mid r\in \SBR_h,\BRBA{i}{\Bel'}\in\BRBd{r}\lor \naf\ \BRBA{i}{\Bel'}\in\BRBd{r},h\in\{1\ldots,\CtxtIn\}}\cup\\&\{\Bel\mid k=i\}.
\end{array}
\]
Then, a \define{projected belief state} for $M$ and $\BRBA{k}{b}$ is a tuple 
\[\TBelS_{\mid M}^{\BRBA{k}{b}}=\tuple{\BelS_1\cap \RBS_1(M,\BRBA{k}{\Bel}),\dots,\BelS_n\cap \RBS_n(M,\BRBA{k}{\Bel})}\] 
where $\TBelS=\tuple{\BelS_1,\dots,\BelS_n}$ is a belief state for $M$.
If $\TBelS$ is an equilibrium, then we call this tuple \emph{projected equilibrium}.
The \define{context complexity} of $\Ctxt_i$ in $M$ \wrt $\BRBA{k}{\Bel}$ for a fixed input $\Tinpt$ 
is the complexity of deciding the \define{context problem} of $\Ctxt_i$, that is, whether
for a given projected belief state $\TBelS=\tuple{\BelS_1,\dots,\BelS_n}$ for $M$ and $\BRBA{k}{\Bel}$,
there is some belief set $\BelS'_i$ for $\Ctxt_i$
with $\BelS_i=\BelS'_i\cap \RBS_i(M,\BRBA{k}{\Bel})$ and $\BelS'_i \in \acc_i(\mng_i(\app_i(\Tinpt,\TBelS),\KB_i))$.
The \emph{context complexity} $\mathcal{CC}(M,\BRBA{k}{\Bel})$ of an entire \rMCS is a (smallest) upper bound for the context complexity classes of its contexts.

\begin{theorem}\label{theo:genCompl}
Table~\ref{tab:complexity} summarizes the complexities of membership of problems $\problemE$ and $\problemA$ for finite input steams (until some $\tau\in\mathbb{N}$) depending on the context complexity.
Hardness also holds if it holds for the context complexity.
\end{theorem}

\begin{table}
\begin{center}
\renewcommand*{\arraystretch}{1.1}
\[
\begin{array}{|l|ll|}
\hline
\mathcal{CC}(M,\BRBA{k}{\Bel})&\problemE&\problemA\\
\hline
\mathbf{P} & \mathbf{NP} & \mathbf{coNP}\\
\mathbf{\Delta^{P}_{i}} (i\ge 2) & \mathbf{\Sigma^{P}_{i}} & \mathbf{\Pi^{P}_{i}}\\
\mathbf{\Sigma^{P}_{i}} (i\ge 1) & \mathbf{\Sigma^{P}_{i}} & \mathbf{\Pi^{P}_{i}}\\
\mathbf{PSPACE} & \mathbf{PSPACE} & \mathbf{PSPACE}\\
\mathbf{EXPTIME} & \mathbf{EXPTIME} & \mathbf{EXPTIME}\\
\hline
\end{array}
\]
{\caption{Complexity results of checking $\problemE$ and $\problemA$}
\label{tab:complexity}}
\end{center}
\end{table}

These results can also be used to show the complexity results of a strongly related problem, namely ensuring that a certain belief does always hold over all time instants for some or all equilibria streams.
This can be achieved by simply considering the co-problem of ensuring that the negated belief does not hold at any time instant for some or all equilibria streams.
In case the considered belief set itself does not allow us expressing this negated belief, an appropriate additional auxiliary context with adequate  bridge rules can be used to represent it and query for it. 

Note that allowing for input streams of infinite length leads to undecidability.
\begin{proposition}\label{prop:undecidability}
Given a finite \rMCS\ $M$, the problems $\problemE$ and $\problemA$ are undecidable for infinite input streams (when $\tau=\infty$).
\end{proposition}
The reason is that \rMCSs are expressive enough (even with very simple context logics) to simulate a Turing machine
such that deciding $\problemE$ or $\problemA$ for infinite runs solves the halting problem.
We provide an \rMCS that implements a Turing machine service in~\ref{app:tm}: 
a fixed \rMCS can read the configuration of a Turing machine $TM$ as well as the corresponding input and then simulate a run of $TM$ on that input.

For persistently reducible \rMCSs, it turns out that the complexity results for the analog problems $\problemE_g$ and $\problemA_g$ on grounded equilibria streams are identical to $\problemE$ and $\problemA$ on equilibria streams.

\begin{theorem}\label{theo-reducMCScompl}
Let $M=\tuple{\TCtxt, \TIL, \TBR}$ be a finite, persistently reducible \rMCS given $\TKB$.
Then membership and hardness results for $\problemE_g$ and $\problemA_g$ on grounded equilibria streams for finite input streams coincide with those for $\problemE$ and $\problemA$ in Theorem~\ref{theo:genCompl}. 
\end{theorem}

Note that this only holds if checking whether $M$ is persistently reducible is already known/can be neglected. 
While such assumption is a strong one in general, in Section~\ref{sec-wfs} it is argued that this can be done easily for contexts of certain kinds, including the ones used in the example scenario in Section~\ref{subsec:scenario} (and in Example \ref{ex-reducible}). 

Regarding the well-founded stream, we can restrict our attention to the problem $\problemE_{wf}$ for the well-founded stream, since it coincides with $\problemA_{wf}$ for the unique well-founded stream, and to polynomial contexts given the motivation for the well-founded stream. 

\begin{theorem}\label{theo-WFScompl}
Let $M=\tuple{\TCtxt, \TIL, \TBR}$ be a finite, normal, persistently reducible \rMCS given $\TKB$ such that $\mathcal{CC}(M,\BRBA{k}{\Bel})=\mathbf{P}$ for $\problemE_{wf}$.
Then, $\problemE_{wf}$ is in $\mathbf{P}$.
In addition, hardness holds provided $\mathcal{CC}(M,\BRBA{k}{\Bel})=\mathbf{P}$ is hard.
\end{theorem}
This result, together with the observation that the well-founded stream coincides with the unique grounded equilibrium stream until $s$, allows us to verify that computing the results in our use case scenario in Section~\ref{subsec:scenario} can be done in polynomial time.
 
\section{Related Work}\label{sec-related}

Several systems and frameworks for modeling the dynamics of knowledge and the flow of information have been developed. These systems are obviously related to reactive multi-context systems. In this section, we present those approaches we consider most relevant, stressing differences as well as commonalities with \rMCSs.
Note that we focus entirely on dynamic approaches and do not include systems which handle heterogeneous information in a static setting. An interesting example of the latter type are Lierler and Truszczy\'nski's abstract modular systems \cite{LierlerT16}. In a nutshell, modular systems realize the communication between different reasoning modules through joint vocabularies rather than bridge rules. These systems are best viewed as alternatives to classical MCSs. It is an open question how to adapt them to a dynamic setting.

Since \rMCSs have been designed for applications involving stream reasoning, we will first consider two recent approaches for stream reasoning, namely
LARS~\cite{BeckDEF15} and STARQL~\cite{OptiqueD5.1}.
Then, we consider EVOLP~\cite{AlferesBLP02}, a framework that focuses on dynamics in the form of updates in the restricted setting of generalized logic programs building on similar notions as the operator $\Nxt{}$ used for \rMCSs.
Finally, we consider asynchronous multi-context systems~\cite{EllmauthalerP14}, and also reactive ASP~\cite{GebserGKS11,Brewka13}.\footnote{In \ref{app:related}, we provide detailed comparisons between \rMCSs\
and the stream reasoning approaches LARS and STARQL (\ref{app:relatedStream}) as well as EVOLP (\ref{app:relatedEvolp}).}

\subsection{Reactive Multi-Context Systems and Stream Reasoning}\label{sec:relatedStream}
Reasoning over streaming data is a topic of increasing interest.
Key driver is the central role that streams play in current endeavors towards
the Internet of Things, the Internet of Services, as well as the vision of a fourth industrial revolution.
Stream reasoning has been a prominent issue in the Semantic Web community for several years,\footnote{\cf \url{http://streamreasoning.org}}
and has also received substantial attention by researchers from areas such as Knowledge Representation and Reasoning and Databases lately~\cite{reactknow2014}.

We consider \rMCSs to be a well-suited formalism for stream reasoning
that addresses important challenges that naturally arise in this context.
One important benefit of \rMCSs is that their managing capabilities provide a dynamic way to decide which data to keep for future reasoning and which data to drop.
In this respect, \rMCSs offer greater flexibility than sliding-window approaches.
Nevertheless, as also demonstrated in Section~\ref{sec-modeling}, \rMCSs can be used to implement windowing techniques.

In the following, we consider two recently proposed frameworks for stream reasoning, LARS~\cite{BeckDEF15} and STARQL~\cite{OptiqueD5.1}.

\subsubsection*{LARS}The Logic-based framework for Analyzing Reasoning over Streams (LARS)~\cite{BeckDEF15} aims at providing a formal declarative logical language for reasoning with streams. LARS is a rule-based formalism, whose language features not only provide different means to refer to or abstract from time, but also a novel window operator, thus providing a flexible mechanism to represent and reason with views on streaming data.
The semantics of LARS is based on the FLP semantics of logic programs~\cite{FaberLP04}.

Since the semantics of a LARS program is defined for a fixed input data stream and for a particular time point, it is in fact mainly static.

Given their generality, \rMCSs\ can be used to add a dynamic layer to LARS. In \ref{app:relatedStream} we show how that can be done by defining an \rMCS\ with a LARS context.
The system continuously feeds this context with the relevant (recent) substream of the input stream such that the LARS semantics can be computed for each time point.

\subsubsection*{STARQL}
STARQL (pronounced Star-Q-L)~\cite{OptiqueD5.1}, a framework for ontology-based stream reasoning, is developed within the Optique Project~\cite{optique} that comes with a stream query language inspired by the RDF query language SPARQL~\cite{sparql13}.
Streams in this framework come in the form of timestamped Description Logic assertions.
Both input as well as answers of STARQL queries are streams of this kind.
Unlike the related language continuous SPARQL (C-SPARQL)~\cite{BarbieriBCVG10}, STARQL goes beyond RDF semantics and allows for DL reasoning.

Due to the abstract nature of \rMCSs, there are many ways to realize STARQL queries as \rMCSs.
One such way would be to assume a correspondence of one equilibrium computation per STARQL query evaluation.
In that case, STARQL input stream and time stamps can be represented using \rMCSs\ input streams, and \rMCSs\ contexts can be devised to handle the different components of a STARQL query.
We illustrate the realization in detail in \ref{app:relatedStream}.

\subsection{EVOLP}\label{sec:evolp}

The framework of evolving logic programs EVOLP~\cite{AlferesBLP02} is a powerful approach for modeling updates of (propositional) generalized logic programs. Evolving logic programs are defined as general logic programs built over a special language which allows them to express self-evolution. For that,  the language includes a reserved unary predicate, \pred{assert}, whose argument may itself be a full-blown rule, thus making arbitrary nesting possible. The idea of EVOLP is that programs can update their own rules thus describing their possible self-evolution.
Moreover, besides self-evolution, evolving logic programs also consider evolution caused by the addition of external rules.

The semantics of evolving logic programs is based on sequences of interpretations, called \emph{evolution stable models}.

Given their general nature, \rMCSs\ can capture EVOLP in such a way that equilibria streams of an \rMCS\ correspond to the evolution stable models of evolving logic programs as we demonstrate in \ref{app:relatedEvolp}. At the heart of this correspondence between evolution stable models in EVOLP and equilibria streams for \rMCSs is the fact that, conceptionally, the operators $\Nxt{}$ and $\pred{assert}$ are rather similar.

\subsection{Reactive ASP}

Closely related to EVOLP are the two frameworks of Reactive ASP, one implemented as a solver \emph{oclingo}~\cite{GebserGKS11} and one described in~\cite{Brewka13}.  The system \emph{oclingo} extends an ASP solver for handling external modules provided at runtime by a controller. The output of these external modules can be seen as the observations of EVOLP. Unlike the observations in EVOLP, which can be rules, external modules in \emph{oclingo} are restricted to produce atoms so the evolving capabilities are very restricted. On the other hand, \emph{oclingo} permits committing to a specific answer set at each state for the sake of efficiency, a feature that is not part of EVOLP. Reactive ASP as described in \cite{Brewka13} can be seen as a more straightforward generalization of EVOLP where operations other than $\pred{assert}$ for self-updating a program are permitted. Since EVOLP can be captured by \rMCSs, and since \rMCSs permit several (evolution) operations in the head of bridge rules, it should be clear that Reactive ASP as described in \cite{Brewka13} can be captured by \rMCSs.

\subsection{Asynchronous Multi-Context Systems}

Asynchronous multi-context systems (\aMCSs) are a framework for loosely coupling knowledge representation formalisms and services~\cite{EllmauthalerP14}.
Like \rMCSs, they consist of heterogeneous contexts and are aware of continuous streams of information.
However, the semantics of \aMCSs is not defined in terms of equilibria 
but every context delivers output whenever available.
Intuitively, equilibria realize a tight integration approach in which the semantics of the individual contexts are interdependent at each time point, while \aMCSs do not reach this high level of integration.
Rather, contexts communicate with each other by means of input and output streams over time.
Consequently, instead of bridge rules that depend on system-wide equilibria, \aMCSs use output rules
that define which information should be
sent to another context or an output stream of the overall system based on a result of a single context.

A further difference is the role of non-determinism in the semantics of \aMCSs and \rMCSs.
While multiple equilibria give rise to non-determinism at each step in a run,
for \aMCSs, all accepted belief sets of a context are computed in a consecutive way.
Nevertheless, there is also a source of non-determinism in the case of \aMCSs.
The durations of computations and communication are taken into consideration in \aMCS but their lengths are left open.
Thus, the order in which contexts finish their computation can influence the system and  resulting knowledge bases.

Finally, both \aMCSs and \rMCSs are very general frameworks that 
allow for simulating Turing machines (\cf~\ref{app:tm}) and thus for performing multi-purpose computations.
A setup to simulate an \rMCS by an \aMCS has been presented in~\cite{EllmauthalerP14}.

\section{Conclusions}\label{sec-conclusions}

In this paper, we have introduced \emph{reactive Multi-Context Systems (rMCSs)}, an adaptation of multi-context systems suitable for continuous reasoning in dynamic environments, with the objective to achieve two goals at the same time: integrating heterogeneous knowledge sources and opening knowledge-based systems for dynamic scenarios in the presence of incoming information.
For addressing the first goal, we have built \rMCSs upon managed multi-context systems, inheriting their functionality for integrating heterogeneous knowledge sources, admitting also relevant operations on knowledge bases. 
To accommodate the dynamic aspects, \rMCSs are equipped with several extensions. 
For one, bridge rules in our framework allow for input atoms to incorporate the information of multiple external streams of data.
Moreover, contrary to standard MCSs which possess only one type of bridge rules modeling the information flow which needs to be taken into account when equilibria are computed (or the operations that need to be applied in case of mMCSs), \rMCS have an additional, different type of bridge rules, distinguished by the occurrence of the operator \textbf{next} in the head. These rules are used to specify how the configuration of knowledge bases evolves whenever an equilibrium was computed providing the definition of equilibria streams which define the semantics of \rMCSs over time.

The resulting formalism is indeed very expressive offering the capabilities to model the integration of heterogeneous knowledge in the presence of incoming information in different dynamic scenarios.
Based on a running example dealing with assisted living, we have demonstrated how to model such different dynamic scenarios using \rMCSs and addressed several temporal aspects of modeling such as incorporating time on the object level and forgetting.

Other real world use cases of an even larger scale can be handled as well, such as the one described in \cite{SlotaLS15} where the customs service needs to assess imported cargo for a variety of risk factors including terrorism, narcotics, food and consumer safety, pest infestation, tariff violations, and intellectual property rights.
Assessing this risk, even at a preliminary level, involves extensive knowledge about commodities, business entities, trade patterns, government policies and trade agreements. 
Some of this knowledge may be external to a given customs agency: for instance the broad classification of commodities according to the international Harmonized Tariff System (HTS), or international trade agreements. 
Other knowledge may be internal to a customs agency, such as lists of suspected violators or of importers who have a history of good compliance with regulations.
In \cite{SlotaLS15}, all this extensive knowledge is encoded in ontologies based on description logics and logic programming rules under the answer set semantics, and \rMCSs naturally cover these formalisms as shown, e.g., in the running example.
In addition, they easily allow for the direct modular integration of information given in databases or for example business rules without the need for any prior conversion (which may not be readily available for arbitrary formalisms).
While some of the knowledge in this risk assessment scenario is relatively stable, much of it changes rapidly. 
Changes are made not only at a specific level, such as knowledge about the expected arrival date of a shipment; but at a more general level as well. For instance, while the broad HTS code for tomatoes (0702) does not change, the full classification and tariffs for cherry tomatoes for import into the US changes seasonally.
Here again, \rMCSs provide mechanisms to easily make changes no matter if they are of a mere temporary nature or more persistent by using sensor data and the incorporation of knowledge via \textbf{next} respectively.
And, unlike \cite{SlotaLS15}, this flexibility is achieved without having to ensure or to test whether the integrated heterogeneous knowledge is organized in so-called layers. 

Naturally, dealing with inconsistency is an important issue in dynamic settings where knowledge changes over time.
Indeed, we may face different kinds of inconsistencies when using \rMCSs, e.g., in the form of inconsistent sensor input for which we discuss modeling-based solutions.
Another type of inconsistency is the absence of equilibria at certain time points which could potentially render an entire system useless.
We have addressed this problem first by showing sufficient conditions on the contexts and the bridge rules that ensure the existence of an equilibria stream.
In the cases where these conditions are not met, we have presented two possible solutions, one following an approach based on repairs -- essentially the selective removal of bridge rules to regain an equilibria stream -- and a second by relaxing the notion of equilibria stream to ensure that intermediate inconsistent states can be recovered.
Thus, \rMCSs remain usable in the face of possible inconsistencies which may always occur in real use cases where knowledge and sensor data from different sources is integrated.

We have also addressed the non-determinism of \rMCSs and discussed possibilities for avoiding it in situations where it is not desired.
To this end, we have also provided a well-founded stream semantics for \rMCSs.
The study of computational complexity we conducted confirms worst case complexities similar to managed multi-context systems.
Hence, the additional integration of sensor data does not raise the worst case complexity of reasoning with heterogeneous knowledge integrated in multi-context systems. 
In addition, the results confirm that if very efficient reasoning with such rich knowledge is essential, we can restrict to the well-founded stream.
This implies some restrictions on the permitted expressiveness, but we have argued that often these restrictions can be accommodated.  

Finally, we compared our approach to related work.
Most importantly, we have shown that \rMCSs can capture two relevant approaches in stream reasoning, namely LARS \cite{BeckDEF15} and STARQL~\cite{OptiqueD5.1}, thus showcasing that \rMCSs are indeed ``defined in the right way''  for the intended integration of constant streams of data. 

Regarding future work, several interesting lines of research can be considered.
First, we may want to extend \rMCSs to enhance their applicability in an even wider set of scenarios.
Namely, one might adopt language constructs for expressing preferences in ASP such as optimization statements~\cite{GKKOST10} or weak constraints~\cite{BuccafurriLR97}, which assign a quality measure to equilibria. 
This would allow us, e.g., to improve on the potentially high non-determinism when faced with several equilibria at one time point and thus avoid having to consider a possibly huge number of different equilibria.
We would also like to explore giving bridge rules access to the entire belief set of another context or an input and not just to one element. 
A potential application would be counting, e.g., to ask if there was no sensor reading.

An alternative to deal with inconsistent states is to follow a paraconsistent approach, as proposed for hybrid knowledge bases in \cite{DBLP:conf/kr/Fink12,DBLP:conf/ijcai/KaminskiKL15}. Also, following the idea of EVOLP~\cite{AlferesBLP02} as explored in \cite{DBLP:conf/clima/GoncalvesKL14}, we could allow the bridge rules to change with time, strengthening the evolving and adaptation capabilities of \rMCSs.
We would also like to establish bridges to asynchronous MCSs \cite{EllmauthalerP14}, a framework for loosely coupling knowledge representation formalisms whose semantics assumes that every context delivers output whenever available. 
Finally, we may build on existing implementations of distributed (static) MCSs \cite{DaoTranEFK15} for providing an implementation of rMCSs that handles the integration and evolution of distributed, heterogeneous knowledge with incoming streams of data and reasoning over such integration.

The framework introduced in this paper is highly abstract. Needless to say this was our intention. We wanted to capture a wide range of situations without imposing any restrictions on the KR formalisms used. It is a matter of the application to choose the best suited formalisms to solve a specific task. The downside of this generality, of course, is that the framework as such needs to be instantiated in various ways before it is ready-to-use. In particular, we need to select the context logics and define the ways in which contexts and dynamic information interact  by specifying adequate bridge rules. We still believe our approach provides a valuable - and to the best of our knowledge unique - solution to the problems outlined in the introduction, that is, problems which originate in the heterogeneous and dynamic nature of the information available in many challenging applications. 
\section*{Acknowledgements}\label{sec-ackn}

We would like to thank K.\ Schekotihin and the anonymous reviewers for their comments, which helped improving this paper.
G.\ Brewka, S.\ Ellmauthaler, and J. P\"uhrer were partially supported by the German Research Foundation (DFG) under grants BR-1817/7-1/2 and FOR~1513.
R.\ Gon\c{c}alves, M.\ Knorr and J.\ Leite were partially supported by Funda{\c c}\~ao para a Ci\^encia e a Tecnologia (FCT) under project NOVA LINCS ({UID}/{CEC}/{04516}/{2013}).
Moreover, R.\ Gon\c{c}alves was partially supported by FCT grant SFRH/BPD/100906/2014 and M.\ Knorr by FCT grant SFRH/BPD/86970/2012.

\bibliographystyle{elsarticle-harv}

\begin{thebibliography}{40}
\expandafter\ifx\csname natexlab\endcsname\relax\def\natexlab#1{#1}\fi
\expandafter\ifx\csname url\endcsname\relax
  \def\url#1{\texttt{#1}}\fi
\expandafter\ifx\csname urlprefix\endcsname\relax\def\urlprefix{URL }\fi

\bibitem[{Alferes et~al.(2000)Alferes, Leite, Pereira, Przymusinska, and
  Przymusinski}]{AlferesLPPP00}
Alferes, J., Leite, J., Pereira, L., Przymusinska, H., Przymusinski, T., 2000.
  Dynamic updates of non-monotonic knowledge bases. J. Log. Program. 45~(1-3),
  43--70.

\bibitem[{Alferes et~al.(2002)Alferes, Brogi, Leite, and
  Pereira}]{AlferesBLP02}
Alferes, J.~J., Brogi, A., Leite, J.~A., Pereira, L.~M., 2002. Evolving logic
  programs. In: Flesca, S., Greco, S., Leone, N., Ianni, G. (Eds.), Logics in
  Artificial Intelligence, European Conference, {JELIA} 2002, Cosenza, Italy,
  September, 23-26, Proceedings. Vol. 2424 of LNCS. Springer, pp. 50--61.

\bibitem[{Anicic et~al.(2012)Anicic, Rudolph, Fodor, and
  Stojanovic}]{AnicicRFS12}
Anicic, D., Rudolph, S., Fodor, P., Stojanovic, N., 2012. Stream reasoning and
  complex event processing in {ETALIS}. Semantic Web 3~(4), 397--407.

\bibitem[{Arenas et~al.(1999)Arenas, Bertossi, and
  Chomicki}]{DBLP:conf/pods/ArenasBC99}
Arenas, M., Bertossi, L.~E., Chomicki, J., 1999. Consistent query answers in
  inconsistent databases. In: Vianu, V., Papadimitriou, C.~H. (Eds.),
  Proceedings of the Eighteenth {ACM} {SIGACT-SIGMOD-SIGART} Symposium on
  Principles of Database Systems, May 31 - June 2, 1999, Philadelphia,
  Pennsylvania, {USA}. {ACM} Press, pp. 68--79.

\bibitem[{Baader et~al.(2007)Baader, Calvanese, McGuinness, Nardi, and
  Patel-Schneider}]{DLhandbook}
Baader, F., Calvanese, D., McGuinness, D.~L., Nardi, D., Patel-Schneider, P.~F.
  (Eds.), 2007. {The Description Logic Handbook: Theory, Implementation, and
  Applications}, 2nd Edition. Cambridge University Press.

\bibitem[{Barbieri et~al.(2010)Barbieri, Braga, Ceri, Valle, and
  Grossniklaus}]{BarbieriBCVG10}
Barbieri, D., Braga, D., Ceri, S., Valle, E.~D., Grossniklaus, M., 2010.
  {C-SPARQL}: a continuous query language for {RDF} data streams. Int. J.
  Semantic Computing 4~(1), 3--25.

\bibitem[{Beck et~al.(2015)Beck, Dao{-}Tran, Eiter, and Fink}]{BeckDEF15}
Beck, H., Dao{-}Tran, M., Eiter, T., Fink, M., 2015. {LARS:} {A} logic-based
  framework for analyzing reasoning over streams. In: Bonet, B., Koenig, S.
  (Eds.), Proceedings of the Twenty-Ninth {AAAI} Conference on Artificial
  Intelligence, January 25-30, 2015, Austin, Texas, {USA}. {AAAI} Press, pp.
  1431--1438.

\bibitem[{Brewka(2013)}]{Brewka13}
Brewka, G., 2013. Towards reactive multi-context systems. In: Cabalar, P., Son,
  T.~C. (Eds.), Proceedings of the 12th International Conference on Logic
  Programming and Nonmonotonic Reasoning ({LPNMR}'13). Vol. 8148 of LNCS.
  Springer, pp. 1--10.

\bibitem[{Brewka and Eiter(2007)}]{BrewkaE07}
Brewka, G., Eiter, T., 2007. Equilibria in heterogeneous nonmonotonic
  multi-context systems. In: Proceedings of the Twenty-Second {AAAI} Conference
  on Artificial Intelligence, July 22-26, 2007, Vancouver, British Columbia,
  Canada. {AAAI} Press, pp. 385--390.

\bibitem[{Brewka et~al.(2011)Brewka, Eiter, Fink, and Weinzierl}]{BrewkaEFW11}
Brewka, G., Eiter, T., Fink, M., Weinzierl, A., 2011. Managed multi-context
  systems. In: Walsh, T. (Ed.), {IJCAI} 2011, Proceedings of the 22nd
  International Joint Conference on Artificial Intelligence, Barcelona,
  Catalonia, Spain, July 16-22, 2011. {IJCAI/AAAI}, pp. 786--791.

\bibitem[{Brewka et~al.(2014)Brewka, Ellmauthaler, and
  P{\"{u}}hrer}]{BrewkaEP14}
Brewka, G., Ellmauthaler, S., P{\"{u}}hrer, J., 2014. Multi-context systems for
  reactive reasoning in dynamic environments. In: {ECAI} 2014 - 21st European
  Conference on Artificial Intelligence, 18-22 August 2014, Prague, Czech
  Republic - Including Prestigious Applications of Intelligent Systems {(PAIS}
  2014). pp. 159--164.

\bibitem[{Brewka et~al.(2007)Brewka, Roelofsen, and Serafini}]{brrose07a}
Brewka, G., Roelofsen, F., Serafini, L., 2007. Contextual default reasoning.
  In: Veloso, M.~M. (Ed.), {IJCAI} 2007, Proceedings of the 20th International
  Joint Conference on Artificial Intelligence, Hyderabad, India, January 6-12,
  2007. pp. 268--273.

\bibitem[{Buccafurri et~al.(1997)Buccafurri, Leone, and Rullo}]{BuccafurriLR97}
Buccafurri, F., Leone, N., Rullo, P., 1997. Strong and weak constraints in
  disjunctive {D}atalog. In: Dix, J., Furbach, U., Nerode, A. (Eds.), Logic
  Programming and Nonmonotonic Reasoning, 4th International Conference,
  LPNMR'97, Dagstuhl Castle, Germany, July 28-31, 1997, Proceedings. Vol. 1265
  of LNCS. Springer, pp. 2--17.

\bibitem[{Dao{-}Tran et~al.(2015)Dao{-}Tran, Eiter, Fink, and
  Krennwallner}]{DaoTranEFK15}
Dao{-}Tran, M., Eiter, T., Fink, M., Krennwallner, T., 2015. Distributed
  evaluation of nonmonotonic multi-context systems. J. Artif. Intell. Res. 52,
  543--600.

\bibitem[{Eiter et~al.(2014)Eiter, Fink, Sch{\"{u}}ller, and
  Weinzierl}]{EiterFSW14}
Eiter, T., Fink, M., Sch{\"{u}}ller, P., Weinzierl, A., 2014. Finding
  explanations of inconsistency in multi-context systems. Artif. Intell. 216,
  233--274.

\bibitem[{Eiter et~al.(2008)Eiter, Ianni, Lukasiewicz, Schindlauer, and
  Tompits}]{DBLP:journals/ai/EiterILST08}
Eiter, T., Ianni, G., Lukasiewicz, T., Schindlauer, R., Tompits, H., 2008.
  Combining answer set programming with description logics for the semantic
  web. Artif. Intell. 172~(12-13), 1495--1539.

\bibitem[{Ellmauthaler(2013)}]{Ellmauthaler2013a}
Ellmauthaler, S., 2013. Generalizing multi-context systems for reactive stream
  reasoning applications. In: Jones, A.~V., Ng, N. (Eds.), 2013 Imperial
  College Computing Student Workshop, {ICCSW} 2013, September 26/27, 2013,
  London, United Kingdom. Vol.~35 of {OASICS}. Schloss Dagstuhl -
  Leibniz-Zentrum fuer Informatik, Germany, pp. 19--26.

\bibitem[{Ellmauthaler and P{\"u}hrer(2014)}]{reactknow2014}
Ellmauthaler, S., P{\"u}hrer, J. (Eds.), 2014. Proceedings of the
  {I}nternational {W}orkshop on {R}eactive {C}oncepts in {K}nowledge
  {R}epresentation ({ReactKnow 2014}). Leipzig University.

\bibitem[{Ellmauthaler and P{\"{u}}hrer(2015)}]{EllmauthalerP14}
Ellmauthaler, S., P{\"{u}}hrer, J., 2015. Asynchronous multi-context systems.
  In: Eiter, T., Strass, H., Truszczynski, M., Woltran, S. (Eds.), Advances in
  Knowledge Representation, Logic Programming, and Abstract Argumentation -
  Essays Dedicated to Gerhard Brewka on the Occasion of His 60th Birthday. Vol.
  9060 of LNCS. Springer, pp. 141--156.

\bibitem[{Faber et~al.(2004)Faber, Leone, and Pfeifer}]{FaberLP04}
Faber, W., Leone, N., Pfeifer, G., 2004. Recursive aggregates in disjunctive
  logic programs: Semantics and complexity. In: Alferes, J.~J., Leite, J.~A.
  (Eds.), Logics in Artificial Intelligence, 9th European Conference, {JELIA}
  2004, Lisbon, Portugal, September 27-30, 2004, Proceedings. Vol. 3229 of
  LNCS. Springer, pp. 200--212.

\bibitem[{Fink(2012)}]{DBLP:conf/kr/Fink12}
Fink, M., 2012. Paraconsistent hybrid theories. In: Brewka, G., Eiter, T.,
  McIlraith, S.~A. (Eds.), Principles of Knowledge Representation and
  Reasoning: Proceedings of the Thirteenth International Conference, {KR} 2012,
  Rome, Italy, June 10-14, 2012. {AAAI} Press, pp. 391--401.

\bibitem[{Gebser et~al.(2012)Gebser, Grote, Kaminski, Obermeier, Sabuncu, and
  Schaub}]{GebserGKOSS12}
Gebser, M., Grote, T., Kaminski, R., Obermeier, P., Sabuncu, O., Schaub, T.,
  2012. Stream reasoning with answer set programming: Preliminary report. In:
  Brewka, G., Eiter, T., McIlraith, S.~A. (Eds.), Principles of Knowledge
  Representation and Reasoning: Proceedings of the Thirteenth International
  Conference, {KR} 2012, Rome, Italy, June 10-14, 2012. {AAAI} Press, pp.
  613--617.

\bibitem[{Gebser et~al.(2011)Gebser, Grote, Kaminski, and Schaub}]{GebserGKS11}
Gebser, M., Grote, T., Kaminski, R., Schaub, T., 2011. Reactive answer set
  programming. In: Delgrande, J.~P., Faber, W. (Eds.), Logic Programming and
  Nonmonotonic Reasoning - 11th International Conference, {LPNMR} 2011,
  Vancouver, Canada, May 16-19, 2011. Proceedings. Vol. 6645 of LNCS. Springer,
  pp. 54--66.

\bibitem[{Gebser et~al.(2010)Gebser, Kaminski, Kaufmann, Ostrowski, Schaub, and
  Thiele}]{GKKOST10}
Gebser, M., Kaminski, R., Kaufmann, B., Ostrowski, M., Schaub, T., Thiele, S.,
  2010. {A} user’s guide to gringo, clasp, clingo, and iclingo. Potassco
  Team.

\bibitem[{Gelfond and Lifschitz(1991)}]{GelfondL91}
Gelfond, M., Lifschitz, V., 1991. Classical negation in logic programs and
  disjunctive databases. New Generation Comput. 9~(3/4), 365--386.

\bibitem[{Giunchiglia(1993)}]{giunchiglia93}
Giunchiglia, F., 1993. Contextual reasoning. Epistemologia XVI, 345--364.

\bibitem[{Giunchiglia and Serafini(1994)}]{GiunchigliaS94}
Giunchiglia, F., Serafini, L., 1994. Multilanguage hierarchical logics or: How
  we can do without modal logics. Artif. Intell. 65~(1), 29--70.

\bibitem[{Gon{\c{c}}alves et~al.(2014{\natexlab{a}})Gon{\c{c}}alves, Knorr, and
  Leite}]{DBLP:conf/clima/GoncalvesKL14}
Gon{\c{c}}alves, R., Knorr, M., Leite, J., 2014{\natexlab{a}}. Evolving bridge
  rules in evolving multi-context systems. In: Bulling, N., van~der Torre, L.
  W.~N., Villata, S., Jamroga, W., Vasconcelos, W.~W. (Eds.), Computational
  Logic in Multi-Agent Systems - 15th International Workshop, {CLIMA} XV,
  Prague, Czech Republic, August 18-19, 2014. Proceedings. Vol. 8624 of LNCS.
  Springer, pp. 52--69.

\bibitem[{Gon{\c{c}}alves et~al.(2014{\natexlab{b}})Gon{\c{c}}alves, Knorr, and
  Leite}]{GoncalvesKL14}
Gon{\c{c}}alves, R., Knorr, M., Leite, J., 2014{\natexlab{b}}. Evolving
  multi-context systems. In: {ECAI} 2014 - 21st European Conference on
  Artificial Intelligence, 18-22 August 2014, Prague, Czech Republic -
  Including Prestigious Applications of Intelligent Systems {(PAIS} 2014). pp.
  375--380.

\bibitem[{Harris and Seaborne(2013)}]{sparql13}
Harris, S., Seaborne, A., 2013. {SPARQL 1.1 query language for RDF}. {W3C
  Recommendation}.

\bibitem[{Kaminski et~al.(2015)Kaminski, Knorr, and
  Leite}]{DBLP:conf/ijcai/KaminskiKL15}
Kaminski, T., Knorr, M., Leite, J., 2015. Efficient paraconsistent reasoning
  with ontologies and rules. In: Yang, Q., Wooldridge, M. (Eds.), Proceedings
  of the Twenty-Fourth International Joint Conference on Artificial
  Intelligence, {IJCAI} 2015, Buenos Aires, Argentina, July 25-31, 2015. {AAAI}
  Press, pp. 3098--3105.

\bibitem[{L{\'{e}}cu{\'{e}} and Pan(2013)}]{LecueP13}
L{\'{e}}cu{\'{e}}, F., Pan, J.~Z., 2013. Predicting knowledge in an ontology
  stream. In: Rossi, F. (Ed.), {IJCAI} 2013, Proceedings of the 23rd
  International Joint Conference on Artificial Intelligence, Beijing, China,
  August 3-9, 2013. {IJCAI/AAAI}, pp. 2662--2669.

\bibitem[{Lierler and Truszczy\'nski(2016)}]{LierlerT16}
Lierler, Y., Truszczy\'nski, M., 2016. On abstract modular inference systems
  and solvers. Artif. Intell. 236, 65--89.

\bibitem[{McCarthy(1987)}]{mccarthy87}
McCarthy, J., 1987. Generality in artificial intelligence. Commun. ACM 30~(12),
  1029--1035.

\bibitem[{Motik and Rosati(2010)}]{Motik2010}
Motik, B., Rosati, R., 2010. Reconciling description logics and rules. Journal
  of the ACM 57~(5), 93--154.

\bibitem[{Mu et~al.(2016)Mu, Wang, and Wen}]{MuWW16}
Mu, K., Wang, K., Wen, L., 2016. Preferential multi-context systems. Int. J.
  Approx. Reasoning 75, 39--56.

\bibitem[{{\"Oz\c{c}ep} et~al.(2013){\"Oz\c{c}ep}, M\"oller, Neuenstadt,
  Zheleznayakow, Kharlamov, Horrocks, Hubauer, and Roshchin}]{OptiqueD5.1}
{\"Oz\c{c}ep}, {\"Ozg\"ur}.~L., M\"oller, R., Neuenstadt, C., Zheleznayakow,
  D., Kharlamov, E., Horrocks, I., Hubauer, T., Roshchin, M., 2013. D5.1
  {E}xecutive {S}ummary: {A} semantics for temporal and stream-based query
  answering in an {OBDA} context. Tech. rep., Optique FP7-ICT-2011-8-318338
  Project.

\bibitem[{Rodriguez{-}Muro et~al.(2013)Rodriguez{-}Muro, Jupp, and
  Srinivas}]{optique}
Rodriguez{-}Muro, M., Jupp, S., Srinivas, K. (Eds.), 2013. The Optique Project:
  Towards {OBDA} Systems for Industry (Short Paper). Vol. 1080 of {CEUR}
  Workshop Proceedings. CEUR-WS.org.

\bibitem[{Roelofsen and Serafini(2005)}]{roeser05}
Roelofsen, F., Serafini, L., 2005. Minimal and absent information in contexts.
  In: Kaelbling, L.~P., Saffiotti, A. (Eds.), IJCAI-05, Proceedings of the
  Nineteenth International Joint Conference on Artificial Intelligence,
  Edinburgh, Scotland, UK, July 30-August 5, 2005. Professional Book Center,
  pp. 558--563.

\bibitem[{Slota et~al.(2015)Slota, Leite, and Swift}]{SlotaLS15}
Slota, M., Leite, J., Swift, T., 2015. On updates of hybrid knowledge bases
  composed of ontologies and rules. Artif. Intell. 229, 33--104.

\end{thebibliography}

\newpage
\appendix
\section{Proofs Inconsistency Management}

\begin{proposition*}
{prop:ConditionsForStrongConsistence}
Let $M=\tuple{\tuple{\Ctxt_1,\ldots,\Ctxt_\CtxtIn}, \TIL, \TBR}$ be an acyclic \rMCS such that every $\Ctxt_i$, $1\leq i\leq n$, is totally coherent, and $\TKB$ a configuration of knowledge bases for $M$. Then, $M$ is strongly consistent with respect to $\TKB$.
\end{proposition*}
\begin{pf}
Let $M= \tuple{\tuple{\Ctxt_1,\ldots,\Ctxt_\CtxtIn}, \TIL, \tuple{\SBR_1,\ldots,\SBR_\CtxtIn}}$ be an acyclic \rMCS\ with totally coherent contexts.
We first prove that $M$ has an equilibrium given $\TKB$ and $\Tinpt$, for any knowledge base configuration $\TKB=\tuple{\KB_1,\ldots,\KB_\CtxtIn}$ for $M$ and input $\Tinpt$ for $M$.

We prove this by induction on the number of contexts of $M$, making use of the following simple observation: if $M$ does not have cycles, then there exists some $i\in \{1,\ldots,n\}$ such that $ref_r(j,i)$ does not hold for any $j\in\{1,\ldots,n\}$ and $r\in \SBR_j$, where $ref_r(j,i)$ holds precisely when $r$ is a bridge rule of context $\Ctxt_j$ and $\BRBA{i}{\Bel}$ occurs in the body of $r$. It is quite easy to see that if this condition is violated then a cycle necessarily exists.

Let $n=1$. Then, since there are no cycles, no bridge rule in $\SBR_1$ contains atoms of the form $\BRBA{1}{\Bel}$ in its body. Thus, $\app^{now}_i(\Tinpt,\TBelS)$ does not depend on $\TBelS$. Total coherence then immediately implies that $M$ has an equilibrium given $\TKB$ and $\Tinpt$. 

Let $n=m+1$. We use the above observation, and assume, w.l.o.g., that $\Ctxt_1$ is a context for which $ref_r(j,1)$ does not hold for any $j\in\{1,\ldots,m+1\}$ and $r\in \SBR_j$.
Then, the \rMCS\ $M^*= \tuple{\tuple{\Ctxt_2,\ldots,\Ctxt_{m+1}}, \TIL, \tuple{\SBR_2,\ldots,\SBR_{m+1}}}$ has $m$ contexts and it is still acyclic. By induction hypothesis, we can conclude that $M^*$ has an equilibrium given $\TKB^*=\tuple{\KB_2,\ldots,\KB_{m+1}}$ and $\Tinpt$. 
Let $\TBelS^*=\tuple{\BelS_2,\ldots,\BelS_{m+1}}$ be such equilibrium. Then, since $\Ctxt_1$ is assumed to be a totally coherent context, there exists $\BelS_1\in \SBelS_1$ such that $\TBelS=\tuple{\BelS_1,\BelS_2,\ldots,\BelS_n}$ is an equilibrium of $M$ given $\TKB$ and $\Tinpt$. This follows easily from the fact that no set $\app^{now}_i(\Tinpt,\TBelS)$ depends on the choice of $\BelS_1$.

We have shown that the existence of an equilibrium for $M$ is independent of the given knowledge base configuration $\TKB$ for $M$ and input $\Tinpt$ for $M$. This immediately implies that for any input stream $\inptStr$ for $M$ (until $\tau$), and  any knowledge base configuration $\TKB$ for $M$, there exists an equilibria stream of $M$ given $\TKB$ and $\inptStr$.
\end{pf}

\begin{proposition*}{prop:emptyRepair}
Every equilibria stream of $M$ given $\TKB$ and $\inptStr$ is a repaired equilibria stream of $M$ given $\TKB$, $\inptStr$ and the empty repair $\Repair_{\emptyset}$.
\end{proposition*}
\begin{pf}
 This result follows easily from the observation that $M[\Repair_{\emptyset}]=M$. In this case the conditions in the definition of an equilibria stream of $M$ coincide with those in the definition of a repaired equilibria stream of $M$.
\end{pf}

\begin{proposition*}{prop:existenceRepairedEquilibria}
Let $M=\tuple{\tuple{\Ctxt_1,\ldots,\Ctxt_\CtxtIn}, \TIL, \TBR}$ be an \rMCS such that  each $\Ctxt_i$, $i\in\{1,\ldots,n\}$, is totally coherent, $\TKB$ a configuration of knowledge bases for $M$, and $\inptStr$ an input stream for $M$ until $\tau$. Then, there exists ${\Repair:[1..\tau]\to 2^{br_M}}$ and ${\EqStr:[1..\tau]\to \BelForM{M}}$ such that $\EqStr$ is a repaired equilibria stream given $\TKB$, $\inptStr$ and $\Repair$.
\end{proposition*}
\begin{pf}
Since each context of $M$ is totally coherent, Proposition~\ref{prop:ConditionsForStrongConsistence} guarantees the existence of an equilibrium if $M$ is acyclic. Now just note that if we take ${\Repair:[1..\tau]\to 2^{br_M}}$ such that $\Repair^t=br_M$ for every $t$, then each $M[\Repair^t]$ does not have bridge rules and it is therefore acyclic. Then, for every $t$, $M[\Repair^t]$ is strongly consistent. Therefore we can easily inductively construct ${\EqStr:[1..\tau]\to \BelForM{M}}$ such that $\EqStr$ is a repaired equilibria stream given $\TKB$, $\inptStr$ and $\Repair$. 
\end{pf}

\begin{proposition*}{prop:equilibriaStreamPartialEquilibria}
Every equilibria stream of $M$ given $\TKB$ and $\inptStr$ is a partial equilibria stream of $M$ given $\TKB$ and $\inptStr$.
\end{proposition*}
\begin{pf}
 This result follows easily from the observation that for an equilibria stream $\EqStr$ of $M$ given $\TKB$ and $\inptStr$, and every $t$, $\EqStr^t$ is never undefined. Therefore, in this case the conditions in the definition of partial equilibria stream coincide with those for equilibria stream. 
\end{pf}

\begin{proposition*}{prop:partialEquilibriaExistence}
Let $M$ be an \rMCS, $\TKB$ a configuration of knowledge bases for $M$, and $\inptStr$ an input stream for $M$ until $\tau$. Then, there exists ${\EqStr:[1..\tau]\nrightarrow \BelForM{M}}$ such that $\EqStr$ is a partial equilibria stream given $\TKB$ and $\inptStr$.
\end{proposition*}
\begin{pf}
We just need to note that if we take ${\EqStr:[1..\tau]\nrightarrow \BelForM{M}}$ such that, for every $t$, $\EqStr^t$ is undefined, then $\EqStr$ is trivially a partial equilibria stream given $\TKB$ and $\inptStr$.
\end{pf}

 \section{Proofs Non-Determinism and Well-Founded Semantics}

We first show some properties of the iteration defined in Def.~\ref{def:definiteIteration} that will prove useful subsequently.

\begin{lemma*}
{lemma:iterationProps}
Let $M$ be a definite \rMCS given $\TKB$, and $\Tinpt$ an input for $M$.
The following holds for the iteration in Definition~\ref{def:definiteIteration} for all ordinals $\alpha$:
\begin{enumerate}
\item $M$ is definite given $\TKB^\alpha$;
\item for any ordinal $\beta$ with $\beta\leq\alpha$ we have $\KB^\beta_i\subseteq \KB^\alpha_i$ for $i\in\{1,\ldots,\CtxtIn\}$.
\end{enumerate}
\end{lemma*}
\begin{pf}
We have to show that the lemma holds for all ordinals $\alpha$.
For the (initial) limit ordinal $\alpha=0$, Claims \ref{itProps1} and \ref{itProps2} hold trivially.

Now, suppose we have shown that \ref{itProps1} and \ref{itProps2} hold for all ordinals $\alpha_1\leq \alpha$.
First, consider a successor ordinal $\alpha+1$. 
Regarding Claim \ref{itProps1}, i.e., $M$ being definite given $\TKB^{\alpha+1}$,  Condition 1 of Definition~\ref{def:definiteMCS} holds trivially as the bridge rules are fix, and Condition~2 follows from the fact that all $\KB^\alpha_i$ are already in reduced form, and from Condition~2 of Definition~\ref{def:reducible}, which prevents the introduction of reducible content by any $\mng_i$.
Regarding Claim \ref{itProps2}, we already know by the hypothesis that this holds for $\alpha$, so we only have to show that $\KB_i^\alpha\subseteq \KB_i^{\alpha+1}$ holds for all $i$ as well.
As all $\acc_i$ for the definite \rMCS are monotonic, we also know that $\TBelS^\beta\subseteq \TBelS^\alpha$ holds for all $\beta\leq \alpha$.
In fact, this holds for any of the $\beta$ just as well, i.e., there is an increasing sequence of $\TBelS^\beta$ associated to the iteration.
Now, since the \rMCS is definite, hence no negation occurs in bridge rules, and $\Tinpt$ is fix, the sequence of all $\app_i^{now}(\Tinpt,\TKB^\beta)$ for each $\beta$ is also increasing. 
But then, by Condition~3 of Definition~\ref{def:reducible}, $\KB_i^\alpha\subseteq \KB_i^{\alpha+1}$ holds for all $i$.

Now consider a limit ordinal $\alpha'$ (with $\alpha\leq \alpha'$).
Since in this case, all $\KB_i^{\alpha'}$ are just the union of all $\KB_i^\beta$ with $\beta\leq \alpha'$, the Claims \ref{itProps1} and \ref{itProps2} trivially follow. 
\end{pf}

With the help of Lemma~\ref{lemma:iterationProps}, we can show that the iteration indeed yields a unique minimal equilibrium (of $M$ given $\TKB$ and $\Tinpt$).
\begin{proposition*}
{prop:GEuniqueForDefinite}
Let $M$ be a definite \rMCS given $\TKB$, and $\Tinpt$ an input for $M$. 
Then $\tuple{\BelS^\infty_1,\ldots,\BelS^\infty_n}$ is the unique minimal equilibrium of $M$ given $\TKB$ and $\Tinpt$.
We call it \emph{grounded equilibrium of $M$ given $\TKB$ and $\Tinpt$}, denoted by $\geql(M,\TKB,\Tinpt)$.
\end{proposition*}
\begin{pf}
We first note that $\tuple{\BelS^\infty_1,\ldots,\BelS^\infty_n}$ is indeed an equilibrium.
This follows directly from the definition of equilibria (Definition~\ref{def:Equilibrium}) and the fact that the fixpoint in Definition~\ref{def:definiteIteration} precisely matches it.

Concerning minimality, suppose it is not a minimal equilibrium. 
Then, there is some $\tuple{\BelS',\ldots,\BelS'}$ which is also an equilibrium such that $\BelS'_i\subseteq\BelS^\infty_i$ for all $i\in\{1,\ldots,\CtxtIn\}$ and $\BelS'_i\subset\BelS^\infty_i$ for at least one $i$.
Therefore some belief $\Bel\in \BelS_i^\infty\setminus\BelS'_i$ and a least ordinal $\alpha$ exist such that $\Bel\in \BelS_i^\alpha$. We consider a $\Bel$ where $\alpha$ is least for all such beliefs.
If $\alpha=0$, then $\acc_i(\KB_i^0)=\BelS_i^0$ and since, by Lemma~\ref{lemma:iterationProps}, $M$ is definite given $\TKB^0$, this belief set is unique, hence $\TBelS'$ cannot be an equilibrium (following from monotonicity of the iteration).
If $\alpha>0$, then necessarily some operation in a bridge rule head, applied when creating $\KB_i^\alpha$, triggered the occurrence of $\Bel$ in $\acc_i(\KB_i^\alpha)=\BelS_i^\alpha$.
Now, as $M$ is definite given all $\TKB^\beta$ in the iteration of $\TKB^\infty$ and since $\alpha$ is least, this head is also applicable by monotonicity w.r.t.\ $\TBelS'$.
Hence, we obtain a contradiction to $\BelS'$ being an equilibrium.

Regarding uniqueness, suppose it is a minimal equilibrium, but not unique.
Then, there is some $\tuple{\BelS',\ldots,\BelS'}$ which is also a minimal equilibrium such that at least one of the following holds:
\begin{itemize}
\item neither $\BelS'_i\subseteq\BelS^\infty_i$ nor $\BelS^\infty_i\subseteq\BelS'_i$ for at least one $i\in\{1,\ldots,\CtxtIn\}$;
\item $\BelS'_i\subset \BelS^\infty_i$ and $\BelS'_j\supset \BelS^\infty_j$ for $i\not=j$ with $i,j\in\{1,\ldots,\CtxtIn\}$.
\end{itemize}
In both cases, consider some $\Bel\in \BelS_i^\infty\setminus\BelS'_i$.
Again, there is a least $\alpha$ such that $\Bel\in \BelS_i^\alpha$ and $\alpha$ is least among all these $\Bel$.
We can apply the same argument as used for proving minimality and obtain a contradiction to $\TBelS'$ being an equilibrium.
\end{pf}

\begin{proposition*}
{prop:grEqIsMinEq}
Every grounded equilibrium of a reducible \rMCS $M$ given $\TKB$ and an input $\Tinpt$ is a minimal equilibrium of $M$ given $\TKB$ and $\Tinpt$.
\end{proposition*}
\begin{pf}
We first show that a grounded equilibrium $\TBelS$ is indeed an equilibrium.
For this, note that $\TBelS$ is the unique minimal equilibrium of the definite $M^{(\Tinpt,\TBelS)}$ given $\TKB^{(\Tinpt,\TBelS)}$ and $\Tinpt$ by Definition~\ref{def:groundedRed} and Proposition~\ref{prop:GEuniqueForDefinite}.
Thus, by Definition~\ref{def:Equilibrium} and since $M^{(\Tinpt,\TBelS)}$ is definite, we know for all $i\in \{1,\ldots,\CtxtIn\}$:
\begin{equation}\label{eq:prop9-1}
\BelS_i=\acc_{i}(\KB'),\text{ where }\KB'=\mng_i(\app^{now}_i(\Tinpt,\TBelS),\KB_i^{\BelS_i}).
\end{equation}
with $\KB_i^{\BelS_i}=\redfun_i(\KB_i,\BelS_i)$ by Definition~\ref{def:reduct}.
Now, according to Definition~\ref{def:reduct}, $M^{(\Tinpt,\TBelS)}$ only differs from $M$ on the sets of bridge rules $\SBR_i^{(\Tinpt,\TBelS)}$, and the way these are obtained from the $\SBR_i$ ensures that $\app_i^{now}(\Tinpt,\TBelS)$ is identical for both $M$ and $M^{(\Tinpt,\TBelS)}$.
Moreover, by 2.\ of Definition~\ref{def:reducible}, (\ref{eq:prop9-1}) can be rewritten to (for all $i$):
\begin{equation}\label{eq:prop9-2}
\BelS_i=\acc_{i}(\KB'),\text{ where }\KB'=\redfun_i(\mng_i(\app^{now}_i(\Tinpt,\TBelS),\KB_i),\BelS_i).
\end{equation}
In this case, by the definition of $\redfun$, we know that, (for all $i$), 
\begin{equation}
\BelS_i\in\acc_i(\mng_i(\app^{now}_i(\Tinpt,\TBelS),\KB_i)
\end{equation}
and this shows, by Definition~\ref{def:Equilibrium}, that $\TBelS$ is an equilibrium.

Now suppose that $\TBelS$ is not minimal.
Then there is an equilibrium $\TBelS' = \tuple{\BelS_1', \ldots, \BelS_\CtxtIn'}$ of $M$ given $\TKB$ and $\Tinpt$ such that $\BelS_i' \subseteq \BelS_i$ for all $i$ with $1 \leq i \leq \CtxtIn$ and $\BelS_j' \subsetneq \BelS_j$ for some $j$ with $j\in \{1 \ldots \CtxtIn\}$. 
Since $\redfun$ is antitonic by definition, we know that $\redfun_i(\KB_i, \BelS_i) \subseteq \redfun_i(\KB_i, \BelS'_i)$ holds for all $i\in\{1,\ldots,\CtxtIn\}$.
Also, by Definition~\ref{def:reduct}, $\SBR^{(\Tinpt,\TBelS^{\alpha})}_i\subseteq {\SBR'}^{(\Tinpt,{\TBelS'}^{\alpha})}_i$ for all $i$.
It can thus be shown by induction on $\alpha$ for the monotonic iteration in Definition~\ref{def:definiteIteration} that $\BelS_i^\alpha\subseteq {\BelS'}_i^\alpha$ holds for all $i\in\{1,\ldots,\CtxtIn\}$:
\begin{itemize}
\item $\alpha=0$: this holds right away by monotonicity of $\acc_i$ for reduced knowledge bases.
\item Suppose the claim holds for all ordinals $\alpha_1\leq \alpha$.
\item Consider a successor ordinal $\alpha+1$. 
 As $\SBR^{(\Tinpt,\TBelS^{\alpha})}_i\subseteq {\SBR'}^{(\Tinpt,{\TBelS'}^{\alpha})}_i$ holds, and no bridge rule contains $\naf$ due to the reduction, we have, by the induction hypothesis that $\app_i^{now}(\Tinpt,\TBelS^{\alpha})\subseteq \app_i^{now}(\Tinpt,{\TBelS'}^{\alpha})$.
 Then, since $\mng_i$ in the iteration cannot remove beliefs (see 3.\ of Definition~\ref{def:reducible}), we obtain $\KB^{\alpha+1}_i\subseteq {\KB'}^{\alpha+1}_i$ and thus $\BelS_i^{\alpha+1}\subseteq {\BelS'}_i^{\alpha+1}$.
\item Consider a limit ordinal $\alpha'$ with $\alpha\leq \alpha'$. 
 As the corresponding knowledge bases are simply the union of those of all smaller ordinals, we necessarily have $\KB^{\alpha'}_i\subseteq {\KB'}^{\alpha'}_i$, and thus $\BelS_i^{\alpha'}\subseteq {\BelS'}_i^{\alpha'}$.
\end{itemize}
Since we already know that $\TBelS$ is a grounded equilibrium, we have that $\BelS_i=\BelS_i^\infty$ for all $i$.
We conclude that $\BelS'_j\subset{\BelS'}_j^\infty$ holds for at least one $j\in\{1,\ldots,\CtxtIn\}$.
Thus, $\TBelS'$ itself cannot be a grounded equilibrium of $M$ given $\TKB$ and $\Tinpt$ by Definition~\ref{def:groundedRed}.

So suppose that $\TBelS'$ is an equilibrium, but not grounded.
Then we know that, by Definition~\ref{def:Equilibrium}, for all $i\in\{1,\ldots,\CtxtIn\}$: 
\begin{equation}
\BelS'_i\in \acc_{i}(\KB'),\text{ where }\KB'=\mng_i(\app^{now}_i(\Tinpt,\TBelS'),\KB_i).
\end{equation}
In this case, by the definition of $\redfun$, we have for all $i$:
\begin{equation}
\acc_i(\redfun_i(\KB',\BelS'_i))=\{\BelS'_i\}
\end{equation}
Therefore, by 2.\ of Definition~\ref{def:reducible}:
\begin{equation}
\acc_i(\mng_i(\app^{now}_i(\Tinpt,\TBelS'),\redfun_i(\KB_i,\BelS'_i)))=\{\BelS'_i\}
\end{equation}
Now, the $\KB_i^0$ used in the iteration of $\BelS_i^\alpha$ are precisely the $\redfun_i(\KB_i,\BelS'_i)$ for all $i$.
As argued in the proof of Lemma~\ref{lemma:iterationProps}, the sequence of operations in the iteration is monotonically increasing.
We consider two cases. 
First, the sequence reaches $\app^{now}_i(\Tinpt,\TBelS')$.
Then, for all $\SOp^j\subseteq \app^{now}_i(\Tinpt,\TBelS')$, $\acc_i(\mng_i(\SOp^j,\KB^\alpha_i))\subseteq \BelS'_i$.
As $\BelS'_j\subset{\BelS'}_j^\infty$ holds for at least one $j\in\{1,\ldots,\CtxtIn\}$, the iteration has to continue beyond the step involving $\app^{now}_i(\Tinpt,\TBelS')$, but this contradicts $\TBelS'$ being an equilibrium.
Alternatively, the sequence does not reach a step with precisely the operations in $\app^{now}_i(\Tinpt,\TBelS')$.
In this case, there is a least ordinal such that an operation not contained in $\app^{now}_i(\Tinpt,\TBelS')$ occurs in the iteration, but this again contradicts that $\TBelS'$ is an equilibrium following the argument applied in the proof of Proposition~\ref{prop:GEuniqueForDefinite}.
\end{pf}

\begin{proposition*}
{prop:relWFSvsGrEq}
Let $M=\tuple{\TCtxt, \TIL, \TBR}$ be a normal, reducible \rMCS given $\TKB$, $\Tinpt$ an input for $M$, $\wfs(M,\TKB,\Tinpt)=\tuple{W_1,\ldots W_n}$, and $\TBelS=\tuple{\BelS_1,\ldots,\BelS_n}$ a grounded equilibrium of $M$ given $\TKB$ and $\Tinpt$.
Then $W_i\subseteq \BelS_i$ for $i\in\{1\ldots\CtxtIn\}$.
\end{proposition*}
\begin{pf}
We show the claim by proving that $(\gamma_{M,\TKB,\Tinpt})^2\uparrow \alpha)_i\subseteq \BelS_i$ holds for all grounded equilibria $\TBelS$ of $M$ given $\TKB$ and $\Tinpt$.
For the case $\alpha=0$, this holds trivially, as the initial belief state is $\TBelS^*$.

Suppose the claim holds for all ordinals $\alpha_1\leq \alpha$.
Consider a successor ordinal $\alpha+1$.
If $(\gamma_{M,\TKB,\Tinpt})^2\uparrow \alpha=(\gamma_{M,\TKB,\Tinpt})^2\uparrow (\alpha+1)$ then the claim trivially holds by the induction hypothesis.
Thus, suppose that $(\gamma_{M,\TKB,\Tinpt})^2\uparrow \alpha\subset (\gamma_{M,\TKB,\Tinpt})^2\uparrow (\alpha+1)$, i.e., new beliefs are added in the step $\alpha+1$ for (at least) some $i$.
Using Definition~\ref{def:definiteIteration} and the induction hypothesis, it can be shown that these new beliefs are indeed contained in each grounded equilibrium, while all beliefs that do not occur in the intermediate fixpoint of the squared operator are not true in any grounded equilibrium of $M$ given $\TKB$ and $\Tinpt$, in the very same way as usual for the alternating fixpoint, as e.g., for logic programs, on which this construction is based.

Finally, regarding a limit ordinal $\alpha'$ with $\alpha\leq\alpha'$, the claim holds trivially by the induction hypothesis.
\end{pf} \section{Proofs Complexity}\label{app:tm}
\begin{theorem*}
{theo:genCompl}
Table~\ref{tab:complexity} summarizes the complexities of membership of problems $\problemE$ and $\problemA$ for finite input steams (until some $\tau\in\mathbb{N}$) depending on the context complexity.
Hardness also holds if it holds for the context complexity.
\end{theorem*}
\begin{pf}
The membership results for the $\problemE$ cases (with the exception of $\mathcal{CC}(M,\BRBA{k}{\Bel})=\mathbf{EXPTIME}$) can be argued for as follows: a non-deterministic Turing machine can be used to guess
a projected belief state $\EqStr^t=\tuple{\BelS_1,\dots,\BelS_n}$ for all $\tau$ inputs in $\inptStr$ in polynomial time.
Then, iteratively for each of the consecutive inputs $\inptStr^t$,
first the context problems can be solved either polynomially or using an oracle for the context complexity
(the guess of $\EqStr^t$ and the oracle guess can be combined which explains why we stay on the same complexity level for
higher context complexity).
If the answer is 'yes', $\EqStr^t$ is the projected equilibrium. 
We can check whether $\Bel\in \BelS_i$, compute the updated knowledge bases
and continue the iteration until reaching the last input. 
For $\mathbf{PSPACE}$ the same line of argumentation holds as $\mathbf{PSPACE}=\mathbf{NPSPACE}$.
In the case of $\mathcal{CC}(M,\BRBA{k}{\Bel})=\mathbf{EXPTIME}$, we iterate through the exponentially many projected belief states
for which we solve the context problem in exponential time and proceed as before.
The argument is similar for the co-problem of $\problemA$.
Hardness holds because being able to solve $\problemE$, respectively the co-problem of $\problemA$, one can decide equilibrium existence for managed MCSs which is hard for the same complexity classes~\cite{BrewkaEFW11} given hardness for the context complexity of the managed MCS.
\end{pf}

\begin{theorem*}
{theo-reducMCScompl}
Let $M=\tuple{\TCtxt, \TIL, \TBR}$ be a finite, persistently reducible \rMCS given $\TKB$.
Then membership and hardness results for $\problemE_g$ and $\problemA_g$ on grounded equilibria streams for finite input streams coincide with those for $\problemE$ and $\problemA$ in Theorem~\ref{theo:genCompl}. 
\end{theorem*}
\begin{pf}
As $M$ is assumed to be finite and persistently reducible, the argument is exactly identical to that of Theorem \ref{theo:genCompl}. 
The only difference is here that we also have to guess the intermediate belief states in the iteration, but this does not raise the (worst-case) complexity itself. 
\end{pf}

\begin{theorem*}
{theo-WFScompl}
Let $M=\tuple{\TCtxt, \TIL, \TBR}$ be a finite, normal, persistently reducible \rMCS given $\TKB$ such that $\mathcal{CC}(M,\BRBA{k}{\Bel})=\mathbf{P}$ for $\problemE_{wf}$.
Then, $\problemE_{wf}$ is in $\mathbf{P}$.
In addition, hardness holds provided $\mathcal{CC}(M,\BRBA{k}{\Bel})=\mathbf{P}$ is hard.
\end{theorem*}
\begin{pf}
If $\mathcal{CC}(M,\BRBA{k}{\Bel})=\mathbf{P}$, then deciding all context problems is polynomial. 
Also, no guessing of equilibria is required, as each equilibrium can be computed iteratively in polynomial time starting from the belief state containing all least elements of each belief set.
Since the context complexity also determines the complexity of performing the update function to obtain the new configuration of knowledge bases, we conclude that the result holds.
The result for hardness follows in the same manner as in the proofs of Theorems~\ref{theo:genCompl} and \ref{theo-reducMCScompl}.
\end{pf} 

Finally, to demonstrate the expressiveness of \rMCSs and prove Proposition~\ref{prop:undecidability} 
we provide an
\rMCS $\MTM$ that implements a service that reads
the configuration of a Turing machine (TM) from external sources
and simulates a run of the TM on request.
The fixed \rMCS $\MTM$ comes with inexpressive context logics
and management functions computable in linear time,
thus showing that Turing-completeness arises from the interplay between the contexts over time.
We use the following variant of TMs,
where \TMsym is the set of all tape symbols and \TMstates is the set of all states:
\begin{definition}
A TM is a tuple $\tuple{Q,\TMalph,\TMblank,\TMinput,\TMfunc,q_0,F}$, where
\bi
     \item $Q\subseteq\TMstates$ is a finite, non-empty set of states,
     \item $\TMalph\subseteq\TMsym$ is a finite, non-empty set of tape symbols,
     \item $\TMblank\in\TMalph$ is the blank symbol,
     \item $\TMinput\subseteq\TMalph$ is the set of input symbols,
     \item $q_0$ is the initial state,
     \item $F\subseteq Q$ is the set of final states, and
     \item $\TMfunc:Q\setminus F\times\TMalph\ra Q\times\TMalph\times\set{\la,\ra}$ is the (partial) transition function.
\ei
\end{definition}
We assume familiarity with the computation model of TMs, in particular
what it means that a TM halts and accepts an input word $w\in\TMinput^*$.
$\MTM$ is an \rMCS with four contexts.
Context $\Ctxt_{t}$ simulates a tape of a TM,
$\Ctxt_{q}$ contains information about TM states,
$\Ctxt_{f}$ encodes a transition function, and
$\Ctxt_{c}$ is a control context for operating the TM simulation and presenting results.
All contexts 
use a storage logic as in Example~\ref{ex-logics},
where the set of entries is respectively given by
\bi
  \item $E_t=\bigcup\limits_{p\in\mathbb{Z},s\in\TMsym}\set{\lit{t}{p,s},\lit{curP}{p}}$,
  \item $E_q=\bigcup\limits_{q\in\TMstates}\set{\lit{final}{q},\lit{curQ}{q}}$,
  \item $E_f=\set{\lit{f}{q,s,q',s',m} \mid q,q'\in\TMstates,s,s'\in\TMsym,m\in\set{L,R}}$, and
  \item $E_c=\set{\pred{computing},\lit{answer}{yes},\lit{answer}{no}}$.
\ei  

The input of \MTM is provides by four streams over the languages
\bi
  \item $\IL_{t}=E_t$,
  \item $\IL_{q}=E_q$,
  \item $\IL_{f}=E_f$, and
  \item $\IL_{c}=\set{\pred{start},\pred{reset}}$,
\ei
where $\IL_{t}$ allows for setting an initial configuration for the tape,
$\IL_{q}$ the allowed and final states of the TM to simulate,
$\IL_{f}$ the transition function, and input over $\IL_{c}$ is used to start and reset the simulation.

The bridge rule schemata of the tape context $\Ctxt_{t}$ are given by:
$$
  \begin{array}{rcl}
      \Nxt{\op{add(\lit{t}{P,S'})}}	&\la&\BRBA{{q}}{\lit{f}{Q,S,Q',S',D}},\BRBA{{t}}{\lit{curP}{P}},\BRBA{q}{\lit{curQ}{Q}},\\&&
                                    \BRBA{t}{\lit{t}{P,S}},S\neq S',\BRBA{{c}}{\pred{computing}}.\\[3pt]
     \Nxt{\op{rm(\lit{t}{P,S})}}	&\la&\BRBA{{q}}{\lit{f}{Q,S,Q',S',D}},\BRBA{{t}}{\lit{curP}{P}},\BRBA{q}{\lit{curQ}{Q}},\\&&
                                    \BRBA{t}{\lit{t}{P,S}},S\neq S',\BRBA{{c}}{\pred{computing}}.\\[3pt]
      \op{add(\lit{nextP}{P-1})}&\la&  \BRBA{{q}}{\lit{f}{Q,S,Q',S',\la}},\BRBA{{q}}{\lit{curQ}{Q}},\\&&
                                    \BRBA{{t}}{\lit{t}{P,S}},\BRBA{{t}}{\lit{curP}{P}},\BRBA{{c}}{\pred{computing}}.\\[3pt]
      \op{add(\lit{nextP}{P+1})}&\la&  \BRBA{{q}}{\lit{f}{Q,S,Q',S',\ra}},\BRBA{{q}}{\lit{curQ}{Q}},\\&&
                                    \BRBA{t}{\lit{t}{P,S}},\BRBA{{t}}{\lit{curP}{P}},\BRBA{{c}}{\pred{computing}}.\\[3pt]
      \op{add(\pred{nextPdefined})}&\la&  \BRBA{t}{\lit{nextP}{P}},\BRBA{{c}}{\pred{computing}}.\\[3pt]
     \Nxt{\op{add(\lit{curP}{P})}}&\la&  \BRBA{t}{\lit{nextP}{P}},\BRBA{{c}}{\pred{computing}}.\\[3pt]
     \Nxt{\op{rm(\lit{curP}{P})}} &\la&  \BRBA{t}{\lit{curP}{P}},\BRBA{{t}}{\pred{nextPdefined}},\BRBA{{c}}{\pred{computing}}.\\[3pt]
     \Nxt{\op{add(X)}} &\la&  \BRSA{t}{X},\naf\ \BRBA{{c}}{\pred{computing}}.\\[3pt]
      \Nxt{\op{clear}}&\la&\BRSA{c}{\pred{reset}}.
    \end{array}
$$
\noindent
The bridge rule schemata for the state context $\Ctxt_{q}$ are the following:
$$
  \begin{array}{rcl}
      \Nxt{\op{add(\lit{curQ}{Q'})}}	&\la&\BRBA{{q}}{\lit{f}{Q,S,Q',S',D}},\BRBA{{t}}{\lit{curP}{P}},\BRBA{q}{\lit{curQ}{Q}},\\&&
                                    \BRBA{t}{\lit{t}{P,S}},Q\neq Q',\BRBA{{c}}{\pred{computing}}.\\[3pt]
     \Nxt{\op{rm(\lit{curQ}{Q})}}	&\la&\BRBA{{q}}{\lit{f}{Q,S,Q',S',D}}, \BRBA{{t}}{\lit{curP}{P}},\BRBA{q}{\lit{curQ}{Q}},\\&&
                                   \BRBA{t}{\lit{t}{P,S}},Q\neq Q',\BRBA{{c}}{\pred{computing}}.\\[3pt]
      \Nxt{\op{add(X)}} &\la&  \BRSA{q}{X},\naf\ \BRBA{{c}}{\pred{computing}}.\\[3pt]
      \Nxt{\op{clear}}&\la&\BRSA{c}{\pred{reset}}.
    \end{array}
$$
\noindent
The state $\Ctxt_{f}$ for the transition function has the bridge rules schemata given next:
$$
  \begin{array}{rcl}
      \Nxt{\op{add(X)}} &\la&  \BRSA{f}{X},\naf\ \BRBA{{c}}{\pred{computing}}.\\[3pt]
      \Nxt{\op{clear}}&\la&\BRSA{c}{\pred{reset}}.
    \end{array}
$$
\noindent
Finally, the schemata for $\Ctxt_{c}$ are:
$$
  \begin{array}{rcl}
      \op{add(\lit{answer}{'Y'})} &\la&  \BRBA{q}{\lit{curQ}{Q}}, \BRBA{q}{\lit{finalQ}{Q}},\BRBA{{c}}{\pred{computing}}.\\[3pt]
      \op{add(\lit{answer}{'N'})} &\la&  \naf\ \BRBA{t}{\lit{nextPdefined}{Q}},\BRBA{q}{\lit{curQ}{Q}}, \\
      & & \naf\  \BRBA{q}{\lit{finalQ}{Q}},\BRBA{{c}}{\pred{computing}}.\\[3pt]
      \Nxt{\op{add(\lit{answer}{X})}} &\la&  \BRBA{c}{\lit{answer}{X}},\BRBA{c}{\pred{computing}}.\\[3pt]
      \Nxt{\op{rm(\pred{computing})}} &\la&  \BRBA{c}{\lit{answer}{X}},\BRBA{c}{\pred{computing}}.\\[3pt]
      \Nxt{\op{clear}}&\la&\BRSA{c}{\pred{reset}}.\\[3pt]
      \Nxt{\op{add(\pred{computing})}}&\la&\BRSA{c}{\pred{start}}.
    \end{array}
$$
\noindent
All contexts use the following management function:
  \begin{equation*}
  \mng(\SOp,\KB)=
    \begin{cases}
      \emptyset & \text{if } \op{clear}\in\SOp\\
    \begin{array}{@{}l@{}}
      \KB\setminus\set{X\mid \op{rm(X)}\in\SOp} \cup
      \\ \set{X\mid \op{add(X)}\in\SOp}
      \end{array} & \text{else} 
    \end{cases}
  \end{equation*}

Let $T=\tuple{Q,\TMalph,\TMblank,\TMinput,\TMfunc,q_0,F}$ be a TM and $w\in\TMinput^*$ an input word for $T$.
We want to use $\MTM$ with input stream \inptStr to simulate $T$.
Assume we start at time $t$. We first make sure that all knowledge bases are empty by setting $\inptStr_c^{t}=\set{\pred{reset}}$.
This activates the bridge rules in all contexts of $\MTM$ that derive $\Nxt{\op{clear}}$.
As a consequence, at time $t+1$ the contents of all knowledge bases are deleted.
Next, we feed $T$ and $w$ to $\MTM$ by sending 
\bi
\item $\lit{final}{q}$ for all $q\in F$ and $\lit{curQ}{q_0}$ on the input stream $q$,
\item $\lit{f}{q,s,q',s'}$ iff $\TMfunc(q,s)=\tuple{q',s'}$ on stream $f$, and 
\item $\lit{curP}{0}$ and $\lit{t}{p,s}$ iff $s=s_p$ for $w=s_0,s_1,s_2,\dots$ on the tape stream $t$.
\ei
Note that it does not matter whether we do this all at once at time $t+1$ or scattered over multiple time points greater than $t$.
Assume that we finished to incorporate all this information to the knowledge bases at time $t'$.
Then, we set $\inptStr_c^{j}=\set{\pred{start}}$ to initiate the simulation of $T$.
At time $t'+1$ the entry $\pred{computing}$ is contained in the knowledge base of context $\Ctxt_{c}$,
activating the bridge rules in all contexts that are responsible for the simulation.
From now on, depending on the current state $\lit{curQ}{q}$ and the transition function, the bridge rules of tape context $\Ctxt_{t}$
always change the content of the tape on the current tape position indicated by $\lit{curP}{p}$.
A new position $p'$ of the tape head indicated by the transition function is reflected by deriving $\lit{nextP}{p'}$. 
If such a belief is in the equilibrium so is $\pred{nextPdefined}$ and $\lit{curP}{p'}$ is added at the next time point.
For context $\Ctxt_{q}$ the current state is updated according to the transition function.
Note that the auxiliary belief $\pred{nextPdefined}$ is also used in the bridge rules of context $\Ctxt_{c}$ for indicating
that if the current state is not final and the transition function is undefined for the current state and input symbol, then the answer of the TM is no, indicated by $\lit{answer}{'N'}$.
Conversely, if we arrive at a final state then $\lit{answer}{'Y'}$ is derived.
If $T$ does not halt on input $w$, then also the simulation in $\MTM$ will continue forever, unless we stop the computation by sending 
$\pred{reset}$ on input stream $\inptStr_c$ once more.
 \section{Comparative Studies}\label{app:related}

In this part, selected related approaches are compared to \rMCS in more detail by demonstrating how they can be implemented in our framework.

\subsection{Reactive Multi-Context Systems and Stream Reasoning}\label{app:relatedStream}
In the following, we relate \rMCSs to two the frameworks for stream reasoning discussed in Section~\ref{sec:relatedStream}
by demonstrating how tasks suited for these approaches can be solved by an \rMCS.
Here, we focus at presenting the intuition on how problems can be modeled
in the different approaches rather than presenting formal translations between them. As the involved frameworks are quite general, it would only make sense to look at technical mappings for well-defined restrictions of the general settings, which would limit their value for a general comparison.
Moreover, such formal translations would have to mainly deal with bridging technical differences in semantics (e.g., between the equilibrium semantics of \rMCS and the FLP semantics used in \cite{BeckDEF15})
rather than giving insight into how we can model stream reasoning tasks in the respective approaches.

\subsubsection*{LARS}
The LARS language~\cite{BeckDEF15} is built over a set of atoms $\mathcal{A}$ defined over disjoint sets of predicates $\mathcal{P}$ and constants $\mathcal{C}$ as usual, where $\mathcal{P}$ is divided into two disjoint subsets, the \emph{extensional predicates} $\mathcal{P}^{\mathcal{E}}$, and the \emph{intensional predicates} $\mathcal{P}^{\mathcal{I}}$. The former is intended to be used for input streams, and the latter for intermediate and output streams.

Given $i,j\in \mathbb{N}$, an \emph{interval} is a set of the form $[i,j]=\{k\in \mathbb{N}\mid i\leq k \leq j\}$. An \emph{evaluation function} over an interval $T$ is a function $v:\mathbb{N}\to 2^{\mathcal{A}}$ such that $v(t)=\emptyset$ if $t\notin T$. Then, a \emph{stream} in LARS is a tuple $s=\tuple{T,v}$, where $T$ is an {interval} and $v$ is an evaluation function over $T$. A stream is a \emph{data stream} if it contains only extensional atoms.
A stream $S=\tuple{T,v}$ is a \emph{substream} of a stream $S'=\tuple{T',v'}$, denoted by $S\subseteq S'$, if $T\subseteq T'$ and $v(t)\subseteq v'(t)$ for all $t\in T$.

The formulas of LARS are defined using the following grammar:
\[\alpha:= a\mid \neg\alpha\mid \alpha\wedge \alpha\mid \alpha\vee \alpha\mid \alpha\to \alpha \mid \diamondsuit \alpha\mid \square\alpha \mid @_{t}\alpha\mid \boxplus^{\mathbf{x}}_{\iota}  \]
where $a\in \mathcal{A}$ and $t\in \mathbb{N}$.
The connectives $\neg, \wedge, \vee$ and $\to$ are the usual classical connectives.
Similarly to modal logics, the operators $\diamondsuit$ and $\square$ are used to represent that a formula holds for some and for each time instant within some interval for a given stream, respectively.
The exact operator $@_{t}$ represents the reference to some specific time $t\in \mathbb{N}$, and window operators of the form $\boxplus^{\mathbf{x}}_{\iota}$ allow focusing on more recent substreams, where $\iota$ represents the type of window and $\textbf{x}$ the tuple of parameters for the respective type.
Among the operators presented in \cite{BeckDEF15} are, for example, time-based operators $\boxplus^{n}_{\tau}$, which allow focusing on the last $n$ time instants of a given stream, or the partition-based operators $\boxplus^{{\sf idx},n}_{p}$, which first split the stream into substreams and then allow focusing on the last $n$ tuples of a particular substream.
LARS programs are based on rules composed of such LARS formulas in a way similar to those in logic programming.
More precisely, a LARS rule is of the form
\[\alpha\la \beta_1,\ldots, \beta_j, \naf\ \beta_{j+1},\ldots, \naf\ \beta_n\]
where $\alpha,\beta_1,\ldots, \beta_n$ are formulas and $\alpha$ contains only intentional predicates.

Consider a simplified version of the example in~\cite{BeckDEF15} that models a scenario where we want to reason about a tram network, including, for example, the prediction of the expected arrival time of a tram at some stop.
The key idea is that such information should not only depend on a fixed timetable, but also dynamically on real-time information about the arrival time at a previous stop and the expected travel time between stations (which heavily depends on real-time traffic jam information).
LARS allows such combination of static knowledge with a stream of information.
Here, static knowledge is composed of atoms of the form $\lit{plan}{L,X,Y,T}$, where $L$ is the line identifier, $X$ and $Y$ are consecutive tram stops on line $L$, and $T$ is the expected travel time between $X$ and $Y$, and atoms of the form $\lit{line}{Id,L}$ mean that the tram $Id$ operates on line $L$.
The stream of information contains atoms of the form $\lit{tram}{Id,X}$ meaning that tram $Id$ is at stop $X$, and atoms of the form $\lit{jam}{X}$ meaning that there is a traffic jam near station $X$.
Consider the following example of a LARS rule:
\begin{align*}
@_{T} \lit{exp}{Id,Y}\la \boxplus^{{\sf idx},1}_{p}@_{T_1}\lit{tram}{Id,X},\ &\lit{line}{Id,L},\ \naf \boxplus^{20}_{\tau}\diamondsuit\lit{jam}{X},\\
&\lit{plan}{L,X,Y,Z},\ T=T_1+Z.
\end{align*}
The intuitive idea of the above rule is that the tram is expected to arrive at stop $Y$ at time $T$ whenever five conditions are met: $i)$ the fact that tram $Id$ stopped at $X$ at time $T_1$ is the last information about stops of $Id$ in the stream; $ii)$ $X$ and $Y$ are consecutive stops of line $L$; $iii)$ $T-T_1$ is the travel time between $X$ and $Y$; $iv)$ the tram $Id$ is operating on line $L$; and $v)$ there is no information about traffic jams near station $X$ within the last 20 minutes.
Note that the use of default negation $\mathbf{not}$ allows for the representation of exceptions, whereas the partition-based operator $\boxplus^{{\sf idx},1}_{p}$ allows for the focus on the last item of the form $\lit{tram}{X}$ in the stream, and the time-based operator $\boxplus^{20}_{\tau}$ allows for the focus on the last 20 time instants of the stream.

The semantics of LARS is based on the FLP semantics of logic programs~\cite{FaberLP04}. Given an input stream $D=\tuple{T,v}$ and time point $t\in T$, each LARS program $P$ is associated with a set of streams, the \emph{answer streams} of $P$ for $D$ at $t$.
Since the semantics of a LARS program is defined for a fixed input data stream and for a particular time point, it is in fact mainly static.

We now describe how \rMCSs\ can be used to run a LARS program over a (possibly infinite) input stream.
The idea
is to define an \rMCS\ with a LARS context and a recent substream of a (possibly infinite) input stream (in the sense of rMCSs). At each time point, the knowledge base of the LARS context contains the LARS program and the relevant substream of the input stream. Then, the answer streams of the program given the available data stream and the current time point can be computed.

More precisely, we assume fixed sets of predicates $\mathcal{P}$ and constants $\mathcal{C}$, a fixed window size $w\in \mathbb{N}$ and a LARS program $P$ over $\larsAlph$, the set of atoms obtained from $\mathcal{P}$ and $\mathcal{C}$.
Let $\larsAlph^{\mathcal{E}}$ be the set of atoms that include only extensional predicates from $\mathcal{P}^{\mathcal{E}}$, and $\larsAlph^T$ be the set of time-tagged atoms, i.e., $\larsAlph^T=\{\tuple{a,t}\mid a\in \larsAlph \text{ and } t\in\mathbb{N}\}$.

Consider the \rMCS $M= \tuple{\tuple{\Ctxt_{\rm{LARS}}}, \tuple{\IL_1, Clock}, \tuple{\SBR_{\Ctxt_{\rm{LARS}}}}}$ obtained from $P$ and $w$ in the following way:

\begin{itemize}

\item $\Ctxt_{\rm{LARS}}=\tuple{\logic,\SOp, \mng}$ where

\item $\logic=\tuple{\SKB,\SBelS, \acc}$ is such that
	
\item $\SKB=\{P\cup A\cup \{\lit{now}{t}\}\mid A\subseteq \larsAlph^T\text{ and } t\in \mathbb{N} \}$

\item $\SBelS= \{S \mid S \text{ is a stream for } \larsAlph \}$

\item $\acc(kb)=\{S \mid  S \text{ is an answer stream of } P \text{ for } D^{kb} \text{ at time } t^{kb} \}$

\bi
	\item $t^{kb}=t$ such that $\lit{now}{t}\in kb$
	\item $T^{kb}=[t^{kb}-w,t^{kb}]$
	\item $v^{kb}(t)=\{a \mid  \tuple{a,t}\in kb\}\mbox{, for each } t\in \mathbb{N}$
	\item $D^{kb}=\tuple{T^{kb},v^{kb}}$
	\ei

\item $\SOp= \{\op{add(\delta)} \mid  \delta\in \larsAlph^{T}\}\cup \{\op{del(\delta)}\mid  \delta\in \larsAlph^{T}\}\cup\{\op{add}(\lit{now}{t}) \mid t
\in \mathbb{N}\}$

\item $\mng(kb,op)=(kb\cup \{\delta \mid  \op{add}(\delta)\in op\})\setminus \{\delta \mid  \op{del}(\delta)\in op\}$

\item $\IL_1=\larsAlph^{\mathcal{E}}$

\item $Clock=\mathbb{N}$

\item $\SBR_{\Ctxt_{\rm{LARS}}}$ contains the following rules for managing history:

	\bi
	\item[] $\op{add}(\lit{now}{T})\la \BRSA{clock}{{T}}$
	\item[] $\op{add}(\tuple{A,T})\la \BRSA{1}{A}, \BRSA{clock}{{T}}$
	\item[] $\op{del}(\tuple{A,T})\la \BRSA{clock}{{T'}}, T<  T'-w$
	\item[] $\Nxt{\op{add}(\tuple{A,T})}\la \BRSA{1}{A}, \BRSA{clock}{{T}}$
	\item[] $\Nxt{\op{del}(\tuple{A,T})}\la \BRSA{clock}{{T'}}, T\leq  T'-w$
	\ei
\end{itemize}

Given an input stream $\inptStr$ for $M$ and a time point $t\in \mathbb{N}$, we consider $t^{\inptStr}_t$, the unique element of stream $Clock$ at step $t$, which represents the current time at step $t$. We also consider the LARS input data stream at time $t$, $D^{\inptStr}_t=\tuple{T,v}$, such that $T=[t^{\inptStr}_t-w,t^{\inptStr}_t]$ and $v(t')=\{a\in \larsAlph^{\mathcal{E}} \mid  \text{ there exists } t''\leq t \text{ such that } t'= t^{\inptStr}_{t''} \text{ and } a\in \inptStr^{t''}_1\}$ for $t'\in T$, and $v(t')=\emptyset$ otherwise. Then, given an input stream $\inptStr$ for $M$, at each time point $t\in \mathbb{N}$, each equilibria stream $\EqStr$ for $M$ given $\TKB=\tuple{\{P\}}$ and $\inptStr$ is composed of an answer stream of $P$ for $D^{\inptStr}_t$ at time $t^{\inptStr}_t$.

Note that at each time instant the knowledge base contains only the relevant part of the (possibly infinite) input stream, meaning that information no longer valid is really discarded, and that the current time, given by the stream $Clock$, is decoupled from the time steps at which equilibria are evaluated.
For the sake of presentation, we have assumed a fixed time window $w$, yet an extension in the spirit of what we presented in Section~\ref{sec:forget} can easily be imagined.

\subsubsection*{STARQL}
Streams in the STARQL framework~\cite{OptiqueD5.1} come in the form of timestamped Description Logic assertions (called ABox assertions).
Both, input as well as answers of STARQL queries are streams of this kind.
A STARQL select expression is structured as follows.
\begin{Verbatim}[commandchars=\\\{\},codes={\catcode`$=3 \catcode`_=8 \catcode`^=7}]
SELECT \m{\mathit{selectClause}(\vec{x},\vec{y})}
FROM \m{\mathit{listOfWindowedStreamExpressions}}
USING \m{\mathit{listOfResources}}
WHERE \m{\Psi(\vec{x})}
SEQUENCE BY \m{\mathit{seqMethod}}
HAVING \m{\Phi(\vec{x},\vec{y})}
\end{Verbatim}
For a comprehensive description of syntax and semantics of STARQL, we refer to Optique Deliverable~5.1 \cite{OptiqueD5.1}.
We make use of an example (partly taken from~\cite{OptiqueD5.1}) to explain components of STARQL queries and to describe the core aspects that are relevant for us.
It is based on the following STARQL query that creates an output stream  \verb|S_out| that indicates temperature sensors whose readings were monotonically increasing for the past two seconds.
\begin{Verbatim}[framesep=3mm,samepage=false]
CREATE STREAM S_out AS

SELECT {?sens rdf:type MonIncTemp}<NOW>
FROM S 0s<-[NOW-2s, NOW]->1s
USING  STATIC ABOX <http://example.org/Astatic>,
       TBOX <http://example.org/TBox>
WHERE { ?sens rdf:type TempSensor }
SEQUENCE BY StdSeq AS SEQ1
HAVING FORALL i<= j in SEQ1 ,x,y:
       IF ( { ?sens rd ?x }<i> AND { ?sens rd ?y }<j> )
       THEN ?x <= ?y
\end{Verbatim}
The considered input stream contains readings of sensors $s_0$, $s_1$:
\[
\begin{array}{l@{}l}
S=\{&
\mathit{rd}(s_0,90)\langle0s\rangle\\&
\mathit{rd}(s_1,30)\langle0s\rangle\\&
\mathit{rd}(s_0,93)\langle1s\rangle\\&
\mathit{rd}(s_1,32)\langle1s\rangle\\&
\mathit{rd}(s_0,94)\langle2s\rangle\\&
\mathit{rd}(s_0,91)\langle3s\rangle\\&
\mathit{rd}(s_0,93)\langle4s\rangle\\&
\mathit{rd}(s_0,95)\langle5s\rangle
\}
\end{array}
\]
In addition, the TBox at \url{http://example.org/TBox}
contains the axiom
\[
\mathit{BurnerTipTempSensor}\sqsubseteq\mathit{TempSensor}
\]
stating that every $\mathit{BurnerTipTempSensor}$ is a temperature sensor,
and the ABox at \url{http://example.org/Astatic} contains the assertion
\[
\mathit{BurnerTipTempSensor}(s_0)
\]
stating that $s_0$ is of type $\mathit{BurnerTipTempSensor}$.
The other sensor $s_1$ is thus not (derivable to be) a temperature sensor.

Taking $S$ as input, the query returns the output stream, represented here as a sequence of timestamped ABox assertions:
\[
\begin{array}{l@{}l}
S_\mathit{out}=\{&
\mathit{MonIncTemp}(s_0)\langle 0s\rangle\\&
\mathit{MonIncTemp}(s_0)\langle 1s\rangle\\&
\mathit{MonIncTemp}(s_0)\langle 2s\rangle\\&
\mathit{MonIncTemp}(s_0)\langle 5s\rangle
\}
\end{array}
\]
We observe that temperature sensor $s_0$ had two seconds of monotonic readings at time points $0s$, $1s$, $2s$, and $5s$.
The \verb|FROM| part of the query specifies that we consider $S$ as input where
the window expression \verb|0s<-[NOW-2s, NOW]->1s| states that, at time point $t$ of the evaluation, we are interested in assertions with a time stamp between $t-2s$ and $t$.
The slide parameter (specified by \verb|->1s|) expresses that every second the window moves forward in time.
We also assume a pulse of one second, \iec one query evaluation per second.
In STARQL, such a pulse can be defined by a pulse expression declared separately from the \verb|SELECT| query.
The so-called slack parameter for handling out-of-order stream assertions is assumed to be zero (\verb|0s<-|) and not used in the initial version of STARQL.

The input stream data that falls into the window is gathered in a \define{temporal ABox}, \iec a set of timestamped ABox assertions:

{\small
\[
\begin{array}{cl}
\mbox{Time} & \mbox{Temporal ABox}\\
\hline
0s & \{\mathit{rd}(s_0,90)\langle0s\rangle, \mathit{rd}(s_1,30)\langle0s\rangle\}\\
1s & \{\mathit{rd}(s_0,90)\langle0s\rangle, \mathit{rd}(s_1,30)\langle0s\rangle,
       \mathit{rd}(s_0,93)\langle1s\rangle, \mathit{rd}(s_1,32)\langle1s\rangle\}\\
2s & \{\mathit{rd}(s_0,90)\langle0s\rangle, \mathit{rd}(s_1,30)\langle0s\rangle,
       \mathit{rd}(s_0,93)\langle1s\rangle, \mathit{rd}(s_1,32)\langle1s\rangle,
       \mathit{rd}(s_0,94)\langle2s\rangle\}\\
3s & \{\mathit{rd}(s_0,93)\langle1s\rangle, \mathit{rd}(s_1,32)\langle1s\rangle,
       \mathit{rd}(s_0,94)\langle2s\rangle, \mathit{rd}(s_0,91)\langle3s\rangle\}\\
4s & \{\mathit{rd}(s_0,94)\langle2s\rangle, \mathit{rd}(s_0,91)\langle3s\rangle,
       \mathit{rd}(s_0,93)\langle4s\rangle\}\\
5s & \{\mathit{rd}(s_0,91)\langle3s\rangle,
       \mathit{rd}(s_0,93)\langle4s\rangle,\mathit{rd}(s_0,95)\langle5s\rangle
\}
\end{array}
\]
}
Variable bindings for $\vec{x}$ in the condition $\Psi(\vec{x})$ of the \verb|WHERE| clause of a query are determined by finding the certain answers for $\vec{x}$
with respect to the ontology formed by the static ABoxes and TBoxes provided by the \verb|USING| clause.
In our example, that means that the variable \verb|?sens| is bound to the sensor named $s_0$,
because the TBox at \url{http://example.org/TBox} together with ABox \url{http://example.org/Astatic} identify only this sensor to be a temperature sensor.

The \verb|SEQUENCE| directive of the query defines how to cluster the information in the temporal ABox obtained as above into a sequence of ABoxes.
In the basic STARQL version, the only method for doing so is \verb|StdSeq| which separates the assertions based on their timestamp.
For example, at time $2s$ we get a sequence of three ABoxes:
$\{\mathit{rd}(s_0,90), \mathit{rd}(s_1,30)\}\langle1\rangle$,
$\{\mathit{rd}(s_0,93), \mathit{rd}(s_1,32)\}\langle2\rangle$, and
$\{\mathit{rd}(s_0,94)\langle3\rangle\}$.
Note that the absolute timestamps are replaced by natural numbers marking the ordinal position of the ABox in the sequence.

The \verb|HAVING| clause of the query specifies a condition $\Phi(\vec{x},\vec{y})$ that has to hold for this sequence of ABoxes, taking also the ontology parts from the \verb|USING| clause into consideration. The variables in $\vec{x}$ have already been bound by the \verb|WHERE| clause, so here \verb|?sens| is being bound to $s_0$.
The \verb|HAVING| condition can either bind the remaining open variables in $\vec{y}$ or, as in the example, act as a boolean condition when there are no further open variables.
Here, the condition is true if, for all numbers \verb|i|, \verb|j| of the sequence (that is from $\{1,2,3\}$) with $\verb|i|\leq\verb|j|$, the condition holds that if there are
sensor readings of $s_0$ in the ABoxes with sequence positions \verb|i| and \verb|j|,
then the reading in ABox \verb|i| is smaller or equal than that of ABox \verb|j|.

Finally, the expression following the \verb|SELECT| keyword determines the form of the output stream.
For every evaluation time in which the \verb|HAVING| clause holds, an ABox assertion in the form of an RDF triple
\verb|{?sens rdf:type MonIncTemp}<NOW>| augmented with a timestamp is returned where \verb|?sens| is replaced by $s_0$ and \verb|NOW| by the evaluation time.

\begin{figure}[t!]\centering
\includegraphics[width=0.9\textwidth]{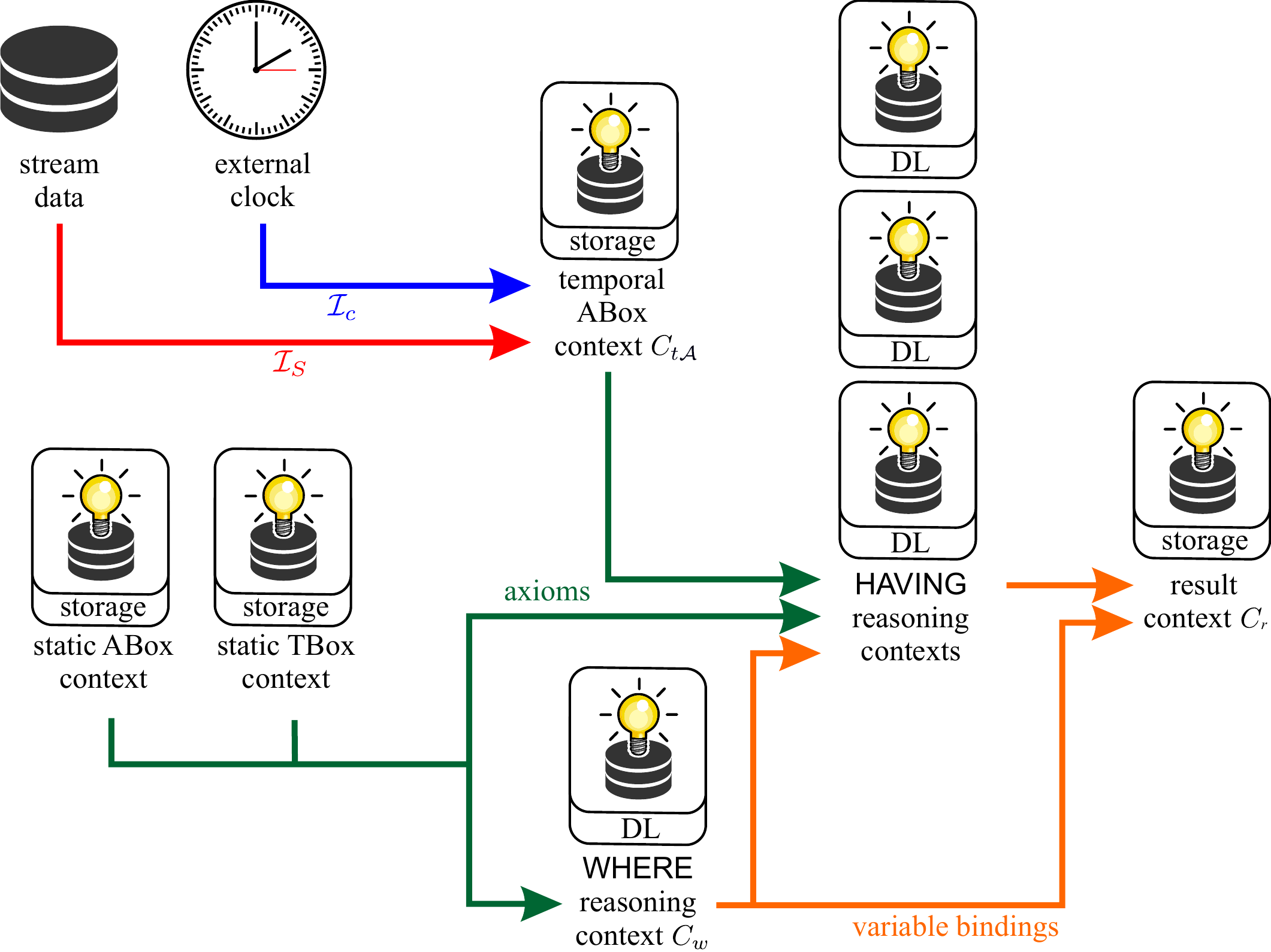}
\caption{Structure of an \rMCS implementing a STARQL query with an external clock.}
\label{fig:starqlrmcs}
\end{figure}
We want to illustrate next how STARQL queries can be realized as \rMCSs (Fig.~\ref{fig:starqlrmcs}).
Certainly, due to the abstract nature of \rMCSs, there are many different ways to do so.
Here, we aim at a setting where we assume a correspondence of one equilibrium computation per STARQL query evaluation,
\iec in our example there is one equilibrium evaluation per second.
We assume one \rMCS input stream $\inptStr_S$ that corresponds to the STARQL input stream
and one \rMCS input stream $\inptStr_c$ that provides the current time as in Section~\ref{sec:time}.
The data in $\inptStr_S$ can either be timestamped assertions such as in the original STARQL approach,
or unstamped raw data (such as pure sensor values) that will be automatically
time-stamped by bridge rules using the current time provided by $\inptStr_c$.
While in the former approach it is possible to have timestamps in the current input that differs from the current time,
such a situation is avoided with unstamped data. Both variants have their place and purpose.
Here, we assume unstamped data.

We reserve one \rMCS context $C_{t\abox}$ for the temporal ABox containing the data currently in the window.
Its bridge rules add new data from $\inptStr_S$ and timestamp it similar as in Section~\ref{sec:time}.
Moreover, other bridge rules ensure that data not ``visible'' in the window is deleted from $C_{t\abox}$.

There is one context $C_{w}$ that computes the variable bindings of the \verb|WHERE| clause.
Its knowledge base is a description logic ontology and its bridge rules import assertions and axioms according to the \verb|USING| clause of the STARQL query.
Thus, we assume that each static ABox and TBox mentioned in the query is available in a corresponding storage context of the \rMCS.
These axioms and assertions are also imported to DL contexts for evaluating the \verb|HAVING| clause:
we assume as many such contexts as there can be ABoxes in the sequence of ABoxes created by the \verb|SEQUENCE| statement.
In our example, we have three contexts of this type because there are assertions with at most three different timestamps in the temporal ABox representing the current window.
Thus, the sequence operator \verb|StdSeq| generates ABox sequences of at most size three.
The reasoning contexts evaluate the different timepoints in the window.
Fore example, the $i$-th context computes assignments for variable \verb|?x| of the expression \verb|{ ?sens rd ?x }<i>|.
We still need to reason on top of the results for the individual DL contexts for fully evaluating the \verb|HAVING| clause:
In the example, this is the evaluation of the \verb|FORALL| condition that ranges over different time points.
This can be done with another context $C_{r}$ that also combines the results of the \verb|WHERE| and \verb|HAVING| clauses and produces output as specified
in the \verb|SELECT| statement. Thus, in an equilibrium for time $t$, the belief set for $C_{r}$ contains timestamped RDF triples of form \verb|{s0 rdf:type MonIncTemp}<t>| whenever
$s_0$ had monotonically increasing readings during the previous two seconds.

\subsection{EVOLP}\label{app:relatedEvolp}

We show that \rMCSs\ are expressive enough to capture EVOLP~\cite{AlferesBLP02} (discussed in Section~\ref{sec:evolp}) as a particular case.
We assume a basic familiarity with logic programs as such, and only point out that a general logic program is a set of logic program rules that allow default negation in the head.
\emph{Evolving logic programs} are then defined as general logic programs built over a special language which allows them to express self-evolution.
The language includes a reserved unary predicate, assert, whose argument may itself be a full-blown rule, thus making arbitrary nesting possible.
Formally, given a propositional language $\mathcal{L}$, the extended language $\mathcal{L}_{\pred{assert}}$ over $\mathcal{L}$ is defined inductively as follows:

\begin{enumerate}

\item All propositional atoms in $\mathcal{L}$ are atoms in $\mathcal{L}_{\pred{assert}}$;

\item if $R$ is a rule over $\mathcal{L}_{\pred{assert}}$ then $\lit{assert}{R}$ is an atom in $\mathcal{L}_{\pred{assert}}$;

\item nothing else is an atom in $\mathcal{L}_{\pred{assert}}$.

\end{enumerate}
 An \emph{evolving logic program} over $\mathcal{L}$ is a generalized logic program over $\mathcal{L}_{\pred{assert}}$.
 We denote by $\mathcal{R}_{\mathcal{L}}$ the set of all rules over $\mathcal{L}_{\pred{assert}}$.

The idea of EVOLP is that programs can update their own rules thus describing their possible self-evolution.
Each self-evolution can be represented by a sequence of programs, each program corresponding to a state, and these sequences of programs can be treated as in Dynamic Logic Programs (DLPs)~\cite{AlferesLPPP00}.
Dynamic logic programs are sequences $P_1\oplus\ldots\oplus P_n$ of generalized logic programs, whose semantics is based on the causal rejection principle.
The idea is that the most recent rules are put in force, (where $P_n$ is to be seen as the most recent set of rules),
and the previous rules are valid as far as possible, i.e., they are only kept if they do not conflict with more recent rules.
Here, these intuitions about DLPs are sufficient, and we point to~\cite{AlferesLPPP00} for more details.

The semantics of evolving logic programs is based on sequences of interpretations. More precisely,
an \emph{evolution interpretation} of length $m$ of an evolving logic program $P$ over a propositional language $\mathcal{L}$ is a finite sequence $I=\tuple{I^1,\ldots,I^m}$ of sets of propositional atoms of $\mathcal{L}_{\pred{assert}}$. The evolution trace associated with an evolution interpretation $I$ is the sequence of programs $P_I=\tuple{P^1,\ldots,P^m}$ such that $P^1=P$ and, for each $2\leq j\leq m$, $P^j=\{r \mid \lit{assert}{r}\in I^{j-1}\}$.

Given an evolving logic program, the main intuition for the construction of a sequence of programs that corresponds to a possible evolution of $P$ is that whenever an atom $\lit{assert}{r}$ belongs to an interpretation in a sequence, then the rule $r$ must belong to the program in the next state.

Besides self-evolution, evolving logic programs also consider evolution caused by the addition of external rules. These rules, called events, are represented as a sequence of evolving logic programs.
Given an evolving logic program $P$ over $\mathcal{L}$, such an \emph{event sequence over $P$} is a sequence of evolving logic programs over $\mathcal{L}$.

This leads to the central model notion of evolving logic programs that also takes into account an incoming event sequence.
Let $P$ be an evolving logic program over $\mathcal{L}$, $I=\tuple{I^1,\ldots,I^m}$ an evolution interpretation of length $m$ of $P$ with evolution trace $P_I=\tuple{P^1,\ldots,P^m}$, and $E=\tuple{E^1,\ldots,E^\ell}$ an event sequence over $P$ such that $\ell\geq m$.
Then, $I$ is an \emph{evolution stable model} of $P$ given $E$ iff for every $1\leq j\leq m$, we have that $I^j$ is a stable model of $P^1\oplus P^2\oplus\ldots\oplus(P^j\cup E^j)$.

We now show how EVOLP can be captured in the framework of \rMCSs. For that, given an evolving logic program $P$ we aim to construct an \rMCS\ $M_P$ whose equilibria streams corresponds to the evolution stable models of $P$.
First of all, note that the events in EVOLP can be modeled by the input streams of \rMCSs.

The intuitive idea for the construction of $M_P$ is that, at each instant, the incoming events are temporarily added to a knowledge base. To keep track of these, we consider a context $C$, whose possible knowledge bases are pairs. The first component of such is to keep track of the sequence of programs $P^1\oplus\ldots\oplus P^{j}$, which corresponds to the trace of an interpretation until the current instant, and the second component is reserved for the current event $E^j$.
Given this intuition, it is clear that $\acc(\tuple{P^1\oplus\ldots\oplus P^{j},E})$ should be defined as the set of stable models of $P^1\oplus\ldots\oplus P^{j-1}\oplus(P^j\cup E)$. To incorporate events, we consider an input language $\IL$ of $M_P$ defined precisely as the language of events, i.e., the set of evolving logic programs. Moreover, context $C$ contains bridge rules to access the current events, which are then used to update the event's component of the knowledge base.
Also, context $C$ has bridge rules designated to guarantee that the formulas that occur under the predicate $\pred{assert}$ at the current instant are used to update the knowledge base in the next instant.
Finally, the management function $\mng$ of $C$ is such that $\mng(op,\tuple{P,E})$ updates the program sequence component or the event component depending on whether $op$ is a set of asserts, or a set of events, respectively.

Formally, let $P$ be an evolving logic program, and consider the following \rMCS\ $M_P=\tuple{\tuple{\Ctxt},\tuple{\IL},\tuple{\SBR_{\Ctxt}}}$ where

\begin{itemize}

\item $\Ctxt=\tuple{\logic,\SOp,\mng}$ such that

		\item $\logic=\tuple{\SKB,\SBelS,\acc}$ is a logic such that

 				\item $\SKB$ is the set of pairs $\tuple{D,E}$ where $D$ is a dynamic logic program over $\mathcal{L}$, and $E$ is an evolving logic program over $\mathcal{L}$
				
				\item $\SBelS$ is the set of all subsets of $\mathcal{L}_{\pred{assert}}$
				
				\item $\acc(\tuple{P^1\oplus\ldots\oplus P^j,E})$ is the set of stable models of $P^1\oplus\ldots\oplus P^{j-1}\oplus(P^j\cup E)$ 		

		\item $\SOp=\{\lit{as}{r}\mid r\in \mathcal{R}_{\mathcal{L}}\}\cup\{\lit{ob}{r}\mid r\in \mathcal{R}_{\mathcal{L}}\}$

		\item $\mng(op,\tuple{D,E})=\{\tuple{D\oplus U,E'}\}$ where $U=\{r\in \mathcal{R}_{\mathcal{L}}\mid \lit{as}{r}\in op \}$ and $E'=\{r\in \mathcal{R}_{\mathcal{L}}\mid \lit{ob}{r}\in op \}$
		
\item $\IL=\mathcal{R}_{\mathcal{L}}$

\item $\SBR_{\Ctxt}=\{\Nxt{\lit{as}{s}}\la \BRBA{1}{\lit{assert}{s}}\  |\ \lit{assert}{s}\in \mathcal{L}_{\pred{assert}}\}\ \cup$\\
\hspace*{1.25cm}$\{\lit{ob}{s}\la \BRSA{1}{s}\ |\ s\in \IL\}$	

\end{itemize}

Evolution stable models of $P$ are strongly related to the equilibria streams of $M_P$. Namely,
given an event sequence $E=\tuple{E_1,\ldots,E_\ell}$ over $P$ consider its associated input stream $\inptStr_{E}=\tuple{\tuple{E_1},\ldots,\tuple{E_\ell}}$.
Then, $I=\tuple{I^1,\ldots,I^\ell}$ is a evolution stable model of $P$ given $E$ iff $I$ is an equilibria stream for $M_P$ given $\TKB=\tuple{\tuple{P,\emptyset}}$ and $\inptStr_{E}$.

This coincidence between evolution stable models in EVOLP and equilibria streams for \rMCSs as defined above can be explained by the fact that, conceptionally, the operators $\Nxt{}$ and $\pred{assert}$ are rather similar.

 \end{document}